\documentclass[pre,superscriptaddress,twocolumn]{revtex4-2}
\usepackage{amsfonts,amsmath,amssymb,mathtools,graphicx,balance}
\usepackage[colorlinks=true,allcolors=blue]{hyperref}
\usepackage[svgnames,dvipsnames,x11names]{xcolor}
\usepackage{bbm}
\usepackage{tikz}
\usepackage{comment}

\newcommand{\G}{\mathcal{G}}

\newcommand{\e}{\varepsilon}
\newcommand{\p}{\mathbf{p}}
\usepackage{placeins}

\definecolor{lightblue}{RGB}{17,97,165}

\newcommand{\inlinetriangle}{%
  \tikz[baseline=-0.0ex,x=1ex,y=1ex]{
    \node[fill=black,draw=black,circle,inner sep=.8pt] (b) at (-0.2,0) {};
    \node[fill=black,draw=black,circle,inner sep=.8pt] (c) at (1.2,0) {};
    \node[fill=black,draw=black,circle,inner sep=.8pt] (a) at (0.6,1.2) {}; 
    \draw (b)--(c)--(a)--cycle;
  }}

\newcommand{\inlineparlines}{%
  \tikz[baseline=-0.0ex,x=1ex,y=1ex]{
    \node[fill=black,draw=black,circle,inner sep=.8pt] (a) at (0,0)   {};
    \node[fill=black,draw=black,circle,inner sep=.8pt] (b) at (1.2,0)   {};
    \node[fill=black,draw=black,circle,inner sep=.8pt] (c) at (0,1.2) {};
    \node[fill=black,draw=black,circle,inner sep=.8pt] (d) at (1.2,1.2) {};
    \draw (a)--(b);
    \draw (c)--(d);
  }}

\makeatletter
\def\@eqnnum{{\normalsize \normalcolor (\theequation)}}
 \makeatother

\newcommand{\numrealworld}[0]{seventeen~}

\begin{document}

\title{Deterministic construction of typical networks in network models}

\author{Narayan~G.~Sabhahit$^\star$}\thanks{\href{mailto:sabhahit.n@northeastern.edu}{sabhahit.n@northeastern.edu}}
\affiliation{Network Science Institute, Northeastern University, Boston, Massachusetts, USA 02115}

\author{Moritz~Laber$^\star$}\thanks{\href{mailto:laber.m@northeastern.edu}{laber.m@northeastern.edu}}
\affiliation{Network Science Institute, Northeastern University, Boston, Massachusetts, USA 02115}
\affiliation{Complexity Science Hub Vienna, Metternichgasse 8, Vienna, 1030 Austria}

\author{Harrison~Hartle}
\affiliation{Santa Fe Institute, Santa Fe, New Mexico, USA 87505}

\author{Jasper~van~der~Kolk}
\affiliation{Department of Condensed Matter Physics, University of Barcelona, Mart\'i i Franqu\'es 1, E-08028 Barcelona, Spain}

\author{Samuel~V.~Scarpino}
\affiliation{Network Science Institute, Northeastern University, Boston, Massachusetts, USA 02115}
\affiliation{Santa Fe Institute, Santa Fe, New Mexico, USA 87505}
\affiliation{Institute for Experiential AI, Northeastern University, Boston, Massachusetts, USA, 02115}
\affiliation{Department of Public Health and Health Sciences, Northeastern University, Boston, USA, 02115}
\affiliation{Khoury College of Computer Sciences, Northeastern University, Boston, USA, 02115}
\affiliation{Vermont Complex Systems Institute, University of Vermont, Burlington, VT, USA, 05405}

\author{Brennan~Klein}
\affiliation{Network Science Institute, Northeastern University, Boston, Massachusetts, USA 02115}
\affiliation{Institute for Experiential AI, Northeastern University, Boston, Massachusetts, USA, 02115}
\affiliation{Department of Physics, Northeastern University, Boston, Massachusetts, USA 02115}

\author{Dmitri~Krioukov}\thanks{\href{mailto:dima@northeastern.edu}{dima@northeastern.edu}\\$^{\star}$ These authors contributed equally}
\affiliation{Network Science Institute, Northeastern University, Boston, Massachusetts, USA 02115}
\affiliation{Department of Physics, Northeastern University, Boston, Massachusetts, USA 02115}
\affiliation{Department of Mathematics, Northeastern University, Boston, Massachusetts, USA 02115}
\affiliation{Department of Electrical \& Computer Engineering, Northeastern University, Boston, Massachusetts, USA 02115}

\begin{abstract}
In network science, one often wants to say that a given real-world network appears to come from a particular network model. In statistical physics, the corresponding problem is about how typical a given state, representing real-world data, is in a particular statistical ensemble. One way to address this problem is to measure the distance between the data and the most typical state in the ensemble. Here, we identify the conditions that allow us to define this most typical state. These conditions hold in a wide class of grand canonical ensembles and their random mixtures. Our main contribution is a deterministic construction of a state that converges to this most typical state in the thermodynamic limit. This construction involves rounds of derandomization procedures, some of which deal with derandomizing point processes, an uncharted territory. We illustrate the construction on one particular network model, deterministic hyperbolic graphs, and its application to real-world networks, many of which we find are close to the most typical network in the model. While our main focus is on network models, our results are very general and apply to any grand canonical ensembles and their random mixtures satisfying certain niceness requirements.
\end{abstract}

\maketitle
\section{Introduction}

Modeling real-world networks is one of the core research activities in network science~\cite{cimini2019_statphysreal}, where it is often tempting to say that a particular real-world network appears ``to come from'' a particular model~\cite{watts1998_wsmodel, barabasi1999_bamodel, watanabe2013pairwise}. Such statements usually mean that the real-world network in question appears to be (close to) a \emph{typical} network in the model under consideration. This is usually supported by generating many random networks in the model, and by comparing their averaged properties of interest to the corresponding properties of the real network. 

However, it is often desirable to ``suppress noise,'' and to work not with a zoo of random networks, but with just one---the ``most representative'' or ``most typical'' one---constructed deterministically. In network science, representative graph constructions are needed to boost reliability of statistical inference, to suppress or control sample-to-sample fluctuations, and for many other purposes~\cite{cimini2019_statphysreal, anand2010gibbs, squartini2011analytical, cimini2022meta-validation, villegas2023laplacian}. More generally, the importance of representative typical states, as opposed to averages over many random realizations, has been emphasized repeatedly in many areas of statistical physics ranging from quantum typicality and thermalization to the breakdown of self-averaging near critical points and in disordered media~\cite{popescu2006entanglement, goldstein2006canonical, reimann2007typicality, rigol2008thermalization, boes2018statistical, schroder2013crackling}.  In more data-driven domains such as machine learning, stochasticity is a key source of irreproducibility, so the push towards more deterministic, reproducible benchmarks has become a major priority~\cite{henderson2018deep,NatMI2021, piantadosi2019_ReproducibilityDeepConvolutional}.

The deterministic construction of typical networks in random network models---a task reminiscent of the construction of expanders~\cite{hoory2006expander} and pseudorandom graphs~\cite{thomason1987pseudo,chung1988_quasirandomgraphs,Krivelevich2006}---requires a definition of typical networks. The definition of network typicality~\cite{park2004_statmech,bianconi2009entropy} is problematic if the probability~$\mathbb{P}(G)$ of networks~$G$ in the model is unknown or intractable, as is typically the case with network models defined by some (complicated) network-construction rules. Typical values of network properties in such models can be computed, but typical networks \textit{per se} cannot be defined because a network with typical values of some properties may have atypical values of some other unchecked properties~\cite{orsini2015quantifying}.

Nor does it make much sense to ask for the ``most typical'' network in models with high degeneracy of typical sets. In the canonical Erd\H{o}s-R\'enyi $\G(n,p)$ model~\cite{erdHos1960evolution}, for instance, the standard definition of typical sets based on the asymptotic equipartition property~\cite{cover1999elements, mackay2003information} says that the most typical networks are the ones with $n$ nodes and $m= \left\lfloor Np\right\rceil$ links, where $N=\binom{n}{2}$ and $\lfloor \cdot \rceil$ is the nearest integer function. However, since $\mathbb{P}(G)$ of all of these many ${N\choose m}$ networks is the same, any of them is equally typical, leaving substantial ambiguity in the selection of a single deterministic representative. In other words, in cases like this, we can only talk about (large) \emph{sets} of typical networks, but we cannot pinpoint a unique \emph{most} typical network.

Here, we develop a general definition of \textit{typical networks} in a model, and discuss a class of network models---equivalent to grand canonical ensembles in statistical physics---where this definition makes sense and allows for the selection of a unique \textit{most typical network}. We also develop methods to construct such networks deterministically in a computationally efficient fashion, and show how these methods work in application to a particular grand canonical network model and to real-world network data.

\section{Network typicality}
\label{ssec:typicality}
\begin{figure}[t]
    \includegraphics[width=1\columnwidth]{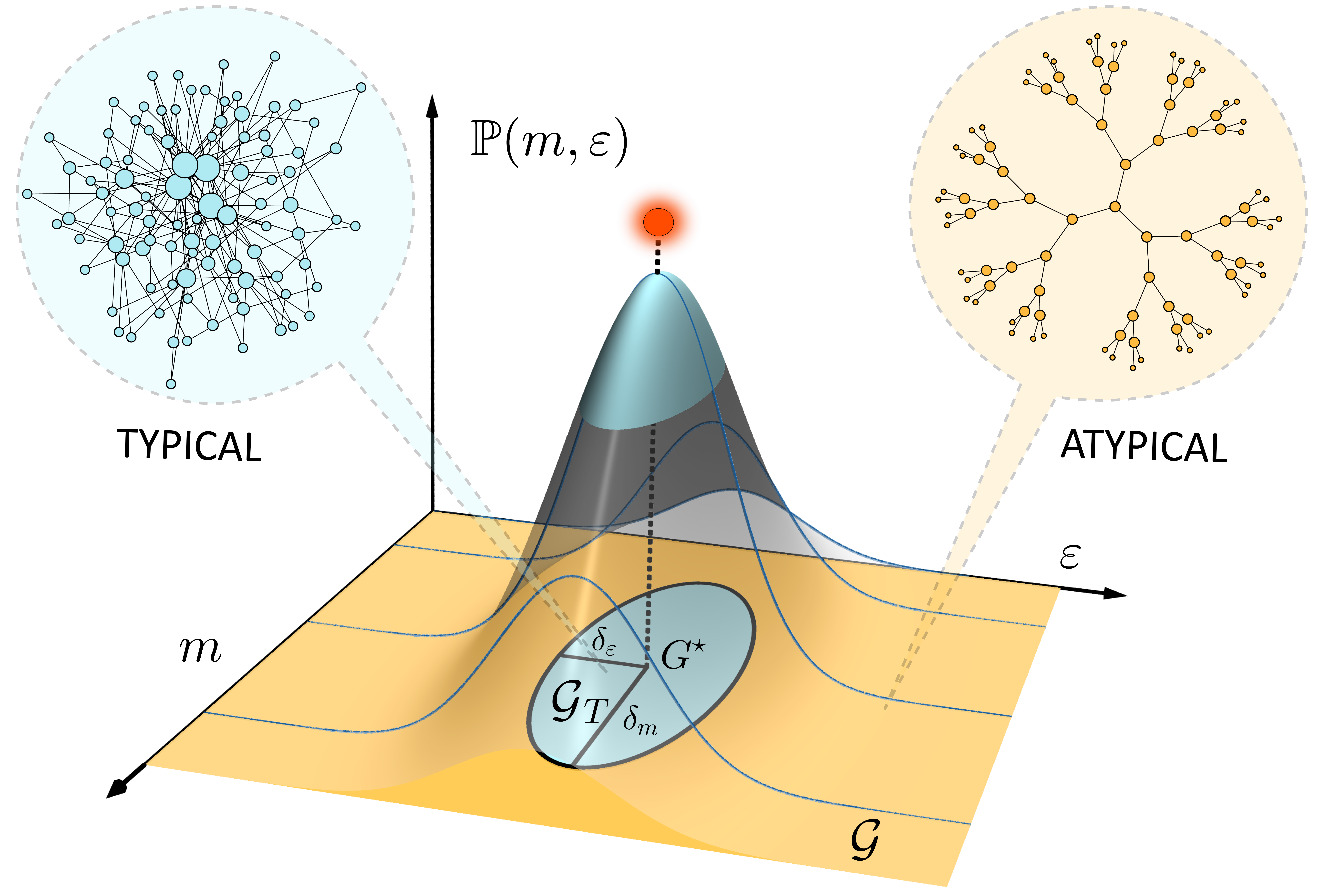} 
    \caption{\textbf{Network typicality definition.} In any grand canonical model of networks, Eq.~\eqref{eq:canonical}, the distribution $\mathbb{P}(m,\varepsilon)$ of the number of links $m(G)$ and energy $\varepsilon(G)$ of network~$G$ is sharply peaked. A typical set ${\G}_{T}$ is the set of networks whose values of $m$ and $\e$ are within the $(\delta_m,\delta_\e)$-neighborhood around their expected values~$(\bar{m},\bar{\e})$, Eq.~\eqref{eq:network typicality}, while the most typical network $G^{\star} = \text{argmin}_{G\in\mathcal{G}}\left(|m(G) - \bar{m}| + |\varepsilon(G) - \bar{\varepsilon}|\right)$ is the network which is closest to the top of the peak. This network is unique with high probability if energy~$\e(G)$ is a real-valued function.}
    \label{fig:figure1}
\end{figure}

The simplest case where a definition of the most typical network is possible is when the following conditions are satisfied:
\begin{enumerate}
\item the probability~$\mathbb{P}(G)$ of networks~$G$ in a model depends on~$G$ only via some network properties~$\boldsymbol{f}$: $\mathbb{P}(G)= \tilde{\mathbb{P}}[\boldsymbol{f}(G)]$;
\item the distribution $\mathbb{P}(v)$ of the values~$v$ of these properties in the model, $\mathbb{P}(v)=\sum_{G \in \mathcal{G}}\mathbb{P}(G)\delta(\boldsymbol{f}(G),v)$, is sharply peaked (self-averaging); and
\item the very top of this peak is populated by just one network~$G^{\star}$.
\end{enumerate}
Such~$G^{\star}$ is then the most typical network in the model, Fig.~\ref{fig:figure1}. Condition~(2) ensures there is a typical set, condition~(3) says there is the most typical element in this set, while condition~(1) allows us to extend the notion of typicality from network properties to networks \textit{per se} because it ensures that nothing other than these properties matters in the model. 

Condition~(3) is satisfied if at least one property among properties~$\boldsymbol{f}(G)$---called \textit{sufficient statistics} in statistics---is real-valued. Indeed, if this is the case, then degeneracy discussed in the introduction is gone. Conditions~(1,2) are satisfied in a multitude of models. Here, we focus on {\it grand canonical} models. This is one of the best-studied---both in statistical physics and network science~\cite{jaynes1957_statphys,park2004_statmech}---class of models in which the conditions above are satisfied, and which is particularly attractive for statistical analysis thanks to its maximum-entropy properties. This is the key feature of \textit{exponential random graph models} (ERGMs), of which grand canonical models are a special case (Appendix~\ref{sec: ERGMs}).

In grand canonical models, links between nodes~$i$ and~$j$ form independently with the Fermi-Dirac probabilities
\begin{equation}\label{eq:1}
  p_{ij}=  \frac{1}{e^{\beta(\e_{ij}-\mu)}+1}=\frac{1}{e^{\alpha+\beta\e_{ij}}+1},
\end{equation} 
where the inverse temperature~$\beta$ and chemical potential~$\mu$ are the model parameters, $\alpha=-\mu\beta$, and $\e_{ij}$ are energies of particle states $\{i,j\}$---particles are links. The models in this class differ in definitions of link energy. The probability~$\mathbb{P}(G)$ of network~$G$ in any grand canonical model is~\cite{park2004_statmech}
\begin{equation}
\label{eq:canonical}
  \mathbb{P}(G)= e^{-\alpha m(G) - \beta \e(G)}/\mathcal{Z},
\end{equation}
so the sufficient statistics~$\boldsymbol{f}(G)$ are the number of particles (links) $m(G)$ in network~$G$, and $G$'s energy, the sum of its link energies $\e(G)=\sum_{i,j}\e_{ij}G_{ij}$, where $\{G_{ij}\}$ is $G$'s adjacency matrix. The partition function $\mathcal{Z}$ is the normalization coefficient. As in any grand canonical ensemble~\cite{jaynes1957_statphys}, the $\mathbb{P}(G)$ in Eq.~\eqref{eq:canonical} maximizes Shannon entropy across all other models with the same expected values of the number of links~$\bar{m}$ and network energy~$\bar{\e}$ (Appendix~\ref{sec: ERGMs}).

The distributions of both the number of particles and energy are known to be self-averaging in grand canonical ensembles~\cite{kardar2007statistical}. Therefore, we define a $(\delta_{m},\delta_{\varepsilon})$-typical set of networks as
\begin{equation}\label{eq:network typicality}
    \G_{T}=\left\{G\in\mathcal{G}_n: \frac{|m(G)-\bar{m}|}{m_n}<\delta_{m},\,\frac{|\varepsilon(G)-\bar{\varepsilon}|}{m_n}<\delta_{\varepsilon}\right\},
\end{equation}
where $\G_n$ is the set of all networks of size~$n$, and $m_n$ is the asymptotic scaling of $\bar{m}$ with~$n$ ($\bar{m}\sim m_n$). The self-averaging properties of grand canonical ensembles ensure that in the thermodynamic limit $n\to\infty$, the bulk of the probability mass is concentrated on the typical set:
for any $\delta_{\varepsilon},\delta_{m}>0$, $\mathbb{P}(G\in\G_{T}) \rightarrow 1$. And if energies $\e_{ij}$ are all different real numbers, then all different networks~$G$ have different energies~$\e(G)$, so the top of the peak is occupied by one network---the most typical one in Fig.~\ref{fig:figure1}. The question is how to find it. 

\section{Constructing typical networks}
\label{ssec:methods_construction}

\subsection{Grand canonical models}
\label{ssec: grand canonical models methods}
Selecting the most typical value $m^\star$ of the number of links in a grand canonical model is easy: $m^\star$ is the closest integer to the expected number of links~$\bar{m}=\sum_{i<j}p_{ij}$, $m^\star=\lfloor\bar{m}\rceil$. Yet if the network size~$n$ is large, then the brute-force search for the network~$G^\star$ minimizing the network energy difference $|\e(G)-\bar{\e}|$ across the resulting~$N \choose m^\star$ possible networks is computationally prohibitive, akin to the $\mathsf{NP}$-complete subset sum problem \cite{10.5555/574848} and knapsack problem \cite{salkin1975_knapsackproblemsurvey}.

\begin{figure}[t]
    \includegraphics[width=1\columnwidth]{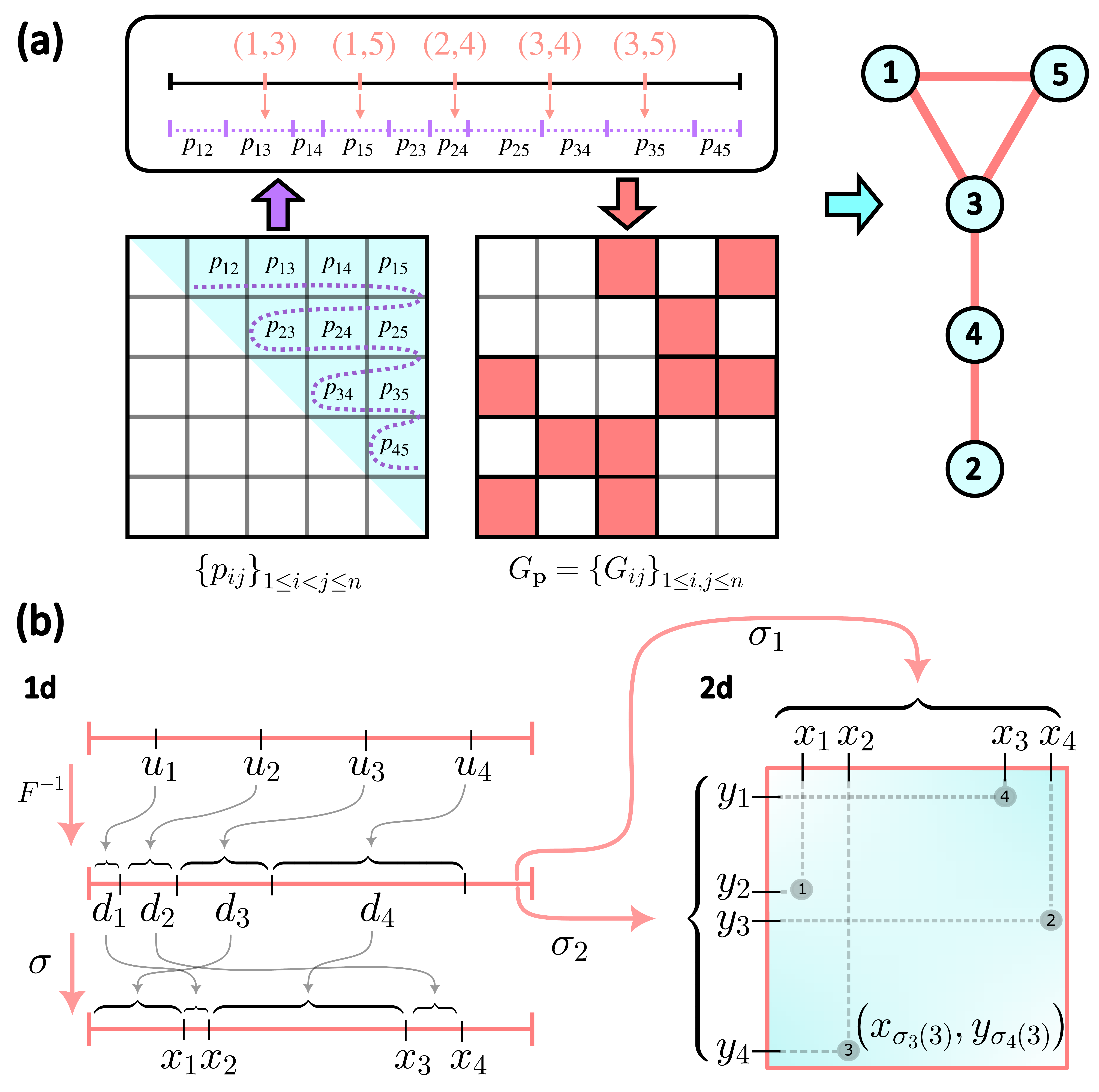}
    \caption{ \textbf{Deterministic construction of typical networks.} \textbf{(a)~Grand canonical derandomization.} The connection probability matrix $\mathbf{p}=\{p_{ij}\}_{i<j}$ is first flattened into an interval of length $\bar{m}=\sum_{i<j}p_{ij}$ composed of subintervals of lengths~$p_{ij}$ as shown in the figure. The whole interval is then overlaid with an equally spaced one-dimensional lattice of the same length and with $m^\star=\lfloor\bar{m}\rceil$ internal vertices. These vertices hit $m^\star$ subintervals~$p_{ij}$. The derandomized network~$G_\p$ consists of those links~$(i,j)$ whose subintervals are hit. \textbf{(b)~Hypercanonical derandomization.} To derandomize the random sprinkling of $n$ points on the unit interval, the distances between consecutive points are first set to $d_i=-(1/n)\ln(1-u_i)$ using the inverse CDF method~\cite{devroye1986_NonUniformRandomVariate} applied to the inverse CDF of the exponential distribution of rate~$n$ evaluated at the regular lattice $u_i=(i-1/2)/n$, $i\in[n]$. These distances are then shuffled using a pseudorandom permutation function~$\sigma$ acting on indices~$i$, so the coordinate of point~$i$ is set to $x_i=x_{i-1}+d_{\sigma(i)}$, where $x_0=0$. In higher dimensions~$d$, the same permutation function is used several times with different seeds---$\sigma_1, \sigma_2$ in $d=2$---to shuffle the distances $d_i$ between consecutive coordinates in different dimensions: $x_i=x_{i-1}+d_{\sigma_1(i)}$, $y_i=y_{i-1}+d_{\sigma_2(i)}$. It is used then again---$\sigma_3, \sigma_4$---to destroy the correlations between the $(x,y)$-coordinates of the same point: $(x_i,y_i)=(x_{\sigma_3(i)},y_{\sigma_4(i)})$. In this paper, permutations~$\sigma_s$, $s\in[2d]$, are defined via the \textit{multiplicative congruential generator}~(MCG)~\cite{knuth1997_taocp2}: $\sigma_s$ is the permutation that sorts the pseudorandom MCG numbers $\left\{b_k\right\}_{k=n(s-1)+1}^{ns}$ in the increasing order, where $b_{k+1}=ab_k\mod c$, $a=7^5$, $b_0=42$, and $c=2^{31}-1$.
    \label{fig:figure2}
    }
\end{figure}

To address this problem, we devise a deterministic link creation process that emulates random sampling of links~$(i,j)$ with probabilities~$p_{ij}$. It is presented in Fig.~\ref{fig:figure2}(a). The process simply overlays the matrix~$\mathbf{p}$ of connection probabilities~$\{p_{ij}\}$ over a regular lattice with $m^\star$ vertices, thus building a unique network~$G_\p$ for any~$\p$, not necessarily grand canonical Eq.~\eqref{eq:1}. However, if $\mathbf{p}$ is grand canonical, then we show in Section~\ref{sec:convergence} that the energy of constructed networks~$\e(G_\p)$ converges to its expected value~$\bar{\e}$, implying that the networks themselves converge to the most typical network, $G_\p \to G^\star$.

To show this convergence, we need to be able to increase the network size~$n$ in a systematic fashion. The easiest way to do so is in the context of hidden variable network models~\cite{caldarelli2002scale,soderberg2002general,boguna2003_hvnm}, an extremely general and well-studied class of models encompassing as special cases many quintessential network models in network science  (Appendix~\ref{sec: HVNMs}). One complication is that in these models, link energies are not fixed, but random
\begin{equation}\label{eq:2}
\e_{ij}=\phi(x_i,x_j),
\end{equation}
where $\phi$ is some fixed function of random hidden variables~$x_i$ of nodes~$i$. The resulting models are thus not grand canonical, but probabilistic mixtures of those, which we call \textit{hypercanonical models} since random model parameters are called hyperparameters in statistics~\cite{evans2009probability}. But then, for our typical network construction to be deterministic, we have to derandomize hidden variables~$x_i$ first. We address this task next.

\subsection{Hypercanonical models}
\label{sec: hypercanonical models}
Without much loss of generality, we can assume that hidden variables~$x_i$ are independent uniformly distributed random variables in the unit $d$-cube, $x_i \sim U\left([0,1]^d\right)$. (If they are not, we can map them to $U\left([0,1]^d\right)$ via their inverse cumulative distribution function~(CDF), adjusting the function~$\phi$ accordingly.) The question is how to derandomize $n$ random points in the unit cube, known as the binomial point process, which converges to the Poisson point process (PPP) as $n\rightarrow\infty$~\cite{last2017lectures}.

This question does not have a unique answer since it depends on what properties of the PPP we want to preserve in our derandomization, reminiscent of the choice of distinguisher in pseudorandomness~\cite{thomason1987pseudo,chung1988_quasirandomgraphs,Krivelevich2006}. Here, we choose to preserve the distances between nearest points in a PPP, since these distances are one of the most basic definitive properties of PPPs~\cite{last2017lectures}. In a $1$-dimensional PPP of rate~$n$, the distances~$d_i$ between consecutive points~$i$ on the real line are independent exponentially distributed random variables with rate~$n$. The inverse CDF of the exponential distribution of rate~$n$ is $F^{-1}(x)=-(1/n)\ln(1-x)$. Therefore, to derandomize the independent sampling of $d_i$'s from the exponential distribution, we simply set them to $d_i=F^{-1}(u_i)$, where $u_i=(i-1/2)/n$, $i\in[n]$, is a regular lattice on~$[0,1]$. The choice of the $1/2$ shift in it is justified in Appendix~\ref{sec:fixing z value}. We then use a pseudorandom function---several times in higher dimensions---as shown in Fig.~\ref{fig:figure2}(b), to scramble all the unwanted correlations that are not present in a PPP. In Appendix~\ref{Sec:Derandomization Heuristics} we show that this deterministic process reproduces the distances not only between nearest points but also between two random points in $1d$ and $2d$ PPPs. This matching ensures that our deterministic construction builds a \emph{typical} PPP sample, as further explained in Appendix~\ref{Sec:Derandomization Heuristics}.

As soon as hidden variables are derandomized, the link energies Eq.~\eqref{eq:2} become fixed, and so are the link existence probabilities Eq.~\eqref{eq:1}. This means that for each network size~$n$, we now have a unique network~$G_{n,\phi}$, which we claim converges to the most typical network~$G^\star$ in the model, $G_{n,\phi} \to G^\star$. We show this next.

\section{Convergence to the most typical network}
\label{sec:convergence}

To demonstrate the aforementioned convergence, we select one particular hypercanonical model---random hyperbolic graphs~(RHGs)~\cite{krioukov2010_rhg}. We choose this model because its energy function is real-valued, so the most typical network is well-defined. Another motivation to consider this model is that it attracted significant research attention over past decades in as diverse disciplines as machine learning, neuroscience, or Internet routing~\cite{boguna2021network}, thanks to its ability to capture a collection of definitive properties of many real-world networks~\cite{boguna2021network}. Yet another reason to consider this model is that several quintessential network models---in particular, the Erd\H{o}s-R\'enyi model, the configuration model, and (soft) random geometric graphs---are included in the RHG model as its limiting regimes~\cite{budel2024random}.

In the model, the hidden variables $x_i$ are the coordinates of points sprinkled randomly over a hyperbolic disc, while energies~$\e_{ij}$ are hyperbolic distances between those points. In addition to the grand canonical parameters~$\alpha,\beta$ controlling the expected number of links and energy---both scaling as $\mathcal{O}(n)$ in the model ($\bar{m}\sim\bar{\e}\sim m_n=n$)---the model has one more parameter~$\gamma$, the tail exponent of the power-law node degree distribution. Further details are in Appendix~\ref{sec:RHG}.

\begin{figure*}[t]
    \includegraphics[width=1.99\columnwidth]{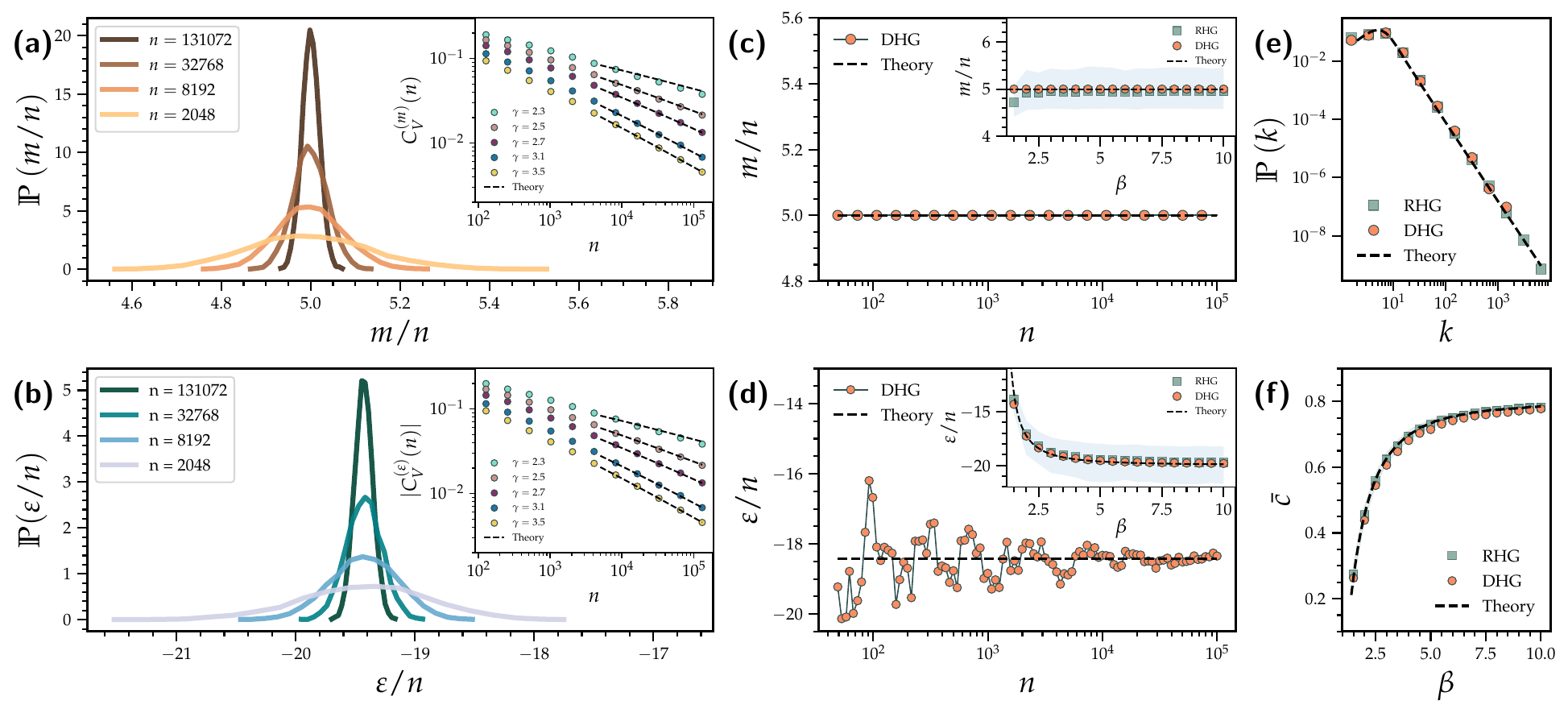}
    \caption{ \textbf{Self-averaging and convergence.} \textbf{(a,b)}~The empirical distributions of the rescaled number of links~$m(G)$ and energy~$\varepsilon(G)$ of the random hyperbolic graphs~(RHGs)~$G$ (Appendix~\ref{sec:RHG}) are shown for increasing graph sizes~$n$, and parameter values $\bar{\kappa} = 10$, $\beta = 4$, and $\gamma = 3.7$. The insets show the empirical coefficients of variation $C_{V}(n)$ (colored circles) of the distributions of $m$ and $\e$ for different values of $n$ and $\gamma$, versus the analytical predictions (black dashed lines) in Eq.~\eqref{eq:scaling of cv}. \textbf{(c,d)}~The values of the rescaled number of links~$m(G)$ and energy~$\varepsilon(G)$ of the deterministic hyperbolic graphs~(DHGs)~$G$ (Appendix~\ref{sec:DHG}) are shown for increasing graph sizes~$n$, and parameter values $\bar{\kappa} = 10$, $\beta = 2.5$, and $\gamma = 2.7$. The black dashed lines are the theoretical expected values. The insets show these properties as functions of the inverse temperature~$\beta$ for both DHGs and RHGs with the same parameter values and $n = 10^{4}$. The green squares are the RHG average values, and the shades are the 95 percentiles computed over $10^{3}$ RHG realizations. \textbf{(e,f)}~The degree distribution and the average local clustering coefficient of DHGs and RHGs with the same parameter values as in~(c,d).
    }
    \label{fig:dhg_properties}
    \end{figure*}

Since the model is not grand canonical but hypercanonical, we cannot rely on the known facts~\cite{kardar2007statistical} that sufficient statistics, the number of particles and energy, are self-averaging in any grand canonical ensemble. We have to show that the number of links and energy are self-averaging in our hypercanonical model as well, thus satisfying the second concentration requirement in Section~\ref{ssec:typicality}. We show this in Fig.~\ref{fig:dhg_properties}(a,b) and Appendix~\ref{sec:edges_selfavg}, where we derive the following scaling of the coefficient of variations~$C_{V}(n)$ for these two properties in the model:
\begin{equation}\label{eq:scaling of cv}
    \begin{aligned}
        C_{V}(n) = \mathcal{O}\left(n^{-\min\left\{1/2,\,(\gamma - 2)/(\gamma - 1)\right \} }(\ln(n))^{\mathbbm{1}\{\gamma = 3\}/2}\right),
    \end{aligned}
\end{equation}
where $\mathbbm{1}(\cdot)$ is the indicator function. This implies that $C_V(n)\to0$ for any $\gamma>2$, so the distributions of $m$ and $\e$ are sharply peaked. Since they are sharply peaked, and since our deterministic coordinate assignment is based on a typical PPP sample, the definitions of (most) typical networks in grand canonical models in Section~\ref{ssec:typicality} extend to this hypercanonical model as well.

Finally, Fig.~\ref{fig:dhg_properties}(c,d) shows the aforementioned convergence $G_{n,\phi} \to G^\star$ of the deterministic hyperbolic graphs~(DHGs)~$G_{n,\phi}$ to the most typical graph~$G^\star$ in the RHG model. Indeed, according to the definition in Section~\ref{ssec:typicality}, this $G^\star$ graph is the graph that matches the peak values of the sufficient statistics, and this is exactly what Fig.~\ref{fig:dhg_properties}(c,d) shows. The DHGs are constructed as discussed in Section~\ref{ssec:methods_construction}. The code is available at~\cite{dhg_code}. Appendix~\ref{sec:DHG} documents other model-specific details, one of which is that the number of links in DHGs matches its expected value exactly for any~$n$ as Fig.~\ref{fig:dhg_properties}(c) demonstrates. Figure~\ref{fig:dhg_properties}(d) shows the convergence of the other sufficient statistics, energy, to its expected value as $n$ increases. These two convergences, coupled with the self-averaging of the both sufficient statistics and typicality of coordinates, ensure that the graphs themselves converge to the most typical graph. As a consequence, all sharply peaked properties of DHGs converge to their expected values as well. The two most basic properties---the degree distribution and clustering---are shown as illustrative examples in Fig.~\ref{fig:dhg_properties}(e,f).

\section{Typicality of real-world networks} 

\begin{figure*}[tb]
    \includegraphics[width=0.95\textwidth]{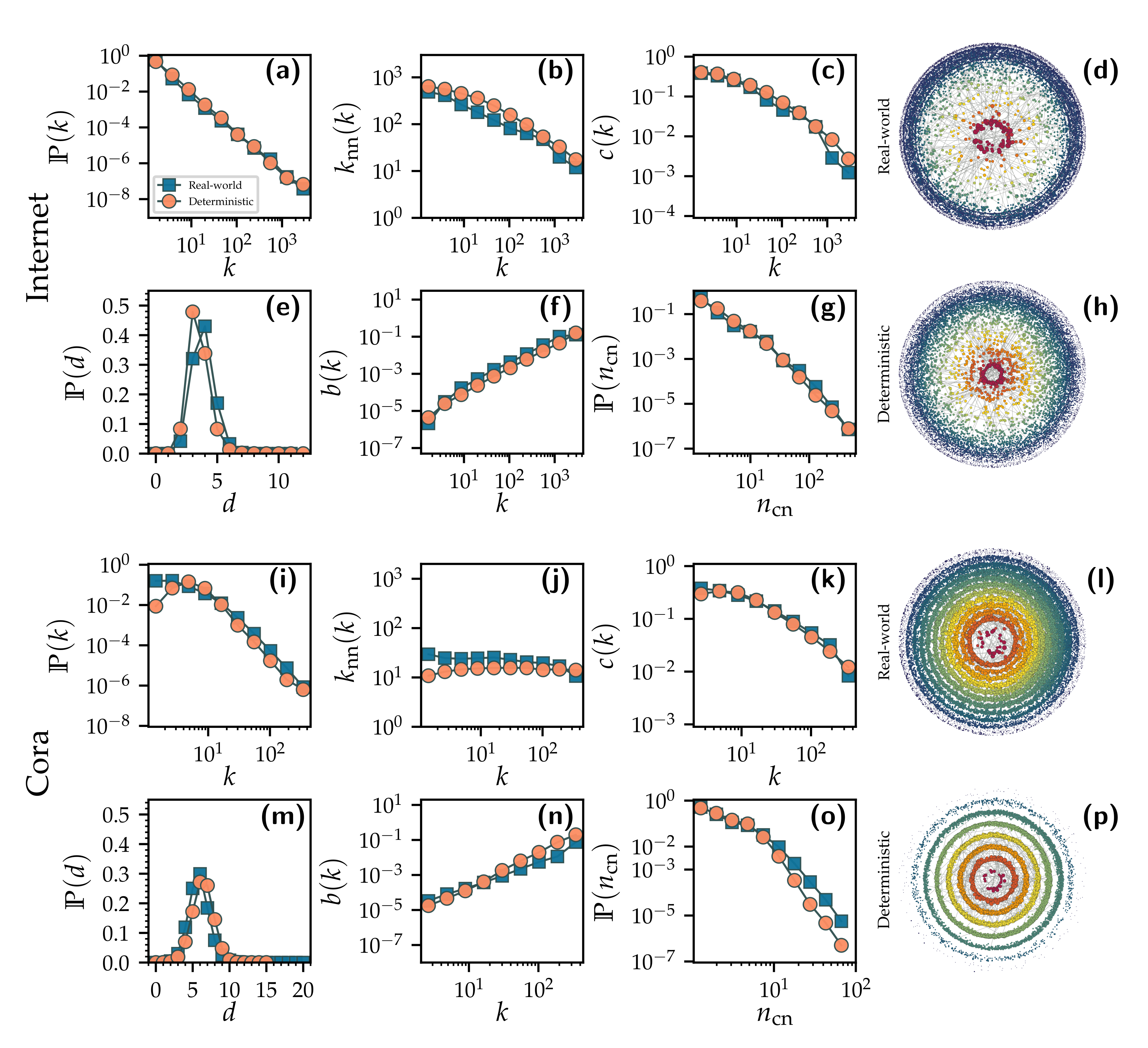}
    \caption{\textbf{Typicality of real-world networks.} The figure displays a collection of network properties---degree distribution~$\mathbb{P}(k)$ (panels \textbf{(a), (i)})), average nearest neighbor degree $k_\mathrm{nn}(k)$ of $k$-degree nodes (panels \textbf{(b), (j)}), average local clustering $c(k)$ of $k$-degree nodes (panels \textbf{(c),(k)}), distribution $\mathbb{P}(d)$ of shortest-path distances~$d$ (panels \textbf{(e),(m)}), average betweenness $b(k)$ of $k$-degree nodes (panels \textbf{(f),(n)}), distribution $\mathbb{P}(n_\mathrm{cn})$ of the number of common neighbors~$n_\mathrm{cn}$ of pairs of connected nodes, a.k.a.\ edge multiplicity~\cite{zlatic2012_networksarbitraryedge} (panels \textbf{(g),(o)}), of two real-world networks (blue squares) and their deterministic hyperbolic counterparts (orange circles). The real-world networks are the Internet at the autonomous system level (the upper two rows) and the Cora citation network (the lower two rows), documented as \texttt{internet\_as} and \texttt{cora} in Appendix~\ref{sec:empirical_methods}, which also details their visualizations in panels~\textbf{(d),(h),(l),(p)}.
    }
    \label{fig:real_world}
\end{figure*}

One natural application of the developed framework, mentioned in the introduction, is the assessment of how typical a given real-world network is in a given model. For such assessments, one needs to compare two networks---a real-world one and the most typical one in the model. There are many ways to compare two networks, none of which is ultimately the best. One option is to evaluate one or many graph similarity metrics on the two networks. For a recent review of such metrics, we refer to~\cite{hartle2020_networkcomparisonwithinensemble, mccabe2021_netrdlibrarynetwork}. Here, instead of evaluating any of such distances, we showcase graph comparisons more directly via measuring a collection of basic properties of two networks, shown in Fig.~\ref{fig:real_world}. We observe that according to these fundamental properties, the Internet and the Cora citation networks are fairly close to the most typical DHG network.

In Appendix~\ref{sec:empirical_methods}, we present the results of such measurements applied to a total of \numrealworld networks from different domains. We find that in many social, technological, and biological networks, the degree distribution, clustering, distance distribution, betweenness centrality, and other properties are close to the ones of the most typical DHG network. In contrast, the distance distribution of transportation networks, such as street and urban public transport networks, that are confined to Euclidean space and therefore are large worlds, are far from the ones of their DHGs, which are small worlds. The DHGs are small worlds because so are RHGs~\cite{boguna2020_Smallworldsclustering}. The observation that geometric DHGs do not match well geometric transportation networks should not come as a surprise because geometries in the two cases are very different: hyperbolic geometry in DHGs versus Euclidean geometry in transportation networks. Across domains, we observe that larger networks are more closely aligned with the most typical DHG network, but even small networks can sometimes have properties surprisingly similar to the ones of the most typical~DHG.

Summarizing, the most typical deterministic hyperbolic graph, with its three parameters $\alpha, \beta, \gamma$, two of which ($\alpha, \beta$) are the standard grand canonical ones, captures well a range of fundamental network properties---some of which, like betweenness or edge multiplicity distributions, are quite non-trivial---in many real-world networks coming from different domains, albeit with notable exceptions mentioned above. We believe that the main reason explaining this finding is the same as in~\cite{orsini2015quantifying}, where a longer list of real-world network properties including motifs was shown to be reproduced as soon as the degree distribution, degree correlations, and degree-dependent clustering are captured. Since DHGs appear to capture these three properties in many real-world networks, they reproduce their other important properties as well.

\section{Discussion}

The last observation in the previous section highlights the following open question: what other properties do we get ``for free'' in the most typical state in a model as defined in Section~\ref{ssec:typicality}? For example, the most typical deterministic hyperbolic graph has the most typical values of the number of links~($\alpha$), energy~($\beta$), and degree distribution tail~($\gamma$), yet what other network properties in this graph have their typical values as well? If the graph was random, this question could be answered by a variety of methods including the large deviation theory. It is less clear how to approach this question given that the graph is deterministic.

Potential applications of the developed framework are definitely not limited to the motivating one discussed in the previous section. One class of such applications can be called \emph{derandomized or deterministic statistical inference}. It would work as follows. Observe that for a given model, our framework establishes a function from the model parameter space to the space of graphs. Any combination of the parameter values corresponds to only one deterministic graph, the most typical one in the model with these parameter values. Therefore, the method is a powerful parameter inference and model selection tool. Indeed, for a given collection of network properties or graph similarity metrics, one can simply find the parameter values that minimize the corresponding graph similarity metric between a given (real-world) network and typical networks in the model. We, in fact, use this approach for parameter inference to recover a set of model parameters in Appendix~\ref{sec:empirical_methods}. The model selection step compares these similarity metrics between a given (real-world) network and most typical networks (with inferred parameters) in different models. The one that minimizes the distance describes the data best, according to this choice of graph distance. The conceptually closest approaches to this programme are the minimum description length principle~\cite{grunwald2007minimum} and algorithmic statistics~\cite{gacs2001algorithmic}.

More importantly, the developed framework is very general, definitely not just limited to the realm of networks, as it applies to any grand canonical or hypercanonical ensembles satisfying the niceness conditions discussed in Section~\ref{ssec:typicality}. These conditions---the presence of sharply peaked and nondegenerate sufficient statistics---are typically satisfied in most paradigmatic systems. While grand canonical ensembles are very common in statistical physics, hypercanonical ones are less so, yet they still appear in many mainstream systems, such as spin glasses with frozen disorder~\cite{edwards1975theory}.

As a final remark, to derandomize hypercanonical ensembles, we had to derandomize the Poisson point process and to ensure that the outcome is typical in Section~\ref{sec: hypercanonical models}. We stress that typicality of hyperparameters is as important as typicality of sufficient statistics conditioned on them, because in hypercanonical mixtures, matching the ensemble-average values of sufficient statistics can be done via atypical configurations of hyperparameters, an example of which is given in Appendix~\ref{sec:atypical-ppp}. Derandomization of point processes and ensuring that outcomes of such derandomization are typical appear to be an unexplored territory that would be interesting to investigate on more rigorous grounds. Potential applications may turn out to be as abundant and fruitful as with derandomization of other discrete random structures in information theory and machine learning~\cite{hoory2006expander, vadhan2012_pseudorandomness}.

\begin{acknowledgments}
N.G.S.~and D.K.~acknowledge the support of NSF Grant No.~CCF-2311160. J.v.d.K.~and B.K.~acknowledge the support of the AccelNet-MultiNet program, a project of the National Science Foundation (Awards \#1927425 and \#1927418). H.H.~acknowledges support from the Santa Fe Institute. H.H., S.V.S., and B.K.~acknowledge the support of a grant from the John Templeton Foundation (61780). S.V.S. acknowledges the support of NSF Grant No. FAIN-2133740. Any opinions, findings, conclusions, or recommendations expressed in this material are those of the author(s) and do not necessarily reflect the views of the funders, including the National Science Foundation and John Templeton Foundation.
\end{acknowledgments}

\appendix 
\section{Exponential random graph models}
\label{sec: ERGMs}

A random graph model is defined as a probability distribution, denoted $\mathbb{P}(G)$, over the set $\mathcal{G}$ of all possible labeled graphs $G$ of size $n$. On many occasions, we are given a graph $G^{\star}$ and tasked with coming up with a model that reproduces a set of graph properties of interest, denoted $\boldsymbol{f}(G^{\star})$, e.g., the number of edges, the degree sequence, or the number of triangles. In such a scenario, it is desirable to develop a model that reproduces the given constraints but is otherwise maximally random; otherwise, the model injects extraneous information. This view is known as the maximum entropy principle \cite{jaynes1957_statphys, jaynes2003_information}, whereby among all distributions satisfying the constraints, the one selected as a model should maximize Shannon entropy,
\begin{equation}
S[\mathbb{P}] = -\sum_{G\in \mathcal{G}}\mathbb{P}(G)\log \mathbb{P}(G).
\end{equation}
Such a model can then be used as a \textit{null model} for $G^\star$, the properties of which, beyond those constrained, can be compared against those of $G^\star$. If major differences are observed, it indicates the presence of statistical mechanisms beyond those implied by the constraints; if not, the constraints chosen are considered sufficiently explanatory.

If the constraints are sharply enforced, then we have a \textit{microcanonical ensemble} and the form of $\mathbb{P}(G)$ is just a uniform measure over the set of all graphs that satisfy the constraint exactly, i.e., drawn uniformly from the set $\{G\in\mathcal{G}: \boldsymbol{f}(G) = \boldsymbol{f}(G^{\star})\}$. If the constraints are enforced to be satisfied only in expectation,
\begin{equation}
\begin{aligned}
     \boldsymbol{\bar{f}} = \sum_{G\in \mathcal{G}}\boldsymbol{f}(G)\mathbb{P}(G) &= \boldsymbol{f}(G^{\star}),
\end{aligned} 
\end{equation}
then we have a \textit{canonical ensemble}; note the requirement of $\sum_{G\in \mathcal{G}}\mathbb{P}(G) = 1$ is an additional constraint on the values of $\{\mathbb{P}(G)\}_{G\in\mathcal{G}}$. This constrained optimization problem is solved using the method of Lagrange multipliers, resulting in Gibbs/Boltzmann exponential family distribution \cite{park2004_statmech},
\begin{equation}
\label{eq: ERGM pdf}
    \mathbb{P}(G) = \dfrac{e^{-\boldsymbol{\lambda}\cdot \boldsymbol{f}(G)}}{\sum_{\tilde{G} \in \mathcal{G}}e^{-\boldsymbol{\lambda}\cdot \boldsymbol{f}(\tilde{G})}},
\end{equation}
where $\boldsymbol{\lambda}$ is a vector of Lagrange multipliers whose values must be set to $\boldsymbol{\lambda} = \boldsymbol{\lambda^{\star}}$ such that the soft constraint $ \boldsymbol{\bar{f}}  = \boldsymbol{f}(G^{\star})$ is realized.

Since the value of $\mathbb{P}(G)$ depends on the graph $G$ only through $\boldsymbol{f}(G)$, the properties $\boldsymbol{f}(G)$ are called {\it sufficient statistics} of the model. Formally, Eq.~\eqref{eq: ERGM pdf} are the entropy maximizing distributions subjected to the constraints $\boldsymbol{f}(G^{\star})$ called \textit{exponential random graph models} (ERGMs) \cite{jaynes1957_statphys, park2004_statmech, holland1981_ergm, Frank1986}. ERGMs and the maximum entropy formalism in general have been successful in modeling patterns of activity in neuronal circuits \cite{jcrn-3nrc}, testing various hypotheses regarding the tie-formation mechanism in social networks \cite{robins2007_IntroductionExponentialRandom, Lusher2013}, and as null models in network science \cite{cimini2019_statphysreal, Squartini2017}. 

A particular class of ERGMs which is of interest to us here is the class of \textit{grand canonical} models~\cite{park2004_statmech, kardar2007statistical, PhysRevLett.102.038701, PhysRevE.73.015101}, where two nodes are connected independently with probability
\begin{equation}\label{eq:fermi-dirac}
  p_{ij}=  \frac{1}{e^{\beta(\e_{ij}-\mu)}+1} =\frac{1}{e^{\alpha+\beta\e_{ij}}+1},
\end{equation} 
where $\beta$ (inverse temperature) and $\mu$ (chemical potential) are the model parameters, $\alpha=-\mu\beta$, and $\varepsilon_{ij}$ are link energies. Letting $G_{ij}\in \{0,1\}$ denote the adjacency matrix elements of $G$, the graph probability mass function is given by,
\begin{equation}
    \begin{aligned}
        \mathbb{P}(G) &= \prod_{1\leq i < j \leq n}p_{ij}^{G_{ij}}(1-p_{ij})^{1 - G_{ij}},\\
        &= \exp{\left(\sum_{1\leq i< j\leq n} G_{ij}\log\left(\dfrac{p_{ij}}{1-p_{ij}}\right) + \log(1-p_{ij})\right)},\\
        &= \dfrac{\exp{\left(-\alpha m(G) - \beta \varepsilon(G)\right)}}{\mathcal{Z}},
    \end{aligned}
\end{equation}
where $m(G) = \sum_{i<j}G_{ij}$ is the number of edges in graph~$G$, $\varepsilon(G) = \sum_{i<j}G_{ij}\varepsilon_{ij}$ is its energy, and $\mathcal{Z} = \prod_{1\leq i<j\leq n} 1/(1-p_{ij})$ is the partition function. Comparing the above equation with Eq.~\eqref{eq: ERGM pdf}, it is clear that the grand canonical model is an entropy-maximizing distribution subjected to the constraint on the expected values of the number of edges and energy. In other words, the grand canonical model is an ERGM in which the sufficient statistics are the number of edges and energy.

\section{Hidden variable network models}
\label{sec: HVNMs}
Hidden variable network models (HVNMs) are a widely studied class of random network models characterizing how heterogeneously distributed node-variables (hidden variables (HVs)) can give rise to various patterns of network connectivity \cite{boguna2003_hvnm, soderberg2002general, caldarelli2002scale}. Graphs sampled from HVNMs arise through two stages of randomness: 
\begin{itemize}
    \item[a)] each node $i \in [n] = \{1,2, \dots, n\}$ gets a random hidden variable $h_i\in\mathcal{X}$ in some space~$\mathcal{X}$ sampled independently from some probability density~$\upsilon(h)$;
    \item[b)] edges form pairwise-independently with probability $p_{ij} = g(h_{i},h_{j})$ where $g: \mathcal{X}^{2} \to [0,1]$ is the connection probability function. 
\end{itemize}
Consequently, the hidden variable configuration $H = (h_{i})_{i=1}^{n} \in \mathcal{X}^{n}$ has a joint probability density of the form $\rho(H) = \prod_{i=1}^{n}\upsilon(h_{i})$. Without much loss of generality, one may set $\mathcal{X}=[0,1]^d$ and $\upsilon=U\left([0,1]^d\right)$ in the case of $d$-dimensional HVs, since we can often map this hidden variable configuration to the intended hidden variable space via the inverse transform sampling \cite{devroye1986_NonUniformRandomVariate,robert1999monte}. In this setting, the probability of a network given $H$ is,
\begin{equation}
\begin{aligned}
\label{eq:P_G_given_H}
    \mathbb{P}(G|H) &= \prod_{1\leq i < j \leq n} p_{ij}^{G_{ij}}(1 - p_{ij})^{1-G_{ij}},\\
    &= \dfrac{e^{-\sum_{i<j}\lambda_{ij}G_{ij}}}{\mathcal{Z}},\\
    \text{where}~\lambda_{ij} &= \ln\left(\dfrac{1-p_{ij}}{p_{ij}}\right),~\mathcal{Z} = \prod_{1\leq i < j \leq n}\left(\dfrac{1}{1-p_{ij}}\right).
\end{aligned}
\end{equation}

Thus, any HVNM conditioned on the hidden-variable configuration $H$ (i.e., an HVNM with a specified ``frozen'' configuration $H$) is an ERGM (Appendix~\ref{sec: ERGMs}) with $\binom{n}{2}$ pairwise constraints of the form
\begin{equation}
 \bar{G}_{ij} = g(h_i,h_j).
\end{equation}
However, some choices of specific functional forms of the connection probability function $g(\cdot)$---including the grand canonical one in Eq.~\eqref{eq:fermi-dirac}---collapse the $\binom{n}{2}$ individual edge indicator sufficient statistics into only a few aggregated sufficient statistics, like the number of edges and energy in the grand canonical case.

The probability of graph $G$ in an HVNM is
\begin{equation}
\label{eq:HVNM_PMF}
    \mathbb{P}(G) = \int_{H}\mathbb{P}(G|H)\rho(H)dH.
\end{equation}
Therefore, HVNMs are probabilistic mixtures of ERGMs, $\mathbb{P}(G|H)$, with mixing distribution $\rho(H)$. We refer to such distributions as {\it hypercanonical ensembles}, so all HVNMs are hypercanonical.

By appropriately choosing the hidden-variable distribution $\rho(\cdot)$ and the connection-probability function $g(\cdot)$, hidden-variable network models (HVNMs) can encompass a broad spectrum of paradigmatic network models, including the Erd\H{o}s-R\'enyi random graph model \cite{erdHos1960evolution, gilbertRandomGraphs1959}, (hypersoft) configuration model \cite{vanderhoorn2018_SparseMaximumEntropyRandom}, stochastic block model \cite{karrer2011stochastic}, latent-space models \cite{penrose2003_RandomGeometricGraphs, penrose2016_Connectivitysoftrandom,krioukov2010_rhg, serrano2008_selfsimilar, boguna2021network}, and many others.

\section{Justification for the lattice shift}
\label{sec:fixing z value}

In this section, we motivate the choice of shifting the regular lattice, $u_{i} = (i-z)/n$, horizontally by $z = 1/2$, in the deterministic sprinkling process (outlined in Sec.~\ref{sec: hypercanonical models}). The coordinate of point $i$ in the $1d$ deterministic sprinkling process on the unit interval is given by,
\begin{equation}
    x_i=-\frac{1}{n}\sum_{l=1}^i\ln \left(1-u_{\sigma(i)}\right)\label{eq:xiAppendix},
\end{equation}
where $\sigma: [n] \to [n]$ is a pseudo-random permutation function. We want these deterministically sprinkled points to have as many of the statistical properties of the Poisson point process (PPP) as possible. To that end, along with matching the (nearest neighbor) distance distribution of the PPP (see Appendix \ref{Sec:Derandomization Heuristics}), we also match the scaling of the maximal coordinate in deterministic sprinkling, $x_{n}$, with the scaling of the mean of the extreme value distribution of $U([0,1])$ \cite{david2004order} by shifting the regular lattice, $u_{i} = (i - z)/n$ horizontally by $z = 1/2$.

\par To show this formally, let us consider $n$ i.i.d.\ uniform random variables, i.e., $\xi_{i} \sim U([0,1])$ and let $\xi_{\max} = \max\left(\xi_{1}, \xi_{2}, \cdots, \xi_{n}\right)$. The cumulative distribution function of the maximal value is, 
\begin{equation}
    \mathbb{P}(\xi_{\max}\leq \xi)=\prod_{i=1}^n\mathbb{P}(\xi_i\leq \xi)=\xi^n.
\end{equation}
This then allows us to derive the expected maximal value as,
\begin{equation}
    \overline{\xi_{\max}} = n\int_0^1 \xi^{n}d\xi =\frac{n}{n+1}=1-\frac{1}{n}+\mathcal{O}\left(\frac{1}{n^2}\right)\label{eq:ximax}.
\end{equation}
In order for our deterministic process to give the same result, we must set $z$ in $u_{i} = (i-z)/n$ appropriately such that $x_{\max} = x_{n}=1-1/n$. From Eq.~\eqref{eq:xiAppendix} it is easy to see that $x_{n}=-1/n\sum_{i=1}^{n}\ln(1-u_i)$, where the pseudo random permutation function, $\sigma(i)$, is irrelevant. The question now is how to choose $z\in[0,1]$ such that the correct scaling is achieved. For some general value of $z$, 
\begin{alignat}{6}
    x_{n} &= - \frac{1}{n}\sum_{i=1}^n\ln\left(1-\frac{i-z}{n}\right)\notag,\\
    &=-\frac{1}{n}\ln\left(\prod_{i=0}^{n-1}(i+z)\right)+\ln n\notag,\\
    &=-\frac{1}{n}\ln\left(\frac{\Gamma(z+n)}{\Gamma(z)}\right)+\ln n.
\end{alignat}
The transcendental equation $x_{n}=\overline{\xi_{\max}}$ cannot be solved analytically, and we must therefore turn to asymptotics. In the setting $n\gg 1$, we can use the following,
\begin{equation}
    \ln\left(\Gamma(w)\right) = \frac{2w-1}{2}\ln(w) - w + \frac{1}{2}\ln(2\pi)+\mathcal{O}\left(\frac{1}{w}\right),
\end{equation}
to obtain,
\begin{equation}
    x_{n} = 1 + \frac{1-2z}{2}\frac{\ln n}{n} + \frac{1}{n}\ln\left(\frac{\Gamma(z)}{\sqrt{2\pi}}\right)+\mathcal{O}\left(\frac{1}{n^2}\right).
\end{equation}
The term of order $\mathcal{O}(\ln n/n)$ is not present in $\overline{\xi_{\max}}$. Therefore, it must be removed by setting $z=1/2$, leading to
\begin{equation}
    x_{n}=1-\frac{\ln2}{2}\frac{1}{n}+\mathcal{O}\left(\frac{1}{n^2}\right).\label{eq:xmaxz1/2}
\end{equation}
This is not exactly the same expression as Eq.~\ref{eq:ximax}, but the constant pre-factor in the second term has no effect on the scaling behavior, so, for instance, the scaling of the largest hidden degree is $\kappa_{\max}\sim n^{1/(\gamma-1)}$ in both the random and derandomized cases.\\

\section{Poisson point process, derandomized}
\label{Sec:Derandomization Heuristics}

In this section, we show that the hypercanonical derandomization in Sec.~\ref{sec: hypercanonical models} results in a deterministic sprinkling of $n$ points on the unit interval and unit square matching the (nearest neighbor) distance distribution of the Poisson point process (PPP).

The nearest neighbor distance distribution in $1d$ and $2d$ PPP of intensity~$n$ is~\cite{last2017lectures}
\begin{equation}\label{eq: nearest neighbor dd PPP}
    \begin{aligned}
        \mathbb{P}^{(nn)}_{1\mathrm{d}}(x) &= 2ne^{-2nx},\\
        \mathbb{P}^{(nn)}_{2\mathrm{d}}(x) &= 2\pi n xe^{-\pi n x^{2}}.
    \end{aligned}
\end{equation}
The PPP distance distribution, on the other hand, is the probability density function of the distance between two random points on the unit interval, $\mathbb{P}_{1\text{d}}(x)$, or unit square, $\mathbb{P}_{2\text{d}}(x)$. It is given by~\cite{mathai1999introduction}
\begin{widetext}
\begin{equation}\label{eq: distance distribution pdfs}
\begin{aligned}
\mathbb{P}_{1\text{d}}(x) &=
  \begin{cases}
    2(1-x), & 0\le x\le 1,\\
    0,      & \text{otherwise};
  \end{cases}\\[1em]
\mathbb{P}_{2\text{d}}(x) &=
  \begin{cases}
    2x\bigl(\pi-4x+x^{2}\bigr), & 0\le x\le 1,\\[6pt]
    2x\!\Bigl[4\sqrt{x^{2}-1}-(x^{2}+2-\pi)-4\arctan\!\sqrt{x^{2}-1}\Bigr], 
        & 1<x\le\sqrt{2},\\[6pt]
    0, & \text{otherwise}.
  \end{cases}
\end{aligned}
\end{equation}
\end{widetext}

\begin{figure}[t]
    \includegraphics[width=1\columnwidth]{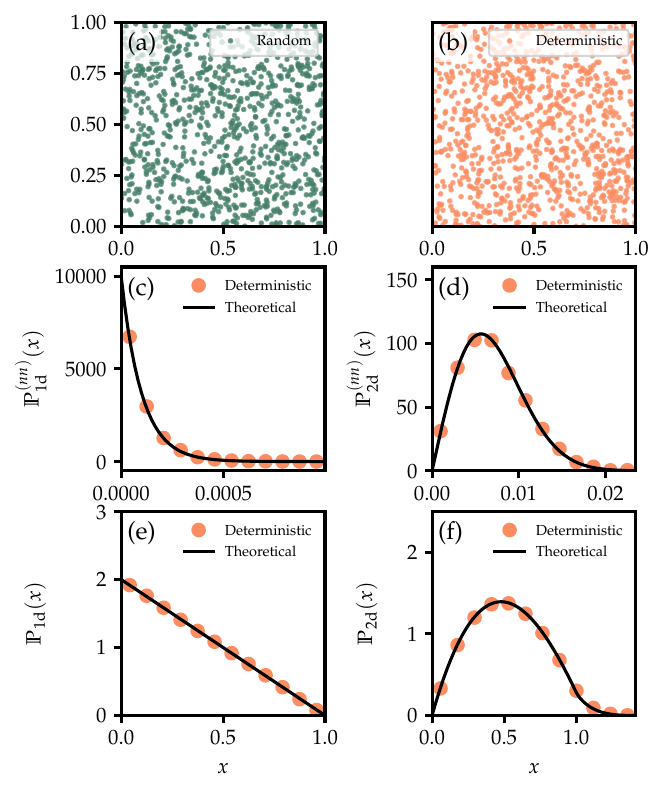}
    \caption{\textbf{Poisson point process versus its derandomization. (a)}~A realization of the PPP of intensity $n=10^{3}$ on the unit square~$[0,1]^{2}$. \textbf{(b)}~The deterministic sprinkling process from Sec.~\ref{sec: hypercanonical models} with $n=10^{3}$. \textbf{(c,d)} The PPP nearest neighbor distance distributions in Eq.~\eqref{eq: nearest neighbor dd PPP}, versus the empirical nearest neighbor distance distribution in the deterministic sprinkling process with $n = 5\times10^{3}$ on the unit interval and unit square, respectively. \textbf{(e,f)} The PPP distance distributions in Eq.~\eqref{eq: distance distribution pdfs}, versus the empirical distance distribution in the deterministic sprinkling with $n=5 \times 10^{3}$ on the unit interval and unit square, respectively.}
    \label{fig:distance_distribution}
\end{figure}

In Fig.~\ref{fig:distance_distribution}(a,b), we show a realization of the $2d$~PPP, and compare it with our deterministic process. In addition to pleasing visual similarity between the two observed there, Figs.~\ref{fig:distance_distribution}(c,d) and Figs.~\ref{fig:distance_distribution}(e,f) confirm that the deterministic sprinkling has the same (nearest neighbor) distance distribution as the PPP.

This matching ensures that our deterministic process constructs a \emph{typical} PPP sample in the following sense. In PPPs, the distance distribution between two random points converges in the continuum limit to the distance distribution of the space in which the PPP takes place~\cite{mathai1999introduction,jammalamadaka1986_LimitTheoremsTriangular,reitzner2013_Centrallimittheorems}. This means that with high probability, approaching~1, a PPP sample has the distance distribution close to the distance distribution in its space. Therefore, PPP samples that do not match the space distance distribution are atypical, and those that do match it are typical.

\section{Random hyperbolic graphs (RHGs)}
 \label{sec:RHG}

An important random graph model belonging to the class of HVNMs (Appendix \ref{sec: HVNMs}) is the degree-corrected latent space model, also known as random hyperbolic graphs (RHG) \cite{krioukov2010_rhg}, which combines the properties of the hypersoft configuration model \cite{vanderhoorn2018_SparseMaximumEntropyRandom} and soft random geometric graphs \cite{penrose2003_RandomGeometricGraphs, penrose2016_Connectivitysoftrandom}. Here, we describe the $\mathbb{H}^{2}$ and $\mathbb{S}^{1}$ equivalent versions of this hypercanonical model~\cite{krioukov2010_rhg, serrano2021shortest, boguna2021network}.

In the $\mathbb{H}^{2}$ formulation, sampling a network first involves sprinkling nodes quasi-uniformly within a disk of radius $R_{\mathbb{H}^{2}}$ on the hyperbolic plane of constant curvature $K = -1$. That is, the hidden variables of nodes~$i$ are their polar coordinates $(r_i,\theta_i)$ which are sampled independently as $\theta_i \sim U[(0,2\pi)]$ and $r_i \sim \rho$, where the support of the radial coordinate distribution~$\rho$ is $[0,R_{\mathbb{H}^{2}}]$, and its PDF is
\begin{equation}\label{eq:cp-H2}
    \rho(r) = \tilde{\alpha}\dfrac{\sinh(\tilde{\alpha}r)}{\cosh{(\tilde{\alpha}R_{\mathbb{H}^{2}}) - 1}}.
\end{equation}
The parameter $\tilde{\alpha} \geq 1/2$ controls the power law exponent of the resulting degree distribution in the model via $\gamma = 2\tilde{\alpha} +1$, while the radius of the disk
\begin{equation}
    R_{\mathbb{H}^{2}} = 2\ln\left(\dfrac{2n}{\bar{k}\beta}\csc{\left(\dfrac{\pi}{\beta}\right)}\left(\dfrac{\tilde{\alpha}}{\tilde{\alpha} - 1/2}\right)^{2}\right),
\end{equation}
fixes the average degree to $\bar{k}>0$ at $n\to\infty$. The parameter $\beta > 1$, the inverse temperature, controls the average clustering in the model.

Upon the assignment of the coordinates, two nodes are connected with the grand canonical Fermi-Dirac connection probability,
\begin{equation}\label{eq: H2 connection probability}
    p_{ij} = \frac{1}{1 + e^{\beta(\tilde{\varepsilon}_{ij} - R_{\mathbb{H}^{2}})/2}}.
\end{equation}
The fermionic energies $\tilde{\varepsilon}_{ij}$ are the hyperbolic distances between nodes $i$ and $j$ given by the hyperbolic law of cosines,
\begin{equation}
    \cosh{\tilde{\varepsilon}_{ij}} = \cosh{r_{i}}\cosh{r_{j}} - \sinh{r_{i}}\sinh{r_{j}}\cos{\Delta \theta_{ij}},
\end{equation}
where $\Delta \theta_{ij} = \pi - |\pi - |\theta_{i} - \theta_{j}||$ is the angle between the two points. This distance is approximately equal to~\cite{krioukov2010_rhg, serrano2021shortest}
\begin{equation}\label{eq: H2 link energy}
    \tilde{\varepsilon}_{ij} \approx r_{i} + r_{j} + 2\ln \left(\dfrac{\Delta \theta_{ij}}{2}\right).
\end{equation}

The $\mathbb{H}^{2}$ model is a hypercanonical mixture of the grand canonical models (Appendix \ref{sec: ERGMs}) with random link energies $\tilde{\e}_{ij}$ defined in Eq.~\eqref{eq: H2 link energy}. Their distribution is the distance distribution between random points on the hyperbolic disk.

In the equivalent $\mathbb{S}^{1}$ formulation of RHGs~\cite{krioukov2010_rhg}, each node is endowed with two random hidden variables: similarity $(x_{i})$ and popularity $(\kappa_{i})$. The similarity space is a circle of radius $R_{\mathbb{S}^{1}} = n/2\pi$ and $x_{i} = R_{\mathbb{S}^{1}}\theta_{i}$, where $\theta_{i} \sim U([0,2\pi])$ (two nodes close on the circle are ``similar''). The radius scales with $n$ so as to maintain the node density on the circle to be one. The popularity hidden variables of nodes are their expected degrees $\kappa_{i} \sim \rho(\kappa|\bar{\kappa}, \gamma)$, where
\begin{equation}\label{eq:cp-S1}
    \rho(\kappa|\bar{\kappa},\gamma) = (\gamma-1)\kappa_{0}^{\gamma - 1}\kappa ^{-\gamma},
\end{equation}
is the Pareto distribution with mean $\bar{\kappa}$, power-law exponent $\gamma$, and support $\kappa \in [\kappa_{0},\infty)$, where $\kappa_{0} = \bar{\kappa}(\gamma - 2)/(\gamma -1)$.

Once the hidden variables are assigned, two nodes are connected with the grand canonical Fermi-Dirac connection probability,
\begin{equation}\label{eq: S1 connection probability}
\begin{aligned}
    p_{ij} &=  \dfrac{1}{1 + \left(\dfrac{d_{ij}}{\hat{\mu} \kappa_{i}\kappa_{j}}\right)^{\beta}} = \dfrac{1}{1+e^{\beta(\varepsilon_{ij} - \mu)}},\text{ where}\\
    d_{ij} &=R_{\mathbb{S}^{1}}\Delta \theta_{ij}, \quad \mu = \log(\hat{\mu}),\text{ and}\\
    \varepsilon_{ij} &= \log\left(\dfrac{d_{ij}}{\kappa_{i}\kappa_{j}}\right)
\end{aligned}    
\end{equation}
are the link energies in the model. The inverse temperature $\beta>1$ is the same as in the $\mathbb{H}^{2}$ formulation, while the chemical potential
\begin{equation}\label{eq: chemical potential}
    \hat{\mu} = \dfrac{\beta}{2\pi\bar{\kappa}} \sin \left(\dfrac{\pi}{\beta}\right)
\end{equation}
fixes the expected average degree in the model to~$\bar{\kappa}>0$ at $n\to\infty$. For completeness, we re-derive this observation in Appendix~\ref{sec:edges_selfavg}.

The $\mathbb{H}^{2}$ and $\mathbb{S}^{1}$ formulations of RHGs are equivalent because they have the same distributions of hidden variables and connection probabilities. Indeed, the angular coordinates in both formulations come from the same uniform distribution, while the radial coordinate and expected degree distributions $\rho(r)$ and $\rho(\kappa)$ in Eqs.~(\ref{eq:cp-H2},\ref{eq:cp-S1}) can be obtained from each other via the following change of variables:
\begin{equation}
    \kappa = \kappa_{0}e^{(R_{\mathbb{H}^2}-r)/2},
\end{equation}
which maps the connection probability in Eq.~\eqref{eq: S1 connection probability} to the one in Eq.~\eqref{eq: H2 connection probability}~\cite{krioukov2010_rhg}.

This change of variables leads to the following relationship between the link energies in the two formulations:
\begin{equation}
    \tilde{\varepsilon}_{ij} = 2\left[\varepsilon_{ij} + \ln{\left(\dfrac{n}{\tau}\right)}\right],
\end{equation}
where $\tau=\pi\kappa_{0}^{2}\hat{\mu}^{2}$. Consequently, the graph energy and expected energy are related via
\begin{equation}\label{eq: energy relation H2-S1}
    \begin{aligned}
        \tilde{\varepsilon}(G) &= 2\left[\varepsilon(G) + \ln{\left(\dfrac{n}{\tau}\right)}m(G)\right],\\
        \bar{\tilde{\varepsilon}} &= 2\left[\bar{\varepsilon} + \ln{\left(\dfrac{n}{\tau}\right)}\bar{m}\right].
    \end{aligned}
\end{equation}

All the analytical and simulation results reported in this paper are obtained using the $\mathbb{S}^{1}$ formulation of the RHG model since it is easier to work with. In particular, to simulate RHGs, we take their size~$n$ and parameters $\bar{\kappa}>0$, $\beta>1$, and $\gamma>2$ as input, and proceed as described the second half of this section.

We note that the RHG model is well defined for $\beta<1$ as well. In particular, in the $\beta\to0$ limit, the RHG model reduces to the hypersoft configuration model if $\gamma$ is finite, and to the Erd\H{o}s-R\'enyi model if $\gamma\to\infty$~\cite{budel2024random}. However, if $\beta<1$, the definition of energy~\eqref{eq: S1 connection probability} breaks down~\cite{boguna2020_Smallworldsclustering}, explaining why we require $\beta>1$ in this paper.

\section{Self-averaging of RHGs' sufficient statistics}
\label{sec:edges_selfavg}

In this section, we derive the closed-form expressions for the mean values of the number of links $m(G) = \sum_{i<j}G_{ij}$, and energy $\varepsilon(G) = \sum_{i<j}\varepsilon_{ij}G_{ij}$ of the random hyperbolic graphs (Appendix \ref{sec:RHG}) in the thermodynamic limit $n \to \infty$. We also show that the distributions of these random variables are self-averaging---their coefficient of variation goes to zero in the thermodynamic limit for any value of the degree exponent~$\gamma>2$.

\subsection{Average number of links $(\bar{m})$}
\label{ssec: average links}
The average number of links in the model is,
\begin{equation}
    \begin{aligned}
        \bar{m} &= \mathbb{E}\left[\sum_{i<j}G_{ij}\right],\\
        &= \binom{n}{2}\mathbb{E}[G_{ij}],\\
        &=   \dfrac{1}{2}\int_{-n/2}^{n/2} \int_{\kappa_{0}}^{\infty} m(\kappa_{i},x_{i}) \rho(\kappa_{i}) dx_{i} d\kappa_{i},\\
        \text{where}~m(\kappa_{i},x_{i}) &= (n-1) \int_{-n/2}^{n/2}\int_{\kappa_{0}}^{\infty} p_{ij}\dfrac{\rho(\kappa_{j})}{n}dx_{j} d\kappa_{j}.
    \end{aligned}
\end{equation}
$m(\kappa_{i},x_{i})$ is the expected number of links attached to node $``i"$ with hidden variable values $(\kappa_{i},x_{i})$. Due to the model being rotationally invariant, we can set $x_{i}=0$ without loss of generality, i.e., $m(\kappa_{i},0) = m(\kappa_{i},x_{i})$, which in the limit $n \to \infty$ is,
\begin{equation}
    \begin{aligned}
        m(\kappa_{i},0)&= 2\int_{0}^{\infty}\int_{\kappa_{0}}^{\infty} \dfrac{1}{1+\left(\dfrac{x_{j}}{\hat{\mu} \kappa_{i}\kappa_{j}}\right)^{\beta}}\rho(\kappa_{j
        })dx_{j} d\kappa_{j},\\
        &=  2\hat{\mu}\kappa_{i}\bar{\kappa}\dfrac{\pi}{\beta}\csc \left(\dfrac{\pi}{\beta}\right).
    \end{aligned}
\end{equation}
Hence, the average number of links in the model is,
\begin{equation}
\begin{aligned}
    \bar{m} &= \dfrac{n}{2} \int_{\kappa_{0}}^{\infty} m(\kappa_{i},0) \rho(\kappa_{i})d\kappa_{i},\\
    &= n\hat{\mu}\bar{\kappa}^{2}\dfrac{\pi}{\beta}\csc \left(\dfrac{\pi}{\beta}\right).
\end{aligned}
\end{equation}

By requiring that $\bar{\kappa}=\bar{k}=2\bar{m}/n$ in the last equation, we get the expression for the chemical potential in Eq.~\eqref{eq: chemical potential}.

\subsection{Average energy $(\bar{\varepsilon})$}
\label{ssec: average energy}
Similarly, the average energy of the model is,
\begin{equation}
    \begin{aligned}
        \bar{\varepsilon} &= \mathbb{E}\left[\sum_{i<j}G_{ij}\varepsilon_{ij}\right],\\
        &= \binom{n}{2}\mathbb{E}[G_{ij}\varepsilon_{ij}],\\
        &= \dfrac{1}{2}\int_{-n/2}^{n/2}\int_{\kappa_{0}}^{\infty} \varepsilon(\kappa_{i},x_{i})\rho(\kappa_{i})dx_{i}d\kappa_{i},\\
        \text{where}~\varepsilon(x_{i}, \kappa_{i}) &= (n-1) \int_{-n/2}^{n/2}\int_{\kappa_{0}}^{\infty} p_{ij}\varepsilon_{ij}\dfrac{\rho(\kappa_{j})}{n}dx_{j} d\kappa_{j}.
    \end{aligned}
\end{equation}
Once again, due to the model being rotationally invariant, we set $x_{i} = 0$ without loss of generality, i.e., $\varepsilon(\kappa_{i},x_{i}) = \varepsilon(\kappa_{i},0)$, which in the $n \to \infty$ limit results in,
\begin{equation}
    \begin{aligned}
        \varepsilon(\kappa_{i},0) &= 2\int_{0}^{\infty}\int_{\kappa_{0}}^{\infty} \dfrac{\log \left(\dfrac{x_{j}}{\kappa_{i} \kappa_{j}}\right)}{1 + \left(\dfrac{x_{j}}{\hat{\mu} \kappa_{i} \kappa_{j}}\right)^{\beta}} \rho(\kappa_{j})d\kappa_{j}dx_{j},\\
        &=2\hat{\mu} \kappa_{i} \bar{\kappa}\csc \left(\dfrac{\pi}{\beta}\right)\dfrac{\pi}{\beta^{2}}\left(\beta\log(\hat{\mu}) - \pi \cot \left(\dfrac{\pi}{\beta}\right)\right).
    \end{aligned}
\end{equation}
Finally,
\begin{equation}
    \begin{aligned}
        \bar{\varepsilon} &= \dfrac{n}{2}\int_{\kappa_{0}}^{\infty} \varepsilon(0, \kappa_{i})\rho(\kappa_{i})d\kappa_{i},\\\
        &= n\hat{\mu}  (\bar{\kappa})^{2}\csc \left(\dfrac{\pi}{\beta}\right)\dfrac{\pi}{\beta^{2}}\left(\beta\log(\hat{\mu}) - \pi \cot \left(\dfrac{\pi}{\beta}\right)\right),\\
        &= \dfrac{n}{2}\bar{\kappa}\left(\log \left(\dfrac{\beta \sin\left(\frac{\pi}{\beta}\right)}{2\pi \bar{\kappa}}\right) - \dfrac{\pi}{\beta} \cot \left(\dfrac{\pi}{\beta}\right)\right),
    \end{aligned}
\end{equation}
where in the last step, we have used the expression for the chemical potential in Eq.~\eqref{eq: chemical potential}.

\subsection{Asymptotics of the coefficients of variation of \\ $m$ and $\e$}
\label{ssec: scaling of coefficient of variation}
To prove that the distribution of the number of links, $m(G)$ and energy, $\varepsilon(G)$ of the $\mathbb{S}^{1}$ model is self-averaging, we need to show that their respective coefficients of variation  $C^{(m)}_{V}(n) = \sqrt{\text{Var}(m)}/\mathbb{E}[m]$ and $C^{(\varepsilon)}_{V}(n) = \sqrt{\text{Var}(\varepsilon)}/\mathbb{E}[\varepsilon]$  go to zero in the limit $n \to \infty$. In Sec.~\ref {ssec: average links} and Sec.~\ref{ssec: average energy} we have already calculated the expected values of the number of links and energy, showing that $\bar{m}$ and $\bar{\varepsilon}$ both scale as $\mathcal{O}(n)$, for some fixed value of $\bar{\kappa}$. We next need to find the variance of the random variables $m(G)$ and $\varepsilon(G)$.
\par The variance of the number of links is,
\begin{equation}\label{eq: edge variance}
\begin{aligned}
    \text{Var}[m] &= \sum_{i<j}\text{Var}[G_{ij}] + 2\sum_{(i,j) <(k,l)}\text{Cov}[G_{ij}, G_{kl}].\\
\end{aligned}
\end{equation}
The first term scales as,
\begin{equation}
    \begin{aligned}
        \sum_{i<j}\text{Var}[G_{ij}] &= \binom{n}{2}\left(\mathbb{E}[G_{ij}^{2}] - \mathbb{E}[G_{ij}]^{2}\right)\\
        &= C_{1}n \cdot (1+o(1)),\\
        \text{where}~C_{1} &= \hat{\mu}\bar{\kappa}^{2}\dfrac{\pi}{\beta}\csc \left(\dfrac{\pi}{\beta}\right),
    \end{aligned}
\end{equation}
since $\mathbb{E}[G_{ij}^{2}] = \mathbb{E}[G_{ij}] = \mathcal{O}(n^{-1})$. To obtain the scaling of the second term, we first point out that non-zero contributions to the sum come from covariance between two edge indicators that share a common node, $\text{Cov}[G_{ij},G_{ik}]$, i.e., \textit{wedges} $\left(\inlinetriangle \right)$. The covariance terms that do not share a common node, $\text{Cov}[G_{ij}, G_{kl}] $, i.e. $\left(\inlineparlines \right)$, are simply zero because $G_{ij}$ and $G_{kl}$ are independent random variables. To that end,
\begin{equation}
    \begin{aligned}
        2\sum_{(i,j)<(k,l)}\text{Cov}[G_{ij}, G_{kl}] &= 6\binom{n}{3}\text{Cov}[G_{ij}, G_{ik}],\\
        &= n^{3}\text{Cov}[G_{ij}, G_{ik}]\cdot \left(1+o(1)\right),\\
    \text{where}~\text{Cov}[G_{ij}, G_{ik}] &=\mathbb{E}[G_{ij}G_{ik}] - \mathbb{E}[G_{ij}]\mathbb{E}[G_{ik}].
         \end{aligned}
\end{equation}
Once we get the asymptotics of the term $\mathbb{E}[G_{ij}G_{ik}]$ we are done, since $\mathbb{E}[G_{ij}]\mathbb{E}[G_{ik}] = (4C_{1}^{2}/n^{2}) \cdot (1+o(1))$. Notice that from the tower rule we have,
\begin{equation}
\begin{aligned}
        \mathbb{E}[G_{ij}G_{ik}] &= \mathbb{E}_{\kappa_{i},x_{i}}[\mathbb{E}[G_{ij}G_{ik}|\kappa_{i},x_{i}]],\\
        &= \mathbb{E}_{\kappa_{i},x_{i}}[\mathbb{E}[G_{ij}|\kappa_{i},x_{i}]\cdot\mathbb{E}[G_{ik}|\kappa_{i},x_{i}]],\\
        &= \mathbb{E}_{\kappa_{i},x_{i}}[\Phi(\kappa_{i},x_{i})^{2}],\\
        \text{where}~\Phi(\kappa_{i},x_{i}) &=  \int_{-n/2}^{n/2} \int_{\kappa_{0}}^{\infty}p_{ij} \dfrac{\rho(\kappa_{j})}{n}dx_{j}d\kappa_{j}.\\
\end{aligned}
\end{equation}
Once again using the fact that the model is rotationally invariant, we set $x_{i} = 0$ without loss of generality, i.e., $\Phi(\kappa_{i},x_{i}) = \Phi(\kappa_{i},0)$. For $n \gg 1$, we get,
\begin{equation}
    \Phi(\kappa_{i},0) = \dfrac{2\hat{\mu} \kappa_{i}\bar{\kappa}}{n}\dfrac{\pi}{\beta}\csc \left(\dfrac{\pi}{\beta}\right).
\end{equation}
Further,
\begin{equation}
    \begin{aligned}
         \mathbb{E}[G_{ij}G_{ik}] &= \mathbb{E}_{\kappa_{i},x_{i}}[\Phi(\kappa_{i},0)^{2}],\\
        &= \int_{\kappa_{0}}^{\infty} \left(\dfrac{2\hat{\mu} \kappa_{i}\bar{\kappa}}{n}\dfrac{\pi}{\beta}\csc \left(\dfrac{\pi}{\beta}\right)\right)^{2}\rho(\kappa_{i})d\kappa_{i},\\
        &= C_{2}\dfrac{\bar{\kappa^{2}}}{n^{2}},\\
        \text{where}~C_{2} &= \left(\dfrac{2\hat{\mu} \bar{\kappa}\pi}{\beta}\csc \left(\dfrac{\pi}{\beta}\right)\right)^{2}.
    \end{aligned}
\end{equation}
Combining these results and plugging them back into Eq.~\eqref{eq: edge variance},
\begin{equation}
    \text{Var}[m] = C_{1}\left(1 - 4C_{1}\right)n + C_{2}\bar{\kappa^{2}}n.
\end{equation}
For any $\gamma>2$, the first moment of the Pareto distribution is $\bar{\kappa} = \mathcal{O}(1)$ but the second moment scales as,
\begin{widetext}
\begin{equation}\label{eq: pareto second moment}
    \begin{aligned}
        \bar{\kappa^{2}} &= \int_{\kappa_{0}}^{\kappa_{0}n^{\frac{1}{\gamma - 1}}} \kappa^{2} \rho(\kappa)d\kappa\\
        &= 
        \begin{cases}
\dfrac{\gamma-1}{3-\gamma}\,\kappa_{0}^{2}\,
                n^{\frac{3-\gamma}{\gamma-1}}, & 2<\gamma<3,\\[6pt]
\kappa_{0}^{2}\,\ln n,                          & \gamma=3,\\[6pt]
\dfrac{\gamma-1}{\gamma-3}\,\kappa_{0}^{2},     & \gamma>3.
\end{cases}
    \end{aligned}
\end{equation}
Hence, depending on the value of $\gamma>2$, the scaling of the coefficients of variation of the distribution of the number of links, $C^{(m)}_{V}(n) = \sqrt{\text{Var}(m)}/\mathbb{E}[m]$, in the $\mathbb{S}^{1}$ model is given by,
\begin{equation}
C_{V}^{(m)}(n)=
\begin{cases}
\dfrac{\sqrt{C_{2}(\gamma-1)/(3-\gamma)}\,\kappa_{0}}{C_{1}}\;
      n^{-\frac{\gamma-2}{\gamma-1}} \cdot (1+o(1)), & 2<\gamma<3,\\[12pt]
\dfrac{\sqrt{C_{2}}\,\kappa_{0}}{C_{1}}\;
      \sqrt{\dfrac{\ln(n)}{n}} \cdot (1+o(1)), & \gamma=3,\\[12pt]
\dfrac{\sqrt{C_{1}(1-4C_{1})+
             C_{2}\kappa_{0}^{2}(\gamma-1)/(\gamma-3)}}{C_{1}}\;
      \dfrac{1}{\sqrt{n}} \cdot (1+o(1)), & \gamma>3.
\end{cases}
\end{equation}
\end{widetext}

\par The same procedure above is repeated verbatim to derive the asymptotics of the variance of energy,
\begin{equation}\label{eq: energy variance}
\begin{aligned}
    \text{Var}[\varepsilon] &= \sum_{i<j}\text{Var}[G_{ij}\varepsilon_{ij}] + 2\sum_{(i,j)<(k,l)}\text{Cov}[G_{ij}\varepsilon_{ij}, G_{kl}\varepsilon_{kl}].\\
\end{aligned}
\end{equation}
Consider the first term,
\begin{equation}
    \begin{aligned}
        \sum_{i<j}\text{Var}\left[G_{ij}\varepsilon_{ij}\right] &= \binom{n}{2}\left(\mathbb{E}[\left(G_{ij}\varepsilon_{ij}\right)^{2}] - \mathbb{E}[G_{ij}\varepsilon_{ij}]^{2}\right).\\
    \end{aligned}
\end{equation}
The expression for $\mathbb{E}[G_{ij}\varepsilon_{ij}]$ is calculated in Sec.~\ref{ssec: average energy}; on the other hand,
\begin{equation}
    \begin{aligned}
        \binom{n}{2}\mathbb{E}[\left(G_{ij}\varepsilon_{ij}\right)^{2}] &= \dfrac{1}{2}\int_{-n/2}^{n/2}\int_{\kappa_{0}}^{\infty} \varepsilon^{\prime}(\kappa_{i},x_{i})\rho(\kappa_{i})dx_{i}d\kappa_{i},\\
        \text{where}~\varepsilon^{\prime}(x_{i}, \kappa_{i}) &= (n-1) \int_{-n/2}^{n/2}\int_{\kappa_{0}}^{\infty} p_{ij}\varepsilon_{ij}^{2}\dfrac{\rho(\kappa_{j})}{n}dx_{j} d\kappa_{j}.
    \end{aligned}
\end{equation}
Once again, using the fact that the model is rotationally invariant, we set $x_{i} = 0$ without loss of generality, i.e., $\varepsilon^{\prime}(x_{i}, \kappa_{i}) = \varepsilon^{\prime}(0,\kappa_{i})$, which in the $n \to \infty$ limit results in,
\begin{equation}
    \begin{aligned}
        \varepsilon^{\prime}(0,\kappa_{i}) &= 2\int_{0}^{\infty}\int_{\kappa_{0}}^{\infty} \dfrac{\log^{2} \left(\dfrac{x_{j}}{\kappa_{i} \kappa_{j}}\right)}{1 + \left(\dfrac{x_{j}}{\hat{\mu} \kappa_{i} \kappa_{j}}\right)^{\beta}} \rho(\kappa_{j})d\kappa_{j}dx_{j},\\
        &= 2\hat{\mu} \kappa_{i}\bar{\kappa}\dfrac{\pi}{\beta}\csc\left(\dfrac{\pi}{\beta}\right) \bigg[\left(\log(\hat{\mu}) - \dfrac{\pi}{\beta} \cot\left(\dfrac{\pi}{\beta}\right)\right)^{2} \\
        &+ \left(\dfrac{\pi}{\beta}\csc\left(\dfrac{\pi}{\beta}\right)\right)^{2}\bigg] =\dfrac{ 2\kappa_{i}
        C_{1}^{\prime}}{\bar{\kappa}}.
    \end{aligned}
\end{equation}
Hence,
\begin{equation}
    \begin{aligned}
        \binom{n}{2}\mathbb{E}[\left(G_{ij}\varepsilon_{ij}\right)^{2}] &= nC_{1}^{\prime} \cdot (1+o(1)),
    \end{aligned}
\end{equation}
which is the asymptotic scaling of the first term in Eq.~\eqref{eq: energy variance}. Following a similar argument as in the previous case, the asymptotics of the second term in Eq.~\eqref{eq: energy variance} is,
\begin{equation}
    \begin{aligned}
        2\sum_{(i,j) < (k,l)}\text{Cov}[G_{ij}\varepsilon_{ij}, G_{kl}\varepsilon_{kl
        }] &= n^{3}\big(\mathbb{E}[G_{ij}\varepsilon_{ij}G_{ik}\varepsilon_{ik}]\\
        &- \mathbb{E}[G_{ij}\varepsilon_{ij}]\mathbb{E}[G_{ik}\varepsilon_{ik}]\big),\\
        \text{where},~\mathbb{E}[G_{ij}\varepsilon_{ij}]\mathbb{E}[G_{ik}\varepsilon_{ik}] &= (C_{2}^{\prime}/n^{2}) \cdot (1+o(1));\\
         \left(\bar{\kappa}\log \left(\hat{\mu}\right) - \bar{\kappa}\dfrac{\pi}{\beta} \cot \left(\dfrac{\pi}{\beta}\right)\right)^{2} &= C_{2}^{\prime}. 
    \end{aligned}
\end{equation}
Further,
\begin{equation}
    \begin{aligned}
        \mathbb{E}[G_{ij}\varepsilon_{ij}G_{ik}\varepsilon_{ik}] &=\mathbb{E}_{\kappa_{i},x_{i}}[\Phi^{\prime}(\kappa_{i},x_{i})^{2}],\\
        \text{where}~\Phi^{\prime}(\kappa_{i},x_{i}) &=  \int_{-n/2}^{n/2} \int_{\kappa_{0}}^{\infty}p_{ij}\varepsilon_{ij} \dfrac{\rho(\kappa_{j})}{n}dx_{j}d\kappa_{j}.\\
    \end{aligned}
\end{equation}
But $\Phi^{\prime}(\kappa_{i},x_{i}) = \Phi^{\prime}(\kappa_{i},0)$, which in the setting $n\gg 1$ is,
\begin{equation}
\begin{aligned}
    \Phi^{\prime}(\kappa_{i},0) & = \dfrac{2\hat{\mu} \kappa_{i}}{n
    }\bar{\kappa}\csc \left(\dfrac{\pi}{\beta}\right)\dfrac{\pi}{\beta^{2}}\left(\beta\log(\hat{\mu}) - \pi \cot \left(\dfrac{\pi}{\beta}\right)\right),\\
    &= \dfrac{\kappa_{i}\sqrt{C_{2}^{\prime}}}{n\bar{\kappa}},
\end{aligned}
\end{equation}
leading to the result,
\begin{widetext}
\begin{equation}
    \begin{aligned}
        \mathbb{E}[G_{ij}\varepsilon_{ij}G_{ik}\varepsilon_{ik}] &= \int_{\kappa_{0}}^{\infty} \Phi^{\prime}(\kappa_{i},0)^{2} \rho(\kappa_{i})d\kappa_{i}\\
        &= C_{2}^{\prime} \dfrac{\bar{\kappa^{2}}}{\bar{\kappa}^{2}n^{2}}.
    \end{aligned}
\end{equation}
Plugging all these results into Eq.~\eqref{eq: energy variance} we have,
\begin{equation}
    \begin{aligned}
        \text{Var}[\varepsilon] &= n\left(C_{1}^{\prime} - C_{2}^{\prime}\right) + nC_{2}^{\prime}\dfrac{\bar{\kappa^{2}}}{\bar{\kappa}^{2}}
    \end{aligned}
\end{equation}

Hence, depending on the value of $\gamma>2$, the scaling of the coefficients of variation of the distribution of the energy, $C^{(\varepsilon)}_{V}(n) = \sqrt{\text{Var}(\varepsilon)}/\mathbb{E}[\varepsilon]$, in the $\mathbb{S}^{1}$ model is given by,
\begin{equation}
|C_{V}^{(\varepsilon)}(n)|=
\begin{cases}
\dfrac{2\kappa_{0}}{\bar{\kappa}}\sqrt{\dfrac{\gamma - 1}{3 - \gamma}}\;
      n^{-\frac{\gamma-2}{\gamma-1}} \cdot (1+o(1)), & 2<\gamma<3,\\[12pt]
\dfrac{2\kappa_{0}}{\bar{\kappa}}\;
      \sqrt{\dfrac{\ln(n)}{n}} \cdot (1+o(1)), & \gamma=3,\\[12pt]
2\sqrt{\dfrac{\left(C_{1}^{\prime} - C_{2}^{\prime}\right)+
             C_{2}^{\prime}\frac{\kappa_{0}^{2}(\gamma-1)}{\bar{\kappa}^{2}(\gamma-3)}}{C_{2}^{\prime}}}\;
      \dfrac{1}{\sqrt{n}} \cdot (1+o(1)), & \gamma>3.
\end{cases}
\end{equation}
\end{widetext}

We conclude that the coefficients of variation of both the distribution of the number of links and energy scale as
\begin{equation}
C_{V}(n) = \mathcal{O}\left(n^{-\text{min} \left \{\frac{1}{2} , \frac{\gamma - 2}{\gamma - 1}\right \} }(\ln(n))^{\frac{\mathbbm{1}\{\gamma = 3\}}{2}}\right)
\end{equation}
and hence are self-averaging distributions for any $\gamma>2$ since $\lim_{n\to \infty}C_{V}(n) = 0$.

\subsection{Clustering}

\begin{figure}[t]
    \centerline{\includegraphics[width=0.95\columnwidth]{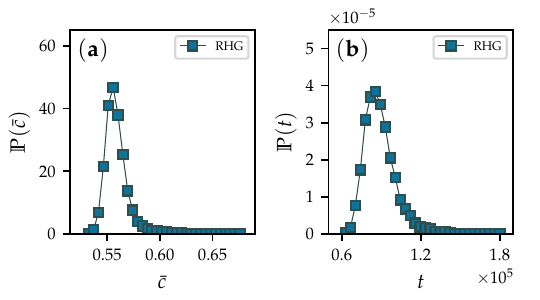}}
    \caption{\textbf{Clustering and number of triangles in RHGs.} The distribution of \textbf{(a)}~the average local clustering~$\bar{c}$ and \textbf{(b)}~the number of triangles~$t$ in $10^4$ RHGs with parameters $n=10^4$, $\bar{\kappa} = 10$, $\beta = 2.5$, and $\gamma = 2.7$.}
    \label{fig:clustering-singlepeak}
\end{figure}

It is well known that the underlying geometry in RHGs is responsible for nonvanishing clustering in the model~\cite{krioukov2010_rhg}, so one may wonder whether such an abundance of triangles leads to any degeneracy akin to the one in the Strauss model, where the distribution of the number of triangles is not single-peaked but double-peaked~\cite{park2005_Solutionpropertiesclustered,horvat2015_ReducingDegeneracyMaximum,Handcock2003}. Figure~\ref{fig:clustering-singlepeak} shows that the distributions of clustering and the number of triangles in RHGs are single-peaked. The main reason for that is the lack of degeneracy of RHGs proven in the previous section.

\section{Deterministic hyperbolic graphs (DHGs)}
\label{sec:DHG}
In this section, we outline the exact details involved in the deterministic construction of typical networks in the random hyperbolic graph (RHG) model (Appendix \ref{sec:RHG}), which we call \textit{deterministic hyperbolic graphs} (DHGs).

The input parameters are the same as in the RHG model: $n$, $\bar{\kappa}$, $\beta$, and $\gamma$. The rest is different. First, hidden variables, $(\kappa_{i},x_{i})$, are assigned to each node deterministically via the two-dimensional hypercanonical derandomization in Sec.~\ref{sec: hypercanonical models}. This derandomization returns deterministic coordinates $(\tilde{x}_{i},\tilde{y}_{i})$ on the unit square, which we then use to obtain the RHG hidden variables. We set the similarity space coordinate to $x_{i}= n\tilde{x}_{i}$, while the expected degree $\kappa_{i}$ is obtained via the inverse CDF of the Pareto distribution applied to~$\tilde{y}_{i}$:
\begin{equation}\label{eq:DHG-kappa}
    \kappa_{i} = \bar{\kappa}\frac{\gamma - 2}{\gamma - 1}(1-\tilde{y
    }_{i})^{\frac{1}{1-\gamma}}.
\end{equation}

Once the hidden variables are determined, the value of the chemical potential $\hat{\mu}$ in Eq.~\eqref{eq: S1 connection probability} is obtained by finding the root of the equation
\begin{equation}
    \sum_{i<j}p_{ij} - \dfrac{n\bar{\kappa}}{2} = 0
\end{equation}
numerically using Newton's method with analytically calculated derivative of the objective function and absolute tolerance of $\tau=10^{-8}$. This ensures that the DHG exactly matches the desired value of one sufficient statistic, the number of edges~$m^\star=\lfloor n\bar{\kappa}/2 \rceil$. Another option would be to use the analytical thermodynamic-limit expression for $\hat{\mu}$ in Eq.~\eqref{eq: chemical potential}, but the convergence of the number of edges $m$ and energy  $\varepsilon$ of DHGs to the ensemble average of RHGs would be slower in this case. Another reason to avoid this option is that upon hypercanonical derandomization, we end up dealing with a unique single instance of the grand canonical ensemble, so we can use it to find the exact value of the chemical potential.

Finally, having this grand canonical ensemble in place, in which all the pairwise connection probabilities are now determined, we use the grand canonical derandomization procedure in Sec.~\ref{ssec: grand canonical models methods} to deterministically place all the $m^\star$ edges.

We make the software package implementing the described DHG construction algorithm available at~\cite{dhg_code}. Its computational complexity is the same $\mathcal{O}(n^2)$ as RHGs', both dominated by computing the $n \choose 2$ connection probabilities~$p_{ij}$. The steps involved in derandomization are subdominant: sampling $4n$ times from the the multiplicative congruential generator takes $\mathcal{O}(n)$ time, while sorting $4n$ elements is~$\mathcal{O}(n \log n)$.

\section{Real-world networks}
\label{sec:empirical_methods}

\subsection{Parameter estimation}
\label{sec:parameter_estimation}

To compare real-world networks with the corresponding most typical networks in the deterministic hyperbolic graph (DHG) model, we first have to measure the value of the model parameters---the numbers of nodes~$n$ and links~$m$, the degree distribution exponent $\gamma$, and the inverse temperature $\beta$---in the real-world networks. We do this as follows.

Determining the number of nodes and links is trivial. To measure the degree distribution exponent, we follow the prescription in~\cite{voitalov2019_welldone}. We employ the \texttt{tail-estimation} software package~\cite{voitalov2025_ivanvoitalovtailestimation}, and use three different estimators of the extreme value exponent $\xi$, which is related to the degree exponent by $\gamma =1 + 1/\xi$. These estimators are the Kernel estimator $\hat{\gamma}_\mathrm{K}$, Hill estimator $\hat{\gamma}_\mathrm{H}$, and Moment estimator $\hat{\gamma}_\mathrm{M}$. We use the value of $\gamma$ determined with the Moments estimator to construct networks in the DHG model.

To determine the inverse temperature $\beta$ for fixed values of $n$, $m$, and $\gamma$, we first compute the average clustering in a real-world network, excluding nodes of degrees $<2$: 
\begin{equation}
    \bar{c} = \frac{1}{|V_{k\geq 2}|}\sum_{i\in V_{k\geq2}} c_i,
\end{equation}
where $V_{k\geq2} = \{i \in V: k_i\geq 2\}$ is the set of nodes with degree $k$ at least two, $c_i = t_i/\binom{k_i}{2}$ is the clustering coefficient of node~$i$, $k_i$ is its degree, and $t_i$ is the number of triangles it is in. Since the DHG is a deterministic model it generates a single graph with a fixed value of average clustering $\bar{c}_\mathrm{DHG}(n,m,\gamma,\beta)$ for any combination of its parameters. We then use Brent's method~\cite{brent2013_algorithmsminimizationderivatives}, as implemented in \texttt{SciPy}~\cite{virtanen2020_scipy10fundamental} (v1.16.1) to find the root $\beta^\star$ of
\begin{equation}
    f(\beta)= \bar{c} - \bar{c}_\mathrm{DHG}(n,\bar{k},\gamma,\beta)
\end{equation}
in the interval $[1,10]$ and treat it as an estimate $\hat{\beta}=\beta^\star$ of the inverse temperature. Since clustering is an increasing function of~$\beta$, if we do not find a root in this interval and the empirical clustering is larger than the clustering in the model at $\beta=10$, we set $\hat{\beta} = 10$. If we do not find a root in the interval and the empirical clustering is lower than the clustering in the model for $\beta = 1$, we set $\hat{\beta}=1$.

\subsection{Network data and network properties}
\label{sec:network_properties}

We infer the parameters discussed in the previous section for \numrealworld networks in the \texttt{Netzschleuder} network repository~\cite{peixoto2020_netzschleudernetworkcatalogue}, which contains a variety of real-world networks from different domains, including social networks, biological networks, transportation networks, technological networks, and literature/word networks. For each network, as soon as its parameters are inferred, we also construct the unique DHG network with the same parameter values, using the code available at~\cite{dhg_code}.

In the resulting \numrealworld pairs of real-world and DHG networks, we then measure the following network properties:
\begin{itemize}
\item distribution $\mathbb{P}(k)$ of node degrees~$k$, and the average degree $\bar{k}=\sum_k kP(k)=2m/n$;
\item average local clustering coefficient $c(k)$ of $(k>1)$-degree nodes, and the average clustering $\bar{c}=\sum_{k>1} c(k)P(k)$;
\item average degree $k_\mathrm{nn}(k)$ of neighbors of $k$-degree nodes, and the Spearman correlation coefficient $\rho$ of degrees of connected nodes;
\item distribution $\mathbb{P}(d)$ of shortest path distances~$d$ between all pairs of nodes, and the average distance $\bar{d}=\sum_d dP(d)$;
\item average betweenness centrality $b(k)$ of $k$-degree nodes;
\item distribution $\mathbb{P}(n_\mathrm{cn})$ of the number of common neighbors $n_\mathrm{cn}$ of all pairs of connected nodes, also known as edge multiplicity~\cite{zlatic2012_networksarbitraryedge}; and
\item fraction $\nu^\mathrm{lcc}$ of nodes in the largest connected component. 
\end{itemize}

\subsection{Network visualization}
\label{sec:visualization}

We also develop an algorithm inspired by \texttt{LaNet-vi}~\cite{NIPS2005_b19aa25f,beiro2008_lowcomplexityvisualization, alvarez-hamelin2011_understandingedgeconnectivity} to visualize the largest connected component of all networks. Note, that all quantities in this section are computed after subsetting a given network to its largest connected component. The visualization algorithm assigns a radial $r_i$ and angular $\varphi_i\in[0,2\pi)$ coordinate to each node~$i$ as follows.

First, we determine the coreness $u_i$ of node $i$ in the $k$-core decomposition. We encode this value using the color~\cite{velden2020_cmasherscientificcolormaps} of a vertex. Nodes of the same value of coreness $u$ form a shell $Q_u=\{i\in V : u_i=u\}$. We also compute the adjusted coreness $u'_i$ of node $i$ as a convex combination of its own coreness $u_i$ and the coreness of its neighbors,
\begin{equation}\label{eq:adjusted_coreness}
    u_i' = \varepsilon u_i + (1-\varepsilon)\frac{1}{k_i}\sum_{j\in V} G_{ij}u_j, 
\end{equation}
where $k_i$ is the degree of node $i$, $G_{ij}$ the adjacency matrix, and $\varepsilon\in[0,1]$ a parameter of the algorithm that governs the relative influence of a nodes' coreness $u_i$ and the average coreness of its neighbors on the adjusted coreness. We set $\varepsilon=0.85$ for our visualization.

The radial coordinate $r_i$ of node $i$ is then
\begin{equation}
    r_i=(r_1-r_0) \frac{u'_\mathrm{max}-u'_i}{u'_\mathrm{max}-u'_\mathrm{min}}+r_0,
\end{equation}
where $r_1\geq r_0>0$ are constants governing the maximum and minimum radial coordinate in the visualization, and $u'_{\min}=\min\{u'_i : i\in V\}$ and $u'_{\max}=\max\{u'_i : i\in V\}$ are smallest and largest values of adjusted coreness.

The angular coordinate $\varphi_i$ of node $i$ is determined by constructing the node-induced subgraphs $S_u = (Q_u, (Q_u\times Q_u)\cap E)$ of a given shell $Q_u$. We denote the $N_u$ connected components of $S_u$ as $C_{u,j}$ where $j\in\{1,2,\dots,N_u\}$ enumerate them in arbitrary order. Each connected component $C_{u,j}$ is then assigned an angular segment $\Phi_{u,j}$ of arc length proportional to their relative size:
\begin{equation}
    \Phi_{u,j}=
    \left[
        \sum_{i=1}^{j-1}\frac{2\pi}{N_u}\frac{|C_{u,i}|}{|Q_u|},
        \sum_{i=1}^{j}\frac{2\pi}{N_u}\frac{|C_{u,i}|}{|Q_u|}
    \right).
\end{equation}

Finally, the angular coordinate $\varphi_i$ of node $i$ belonging to connected component $C_{u_i,j_i}$ is chosen uniformly at random in $\Phi_{u,j}$, i.e., $\varphi_i\sim \mathrm{U}[\Phi_{u_i,j_i}]$.

All nodes also have sizes~$s_i$, which grows with the logarithm of their degree $k_i$:
\begin{equation}
    s_i = (s_1-s_0) + \frac{\log k_i-\log k_\mathrm{min}}{\log k_\mathrm{max}-\log k_\mathrm{min}}+ s_0
\end{equation}
where $s_1\geq s_0>0$ are constants that govern the maximum and minimum node size respectively, and $k_\mathrm{max},k_\mathrm{min}$ are the maximum and minimum degree.

We only show a subset of all edges. Depending on network size and density we randomly select a subset of edges. Our visualizations retain between $0.25\%$ and $100\%$ of edges.

We make the code implementing the described graph visualization procedure available at~\cite{kcore_viz_code}. 

\FloatBarrier
\subsection{Real-world networks vs.\ their DHG counterparts}
\label{sec:empirical_results}

Tables~\ref{tab:real-world}-\ref{tab:wasserstein} and Figures~\ref{fig:real_world_inploid}-\ref{fig:real_world_game_thrones} present the results of the inferences, measurements, and visualizations described in the preceding sections for all the \numrealworld network pairs. In what follows, we provide some network-specific comments for each network pair.

\begin{table*}
\begin{tabular}{l|c|c|c|c|c|c|c|c|c|c|c|c|c|c|c|c}
\hline
name & $n$ & $\bar{k}_\mathrm{det}$ & $\bar{k}_\mathrm{real}$ & $\hat{\gamma}_\mathrm{H}$ & $\hat{\gamma}_\mathrm{M}$ & $\hat{\gamma}_\mathrm{K}$ & $\hat{\beta}$ & $\bar{c}_\mathrm{det}$ & $\bar{c}_\mathrm{real}$ & $\bar{d}_\mathrm{det}$ & $\bar{d}_\mathrm{real}$ & $\nu^\mathrm{lcc}_\mathrm{det}$ & $\nu^\mathrm{lcc}_\mathrm{real}$ & $\rho_\mathrm{det}$ & $\rho_\mathrm{real}$ & domain \\ \hline
\texttt{inploid}~\cite{gursoy2018_Influencemaximizationsocial} & 39749 & 2.49 & 2.49 & 2.47 & 2.52 & 2.40 & 2.05 & 0.43 & 0.43 & 3.38 & 5.19 & 0.36 & 0.66 & -0.29 & 0.02 & social   \\
\texttt{cora}~\cite{mccallum2000_automatingconstructioninternet} & 23166 & 7.70 & 7.70 & 3.31 & 3.44 & 3.57 & 1.82 & 0.31 & 0.31 & 5.86 & 6.35 & 1.00 & 1.00 & 0.18 & 0.09 & social  \\
\texttt{anybeat}~\cite{fire2013_LinkPredictionHighly} & 12645 & 7.77 & 7.77 & 2.37 & 2.49 & 2.54 & 1.72 & 0.40 & 0.40 & 3.17 & 3.63 & 1.00 & 0.99 & -0.31 & -0.04 & social \\
\texttt{bitcoin\_trust}~\cite{kumar2016_EdgeWeightPrediction} & 5881 & 7.31 & 7.31 & 2.50 & 2.80 & 2.24 & 1.60 & 0.29 & 0.29 & 3.57 & 4.19 & 1.00 & 1.00 & -0.13 & -0.00 & social  \\
\texttt{linux}~\cite{kunegis2013_KONECTKoblenznetwork} & 30837 & 13.83 & 13.83 & 2.60 & 2.03 & 1.95 & 1.00 & 0.14 & 0.31 & 3.25 & 2.82 & 1.00 & 1.00 & -0.02 & -0.27 & technological  \\
\texttt{internet\_as}~\cite{karrer2014_percolationsparsenetworks} & 22963 & 4.22 & 4.22 & 2.09 & 2.10 & 2.13 & 1.20 & 0.35 & 0.35 & 3.84 & 3.47 & 1.00 & 0.83 & 0.14 & -0.26 & technological  \\
\texttt{google}~\cite{palla2007_directednetworkmodules} & 15763 & 18.85 & 18.85 & 2.52 & 2.34 & 2.15 & 2.24 & 0.60 & 0.60 & 2.52 & 2.73 & 1.00 & 1.00 & -0.09 & -0.11 &   technological \\
\texttt{jdk}~\cite{kunegis2013_KONECTKoblenznetwork} & 6434 & 16.68 & 16.68 & 2.28 & 2.22 & 2.13 & 2.39 & 0.68 & 0.68 & 2.12 & 2.50 & 1.00 & 1.00 & -0.25 & -0.19 & technological  \\
\texttt{interactome\_figeys}~\cite{ewing2007_largescalemappinghuman} & 2239 & 5.75 & 5.75 & 2.45 & 2.38 & 2.43 & 1.00 & 0.07 & 0.09 & 3.84 & 3.84 & 0.99 & 0.99 & -0.01 & -0.06 & biological  \\
\texttt{celegans\_metabolic}~\cite{duch2005_communitydetectioncomplex} & 453 & 8.94 & 8.94 & 2.55 & 2.45 & 2.59 & 2.69 & 0.66 & 0.65 & 2.66 & 2.60 & 1.00 & 0.99 & -0.20 & -0.17 & biological  \\
\texttt{celegansneural}~\cite{white1997_structurenervoussystem, watts1998_wsmodel} & 297 & 14.46 & 14.46 & 3.36 & 4.08 & 7.54 & 1.73 & 0.31 & 0.31 & 2.45 & 2.61 & 1.00 & 1.00 & -0.11 & 0.09 &  biological \\
\texttt{marvel\_universe}~\cite{alberich2002_marveluniverselooks} & 19428 & 9.83 & 9.83 & 2.55 & 2.27 & 2.14 & 1.00 & 0.00 & 0.10 & 4.45 & 3.42 & 0.99 & 1.00 & -0.03 & -0.08 & literary   \\
\texttt{bible\_nouns}~\cite{peixoto2020_netzschleudernetworkcatalogue} & 1773 & 10.30 & 10.30 & 3.04 & 3.30 & 2.91 & 6.57 & 0.72 & 0.72 & 3.37 & 4.52 & 0.96 & 1.00 & 0.37 & 0.14 & literary  \\
\texttt{game\_thrones}~\cite{beveridge2016_networkthrones} & 107 & 6.58 & 6.58 & 4.01 & 3.08 & 3.20 & 3.97 & 0.65 & 0.65 & 2.88 & 2.69 & 1.00 & 0.99 & 0.09 & 0.04 & literary  \\
\texttt{faa\_routes}~\cite{peixoto2020_netzschleudernetworkcatalogue} & 1226 & 3.93 & 3.93 & 3.79 & 12.09 & 7.35 & 1.16 & 0.07 & 0.07 & 5.92 & 6.25 & 1.00 & 1.00 & 0.04 & 0.10 & transport  \\
\texttt{euroroad}~\cite{subelj2011_robustnetworkcommunit} & 1174 & 2.41 & 2.41 & 6.50 & 23.86 & - & 1.00 & 0.02 & 0.03 & 18.35 & 9.93 & 0.89 & 0.94 & 0.23 & 0.02 & transport   \\
\texttt{london\_transport}~\cite{dedomenico2014_navigabilityinterconnectednetworks} & 369 & 2.33 & 2.33 & 4.17 & 4.60 & 1.00 & 1.00 & 0.03 & 0.04 & 13.70 & 7.42 & 1.00 & 0.89 & 0.19 & 0.04 & transport \\
\end{tabular}
\caption{\textbf{Basic network properties of real-world versus DHG networks.} The table displays the measured values of the deterministic hyperbolic graph~(DHG) parameters (Section~\ref{sec:parameter_estimation})---number of nodes $n$, average degree $\bar{k}$, inverse temperature estimate~$\hat{\beta}$, and degree distribution exponent $\hat{\gamma}_\mathrm{M}$ estimated by the Moments estimator---for the \numrealworld real-world networks from the \texttt{Netzschleuder} network repository (Section~\ref{sec:network_properties}). The table also reports the values of the degree distribution exponent estimated by the the Kernel~$\hat{\gamma}_\mathrm{K}$ and Hill~$\hat{\gamma}_\mathrm{H}$ estimators, and juxtaposes the real-world (subscript~``$\mathrm{real}$'') and DHG (subscript~``$\mathrm{det}$'') networks in terms of average local clustering $\bar{c}$, average shortest path length $\bar{d}$, fraction of nodes in the largest connected component $\nu^\mathrm{lcc}$, and the Spearman correlation coefficient~$\rho$ of the degrees of connected nodes.}
\label{tab:real-world}
\end{table*}

\begin{table*}
\begin{tabular}{l|r|r|r|r|r|r|r|r}
\hline
 name & $n$ & $k$ & $k_\mathrm{nn}$ & $c$ & $b$ & $n_\mathrm{cn}$ & $u$ & $d$ \\ \hline
\texttt{inploid} & 39749 & 1.716119 & 783.883968 & 0.046382 & 0.000038 & 4.336344 & 1.120582 & 1.816208 \\
\texttt{cora} & 23166 & 2.406889 & 10.764882 & 0.066461 & 0.000084 & 0.433359 & 1.329491 & 0.503021 \\
\texttt{anybeat} & 12645 & 3.926295 & 1162.883300 & 0.083438 & 0.000058 & 5.040788 & 2.539897 & 0.461494 \\
\texttt{bitcoin\_trust} & 5881 & 4.088420 & 115.115918 & 0.080320 & 0.000181 & 2.711893 & 2.346370 & 0.614713 \\
\texttt{linux} & 30837 & 3.272627 & 213.043462 & 0.168785 & 0.000030 & 5.942050 & 1.317508 & 0.431453 \\
\texttt{internet\_as} & 22963 & 1.014589 & 135.968854 & 0.074218 & 0.000065 & 0.839561 & 0.628054 & 0.370889 \\
\texttt{google} & 15763 & 4.595699 & 995.292727 & 0.077474 & 0.000076 & 2.914769 & 4.411660 & 0.234965 \\
\texttt{jdk} & 6434 & 4.015853 & 1910.040637 & 0.027206 & 0.000198 & 4.168325 & 2.051445 & 0.385152 \\
\texttt{interactome\_figeys} & 2239 & 2.436802 & 64.209666 & 0.034678 & 0.000541 & 0.289646 & 1.456454 & 0.151930 \\
\texttt{celegans\_metabolic} & 453 & 1.302428 & 5.288066 & 0.027666 & 0.001014 & 0.947654 & 0.313466 & 0.081858 \\
\texttt{celegansneural} & 297 & 2.727273 & 13.497955 & 0.030465 & 0.001502 & 0.557728 & 1.124579 & 0.165879 \\
\texttt{marvel\_universe} & 19428 & 1.981985 & 41.983669 & 0.102819 & 0.000071 & 1.723185 & 1.181954 & 1.029496 \\
\texttt{bible\_nouns} & 1773 & 4.042865 & 19.803855 & 0.120285 & 0.001009 & 1.540576 & 1.684715 & 1.142272 \\
\texttt{game\_thrones} & 107 & 1.233645 & 1.259529 & 0.029693 & 0.008181 & 0.278409 & 0.383178 & 0.184580 \\
\texttt{faa\_routes} & 1226 & 1.340946 & 3.010379 & 0.026420 & 0.002524 & 0.122093 & 0.462480 & 0.500701 \\
\texttt{euroroad} & 1174 & 0.201022 & 0.244888 & 0.014330 & 0.006357 & 0.008469 & 0.090290 & 8.430740 \\
\texttt{london\_transport} & 369 & 0.525745 & 0.649403 & 0.011271 & 0.020670 & 0.011628 & 0.062331 & 6.271198 \\
\end{tabular}
\caption{\textbf{Wasserstein distance between the property distributions in real-world versus DHG networks.} The reported properties are the node degree~$k$, the average nearest neighbor degree~$k_\mathrm{nn}$, the local clustering coefficient~$c$, the betweenness centrality~$b$, the edge multiplicity~$n_\mathrm{cn}$, the coreness~$u$, and the shortest-path distance~$d$ between pairs of nodes at positive finite distance. All the properties are first converted to distributions---in case of clustering, for example, the clustering distribution $\mathbb{P}({c})$ is the probability that a random node in the network has clustering~$c$---and then the Wasserstein-1 distances between the two distributions---one in the real-world network and the other one in its DHG counterpart---are computed. The Wasserstein distance between two distributions with a support of size~$S$ is bounded by~$S$. The sizes of the supports of the distributions in the table are either~$\sim n$---the degree values cannot exceed $n-1$, for instance---or the unit interval in cases like clustering or betweenness, for example. Most numbers in the table are small compared to their respective maxima.}
\label{tab:wasserstein}
\end{table*}

\subsubsection{Social Networks}

\begin{figure*}[tb]
    \centering
    \includegraphics[width=0.95\textwidth]{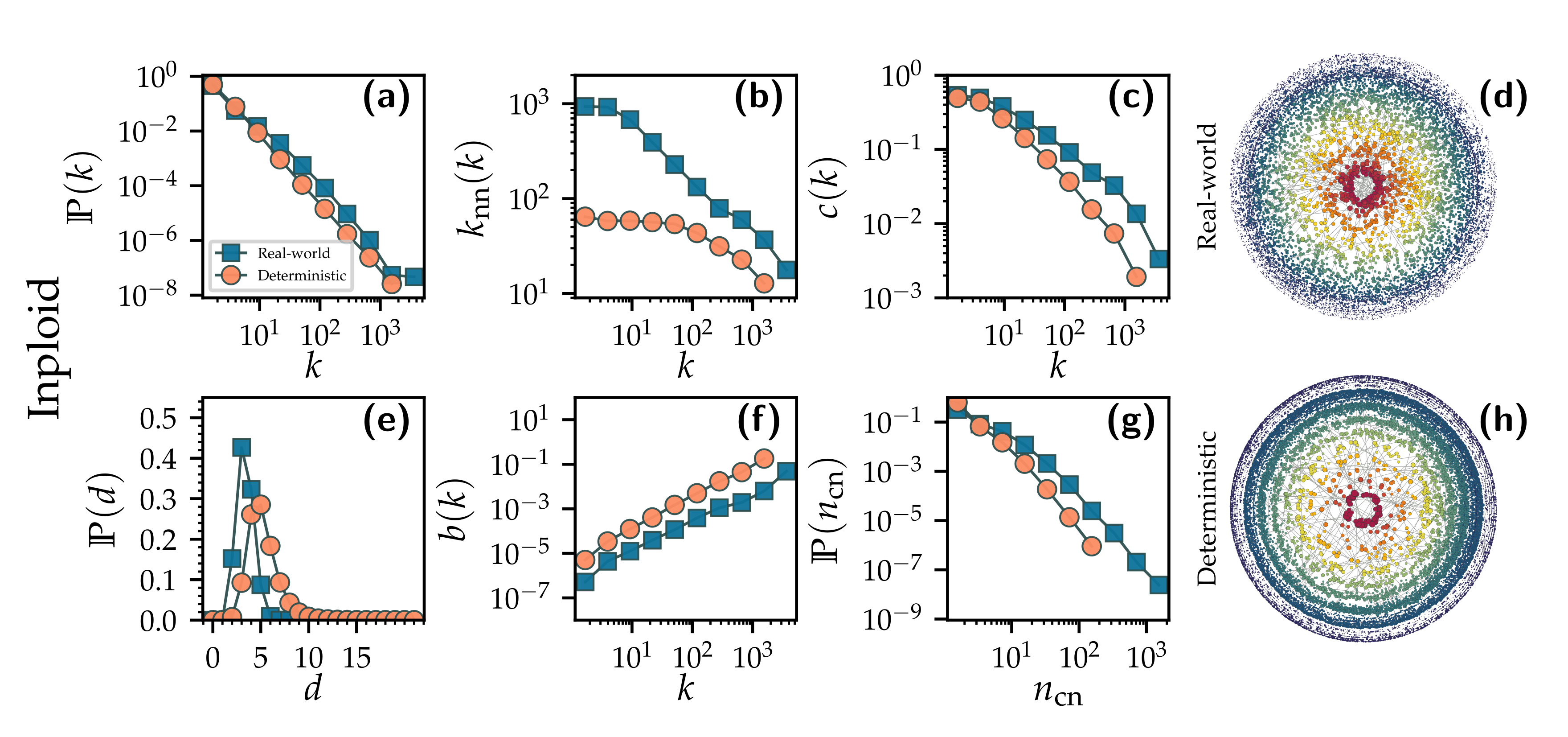}
    \caption{Network properties (degree distribution $\mathbb{P}(k)$ (panel \textbf{(a)}), average nearest neighbor degree $k_\mathrm{nn}(k)$ (panel \textbf{(b)}), local clustering  $c(k)$ (panel \textbf{(c)}), pairwise distance distribution $\mathbb{P}(d)$ (panel \textbf{(e)}), betweenness centrality $b(k)$ (panel \textbf{(f)}), edge multiplicity distribution $\mathbb{P}(n_\mathrm{cn})$ (panel \textbf{(g)})) of the \texttt{inploid} network~\cite{gursoy2018_Influencemaximizationsocial} (blue squares) and a DHG graph (orange circles) with matching parameters, as well as artistic representations of the real-world (panel \textbf{(d)}) and deterministic (panel \textbf{(h)}) network.
    }
    \label{fig:real_world_inploid}
\end{figure*}

The \texttt{inploid} network~\cite{gursoy2018_Influencemaximizationsocial} (Fig.~\ref{fig:real_world_inploid}) represents the Turkish online Q\&A platform. Nodes are users, and edges between nodes indicate that at least one of the two users follows the other. The DHG model (orange circles) with matched parameters is able to reproduce the broad degree distribution (panel \textbf{(a)}), clustering (panel \textbf{(c)}) small worldness (panel \textbf{(e)}) of the real-world network (blue squares). However, the model only qualitatively captures disassortativity (panel \textbf{(b)}) overestimates betweenness centrality (panel \textbf{(f)}), has a less broad distribution of common neighbors (panel \textbf{(g)}). We provide artistic visualizations of the real-world network (panel \textbf{(d)}) and its deterministic counterpart (panel \textbf{(h)}), showing $0.35\%$ of edges.

\begin{figure*}[tb]
    \centering
    \includegraphics[width=0.95\textwidth]{figures/real_world_final/cora_plot.png}
    \caption{Network properties (degree distribution $\mathbb{P}(k)$ (panel \textbf{(a)}), average nearest neighbor degree $k_\mathrm{nn}(k)$ (panel \textbf{(b)}), local clustering  $c(k)$ (panel \textbf{(c)}), pairwise distance distribution $\mathbb{P}(d)$ (panel \textbf{(e)}), betweenness centrality $b(k)$ (panel \textbf{(f)}), edge multiplicity distribution $\mathbb{P}(n_\mathrm{cn})$ (panel \textbf{(g)})) of the \texttt{cora} network~\cite{mccallum2000_automatingconstructioninternet} (blue squares) and a DHG graph (orange circles) with matching parameters, as well as artistic representations of the real-world (panel \textbf{(d)}) and deterministic (panel \textbf{(h)}) network.}
    \label{fig:real_world_cora}
\end{figure*}

In the \texttt{cora} citation network~\cite{mccallum2000_automatingconstructioninternet} (Fig.~\ref{fig:real_world_cora}) nodes are scientific publications and edges encode that one paper cites the other. The DHG model (orange circles) accurately describes many properties of the real-world network (blue squares). This includes the degree distribution (panel \textbf{(a)}), assortativity (panel \textbf{(b)}) clustering (panel \textbf{(c)}), pairwise distance distribution (panel \textbf{(e)}), betweenness centrality (panel \textbf{(f)}), edge multiplicity distribution (panel \textbf{(g)}). We visualize the real-world (panel \textbf{(d)}) and deterministic (panel \textbf{(h)}) network, showing $0.25\%$ of edges.

\begin{figure*}[tb]
    \centering
    \includegraphics[width=0.95\textwidth]{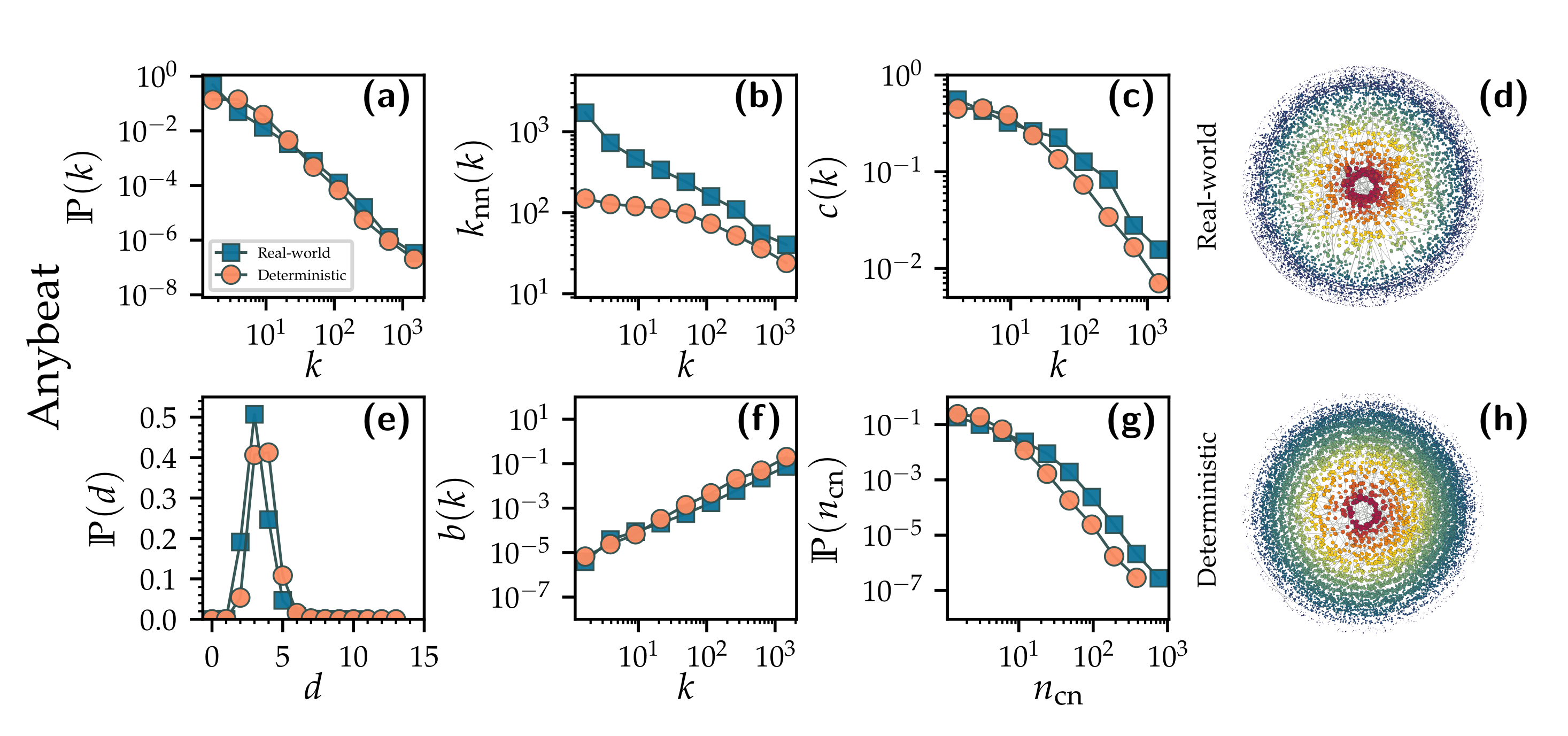}
    \caption{Network properties (degree distribution $\mathbb{P}(k)$ (panel \textbf{(a)}), average nearest neighbor degree $k_\mathrm{nn}(k)$ (panel \textbf{(b)}), local clustering  $c(k)$ (panel \textbf{(c)}), pairwise distance distribution $\mathbb{P}(d)$ (panel \textbf{(e)}), betweenness centrality $b(k)$ (panel \textbf{(f)}), edge multiplicity distribution $\mathbb{P}(n_\mathrm{cn})$ (panel \textbf{(g)})) of the \texttt{anybeat} network~\cite{fire2013_LinkPredictionHighly} (blue squares) and a DHG graph (orange circles) with matching parameters, as well as artistic representations of the real-world (panel \textbf{(d)}) and deterministic (panel \textbf{(h)}) network.}
    \label{fig:real_world_anybeat}
\end{figure*}

Anybeat was an online social network site. Nodes in the \texttt{anybeat} network~\cite{fire2013_LinkPredictionHighly} (Fig.~\ref{fig:real_world_anybeat}) are users of this site. An edge between two nodes indicates that at least one of the nodes follows the other one. The DHG model (orange circles) is able to accurately capture the degree distribution (panel \textbf{(a)}), clustering (panel \textbf{(c)}), pairwise distance distribution (panel \textbf{(e)}), betweenness centrality (panel \textbf{(f)}), and edge multiplicity distribution (panel \textbf{(g)}) of the real-world network (blue squares). The network's disassortativity is reproduced qualitatively (panel \textbf{(b)}) but shows quantitative disagreement. We also provide visualizations of the real-world (panel \textbf{(d)}) and deterministic (panel \textbf{(h)}) network, showing $0.35\%$ of edges.

\begin{figure*}[tb]
    \centering
    \includegraphics[width=0.95\textwidth]{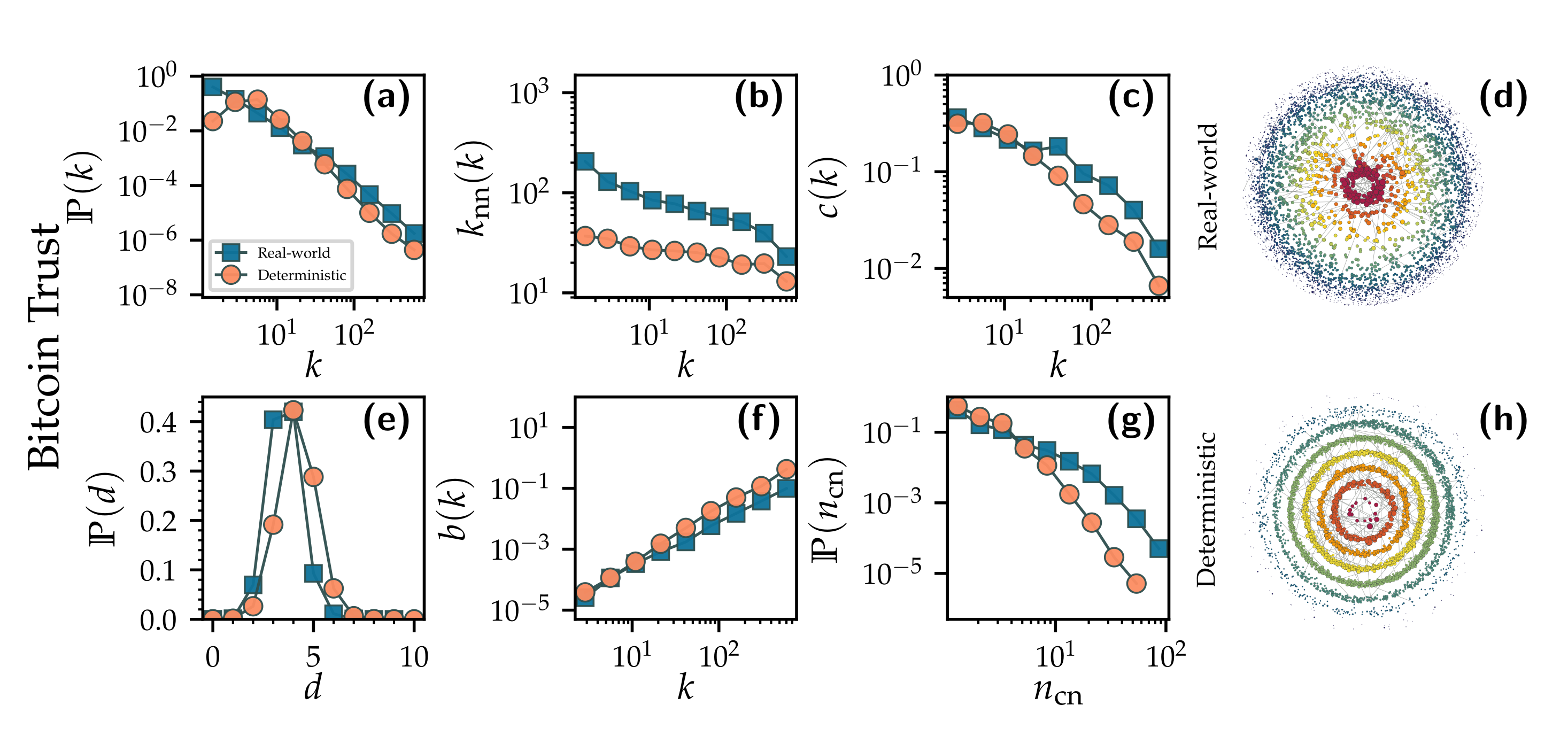}
    \caption{Network properties (degree distribution $\mathbb{P}(k)$ (panel \textbf{(a)}), average nearest neighbor degree $k_\mathrm{nn}(k)$ (panel \textbf{(b)}), local clustering  $c(k)$ (panel \textbf{(c)}), pairwise distance distribution $\mathbb{P}(d)$ (panel \textbf{(e)}), betweenness centrality $b(k)$ (panel \textbf{(f)}), edge multiplicity distribution $\mathbb{P}(n_\mathrm{cn})$ (panel \textbf{(g)})) of the \texttt{bitcoin\_trust} network~\cite{kumar2016_EdgeWeightPrediction} (blue squares) and a DHG graph (orange circles) with matching parameters, as well as artistic representations of the real-world (panel \textbf{(d)}) and deterministic (panel \textbf{(h)}) network.}
    \label{fig:real_world_bitcoin}
\end{figure*}

In the \texttt{bitcoin\_trust} network~\cite{kumar2016_EdgeWeightPrediction} (Fig.~\ref{fig:real_world_bitcoin}) nodes are users of the Bitcoin OTC platform and edges encode trust between them. The DHG model (orange circles) can reproduce many features of the real-world network (blue squares), such as degree distribution (panel \textbf{(a)}), clustering (panel \textbf{(c)}), pairwise distance distribution (panel \textbf{(e)}), betweenness centrality (panel \textbf{(f)}). However, there are discrepancies in the degree-degree correlations (panel \textbf{(b)}) and edge multiplicity distribution (panel \textbf{(g)}). We provide artistic visualizations of the real-world network (panel \textbf{(d)}) and its deterministic counterpart (panel \textbf{(h)}), showing $0.55\%$ of edges.

\subsubsection{Technological Networks}

\begin{figure*}[tb]
    \centering
    \includegraphics[width=0.95\textwidth]{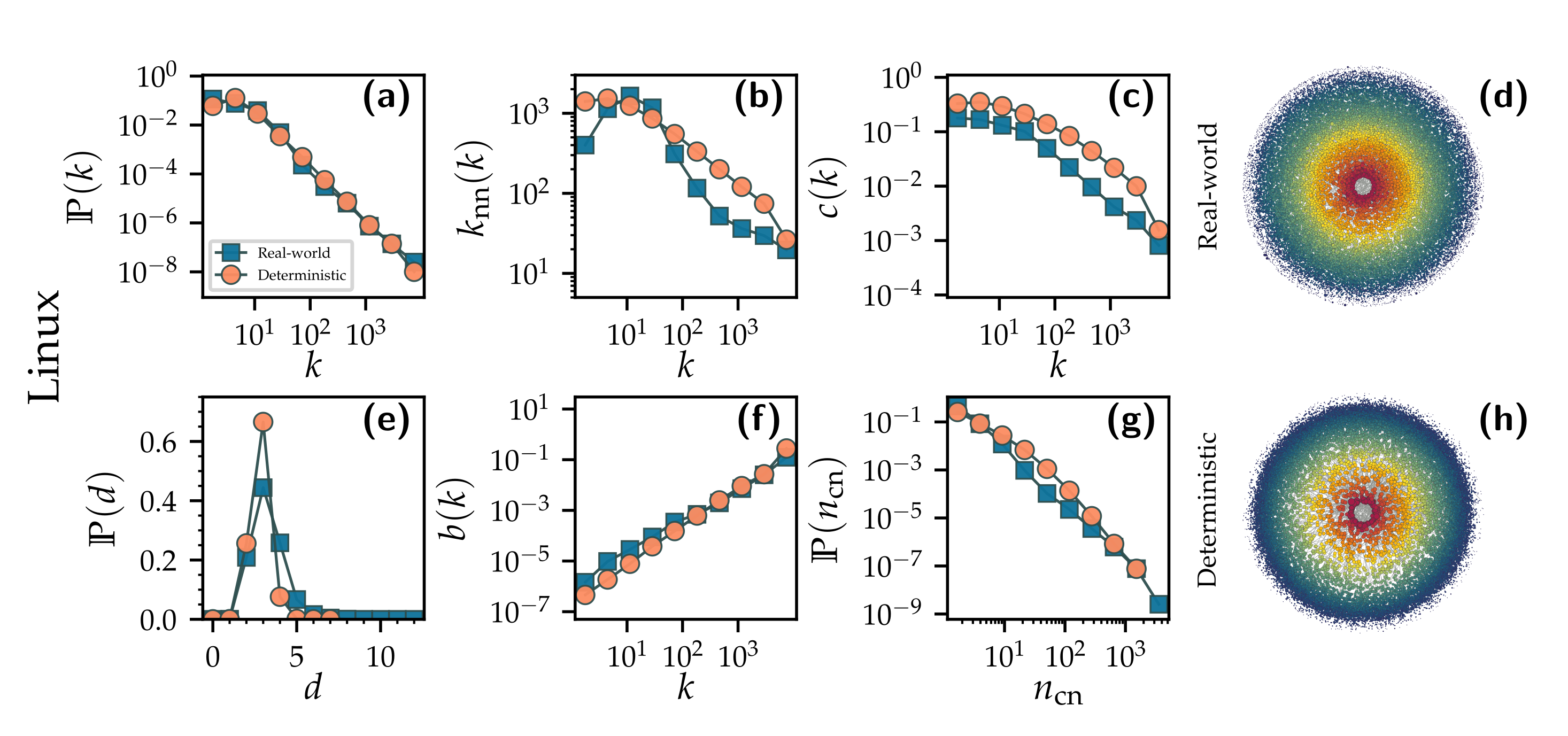}
    \caption{Network properties (degree distribution $\mathbb{P}(k)$ (panel \textbf{(a)}), average nearest neighbor degree $k_\mathrm{nn}(k)$ (panel \textbf{(b)}), local clustering  $c(k)$ (panel \textbf{(c)}), pairwise distance distribution $\mathbb{P}(d)$ (panel \textbf{(e)}), betweenness centrality $b(k)$ (panel \textbf{(f)}), edge multiplicity distribution $\mathbb{P}(n_\mathrm{cn})$ (panel \textbf{(g)})) of the \texttt{linux} network~\cite{kunegis2013_KONECTKoblenznetwork} (blue squares) and a DHG graph (orange circles) with matching parameters, as well as artistic representations of the real-world (panel \textbf{(d)}) and deterministic (panel \textbf{(h)}) network.}
    \label{fig:real_world_linux}
\end{figure*}

In the \texttt{linux} network~\cite{kunegis2013_KONECTKoblenznetwork} (Fig.~\ref{fig:real_world_linux}) nodes are source files of the Linux operating system and there is an edge between two nodes if one of the source files includes the other. A graph from the DHG model (orange circles) with matched parameters reproduces many network properties of the real-world \texttt{linux} network (blue squares) including the degree distribution (panel \textbf{(a)}), average nearest neighbor degree (panel \textbf{(b)}), clustering (panel \textbf{(c)}), pairwise distances (panel \textbf{(e)}), betweenness centrality (panel \textbf{(f)}), and edge multiplicity distribution (panel \textbf{(g)}). We visualize the real-world network in panel \textbf{(d)} and the network from the DHG model in panel \textbf{(h)}, showing $0.25\%$ of edges.

\begin{figure*}[tb]
    \centering
    \includegraphics[width=0.95\textwidth]{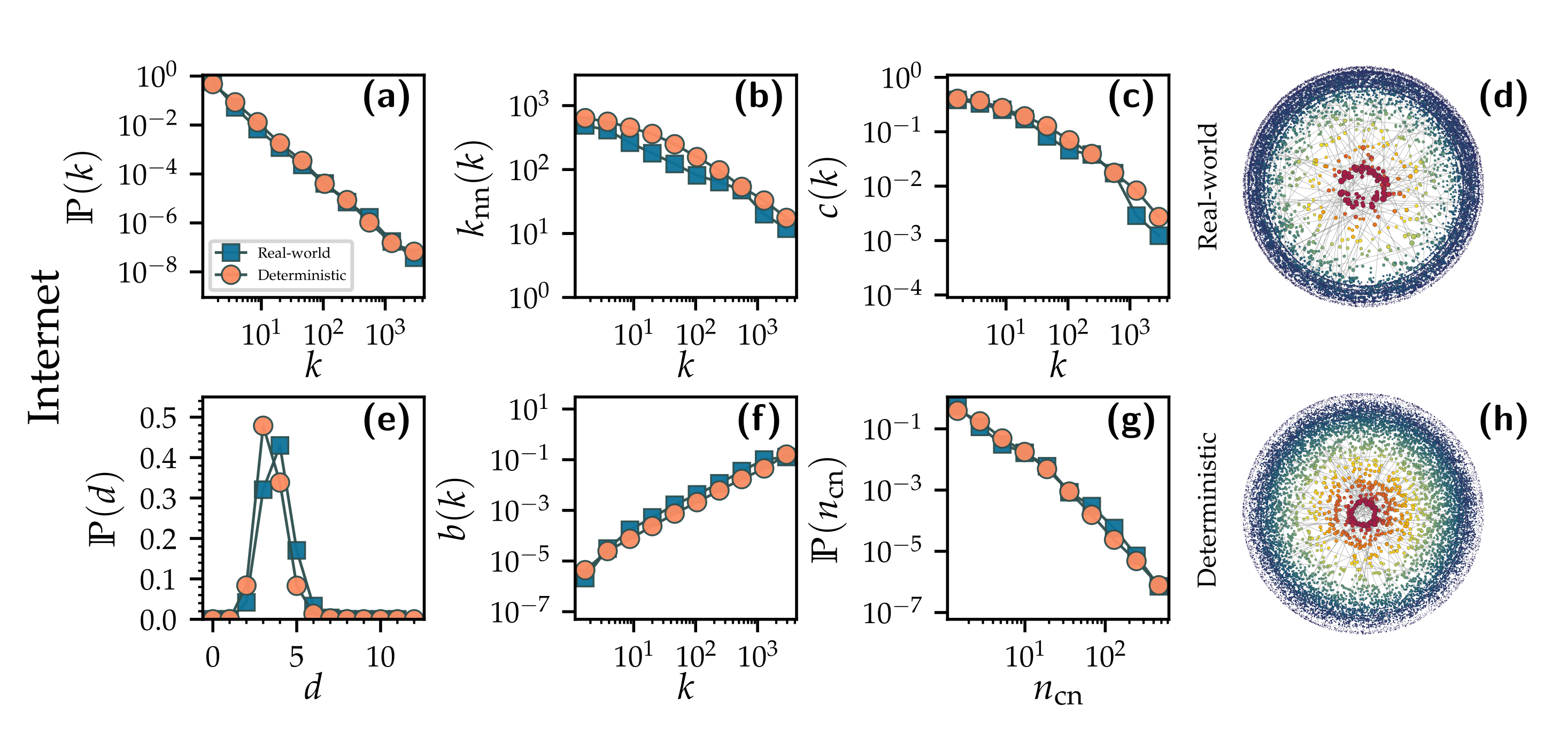}
    \caption{Network properties (degree distribution $\mathbb{P}(k)$ (panel \textbf{(a)}), average nearest neighbor degree $k_\mathrm{nn}(k)$ (panel \textbf{(b)}), local clustering  $c(k)$ (panel \textbf{(c)}), pairwise distance distribution $\mathbb{P}(d)$ (panel \textbf{(e)}), betweenness centrality $b(k)$ (panel \textbf{(f)}), edge multiplicity distribution $\mathbb{P}(n_\mathrm{cn})$ (panel \textbf{(g)})) of the \texttt{internet\_as} network~\cite{karrer2014_percolationsparsenetworks} (blue squares) and a DHG graph (orange circles) with matching parameters, as well as artistic representations of the real-world (panel \textbf{(d)}) and deterministic (panel \textbf{(h)}) network.}
    \label{fig:real_world_internet_as}
\end{figure*}

In the \texttt{internet\_as} network~\cite{karrer2014_percolationsparsenetworks} (Fig.~\ref{fig:real_world_internet_as}) is a map of the internet at the autonomous system level. Nodes are physical routers and two nodes are connected if they can directly exchange information with each other. We observe an excellent agreement between the real-world network (blue squares) and the one generated from the DHG model (orange circles) across many network properties such as the degree distribution (panel \textbf{(a)}), average nearest neighbor degree (panel \textbf{(b)}), clustering (panel \textbf{(c)}), pairwise distances (panel \textbf{(e)}), betweenness centrality (panel \textbf{(f)}), and edge multiplicity distribution (panel \textbf{(g)}). We also provide visualizations of the real-world (panel \textbf{(d)}) and the deterministic network (panel \textbf{(h)}, showing $0.35\%$ of edges.

\begin{figure*}[tb]
    \centering
    \includegraphics[width=0.95\textwidth]{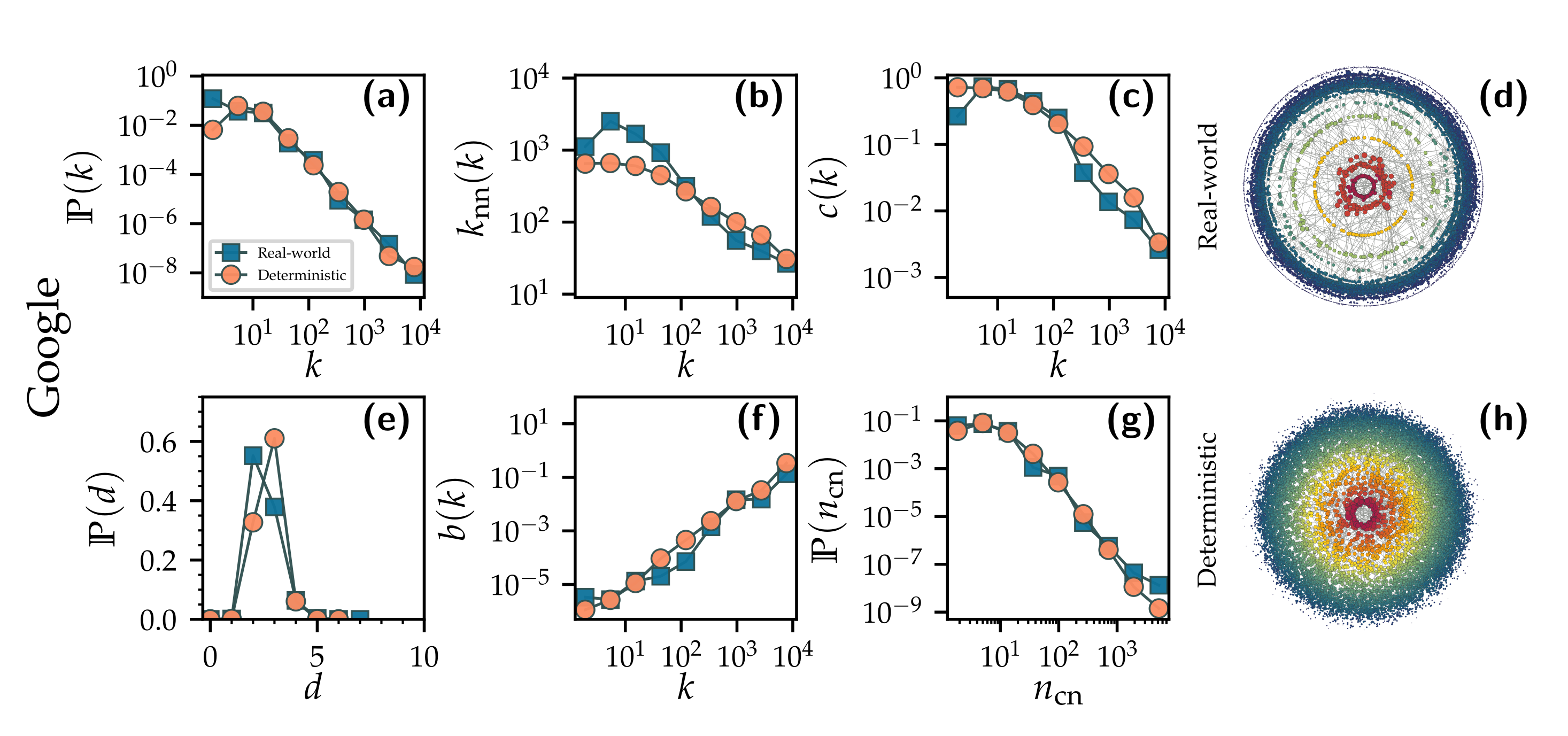}
    \caption{Network properties (degree distribution $\mathbb{P}(k)$ (panel \textbf{(a)}), average nearest neighbor degree $k_\mathrm{nn}(k)$ (panel \textbf{(b)}), local clustering  $c(k)$ (panel \textbf{(c)}), pairwise distance distribution $\mathbb{P}(d)$ (panel \textbf{(e)}), betweenness centrality $b(k)$ (panel \textbf{(f)}), edge multiplicity distribution $\mathbb{P}(n_\mathrm{cn})$ (panel \textbf{(g)})) of the \texttt{google} network~\cite{palla2007_directednetworkmodules} (blue squares) and a DHG graph (orange circles) with matching parameters, as well as artistic representations of the real-world (panel \textbf{(d)}) and deterministic (panel \textbf{(h)}) network.}
    \label{fig:real_world_google}
\end{figure*}

In the \texttt{google} network~\cite{palla2007_directednetworkmodules} (Fig.~\ref{fig:real_world_google}), nodes are google-internal webpages and edges represent hyperlinks between them. For this network dataset a wide range of network properties of the real-world network (blue squares) such as heterogeneous degree distribution (panel \textbf{(a)}), disassortativity (panel \textbf{(b)}), clustering (panel \textbf{(c)}), small-world behavior (panel \textbf{(e)}), betweenness (panel \textbf{(f)}),  and edge multiplicity distribution (panel \textbf{(g)}) are well captured by the DHG model (orange circles). We also visualize the real-world (panel \textbf{(d)}) and deterministic (panel \textbf{(h)}) network, showing $0.25\%$ of edges.

\begin{figure*}[tb]
    \centering
    \includegraphics[width=0.95\textwidth]{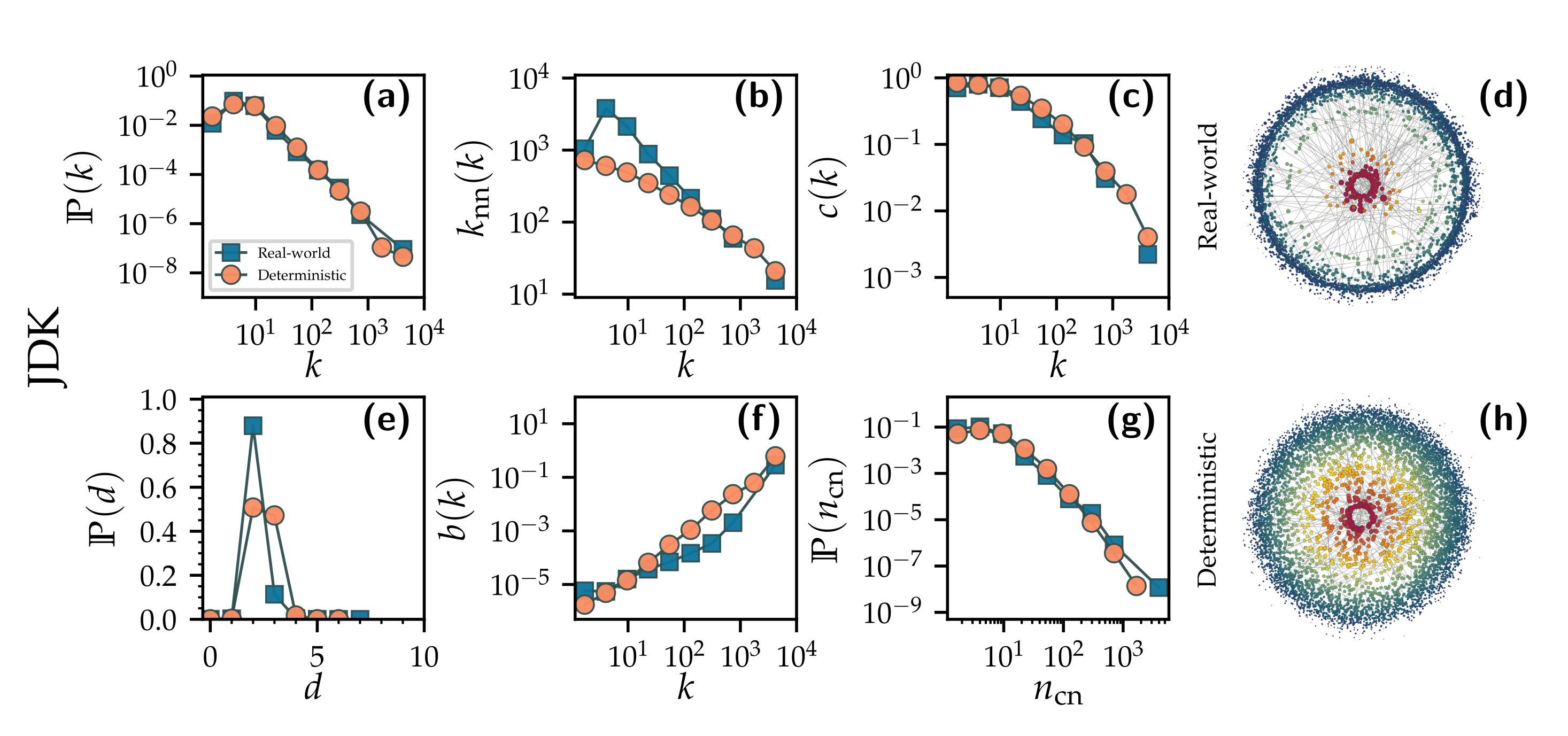}
    \caption{Network properties (degree distribution $\mathbb{P}(k)$ (panel \textbf{(a)}), average nearest neighbor degree $k_\mathrm{nn}(k)$ (panel \textbf{(b)}), local clustering  $c(k)$ (panel \textbf{(c)}), pairwise distance distribution $\mathbb{P}(d)$ (panel \textbf{(e)}), betweenness centrality $b(k)$ (panel \textbf{(f)}), edge multiplicity distribution $\mathbb{P}(n_\mathrm{cn})$ (panel \textbf{(g)})) of the \texttt{jdk} network~\cite{kunegis2013_KONECTKoblenznetwork} (blue squares) and a DHG graph (orange circles) with matching parameters, as well as artistic representations of the real-world (panel \textbf{(d)}) and deterministic (panel \textbf{(h)}) network.}
    \label{fig:real_world_jdk}
\end{figure*}

The \texttt{jdk} network~\cite{kunegis2013_KONECTKoblenznetwork} (Fig.~\ref{fig:real_world_jdk}) represents software dependencies within the JDK framework. Each node in the network is a package and edges encode dependencies. The DHG model (orange circles) captures the real-world (blue squares) networks' broad degree distribution (panel \textbf{(a)}), disassortativity (panel \textbf{(b)}), clustering (panel \textbf{(c)}), small-world behavior (panel \textbf{(e)}), betweenness centrality (panel \textbf{(f)}), and to some extent also the edge multiplicity distribution (panel \textbf{(g)}). We provide visualization of the real-world network (panel \textbf{(d)}) and its deterministic counterpart (panel \textbf{(h)}), showing $0.45\%$ of edges.

\subsubsection{Biological Networks}

\begin{figure*}[tb]
    \centering
    \includegraphics[width=0.95\textwidth]{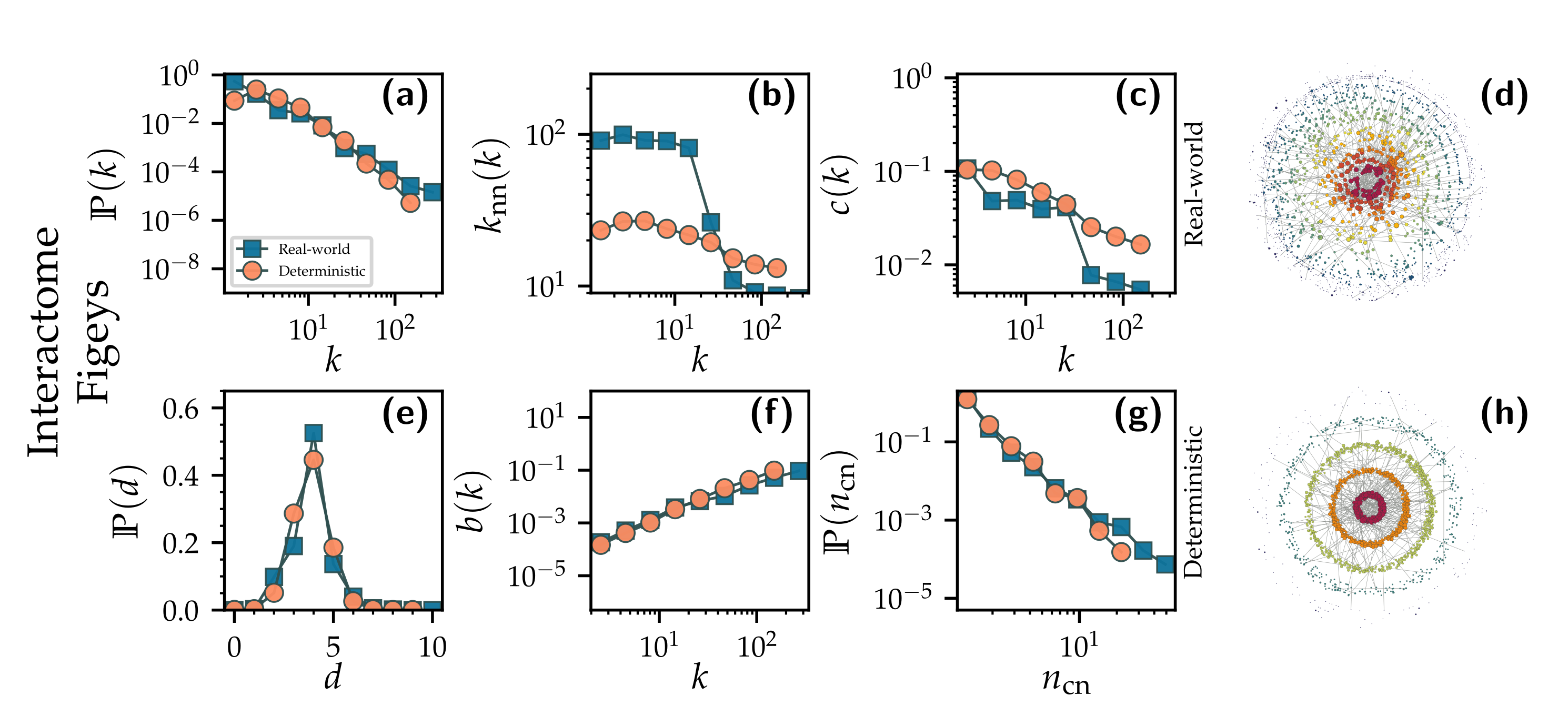}
    \caption{Network properties (degree distribution $\mathbb{P}(k)$ (panel \textbf{(a)}), average nearest neighbor degree $k_\mathrm{nn}(k)$ (panel \textbf{(b)}), local clustering  $c(k)$ (panel \textbf{(c)}), pairwise distance distribution $\mathbb{P}(d)$ (panel \textbf{(e)}), betweenness centrality $b(k)$ (panel \textbf{(f)}), edge multiplicity distribution $\mathbb{P}(n_\mathrm{cn})$ (panel \textbf{(g)})) of the \texttt{interactome\_figeys} network~\cite{ewing2007_largescalemappinghuman} (blue squares) and a DHG graph (orange circles) with matching parameters, as well as artistic representations of the real-world (panel \textbf{(d)}) and deterministic (panel \textbf{(h)}) network.}
    \label{fig:real_world_interactome}
\end{figure*}

The \texttt{interactome\_figeys} network~\cite{ewing2007_largescalemappinghuman} (Fig.~\ref{fig:real_world_interactome}) is a human protein-protein interaction network. Nodes represent proteins, and edges encode binding interactions between them. The DHG model (orange circles) reproduces the degree heterogeneity (panel \textbf{(a)}), distance distribution (panel \textbf{(e)}), and betweenness centrality (panel \textbf{(f)}) of the real-world network (blue squares) well. However, the dependence of average nearest neighbor degree (panel \textbf{(b)}) and clustering (panel \textbf{(c)}) on node degree are less well captured. The model also fails to capture the tail of the edge multiplicity distribution (panel \textbf{(g)}). We visualize the real-world (panel \textbf{(d)}) and deterministic (panel \textbf{(h)}) network, showing $5\%$ of edges.

\begin{figure*}[tb]
    \centering
    \includegraphics[width=0.95\textwidth]{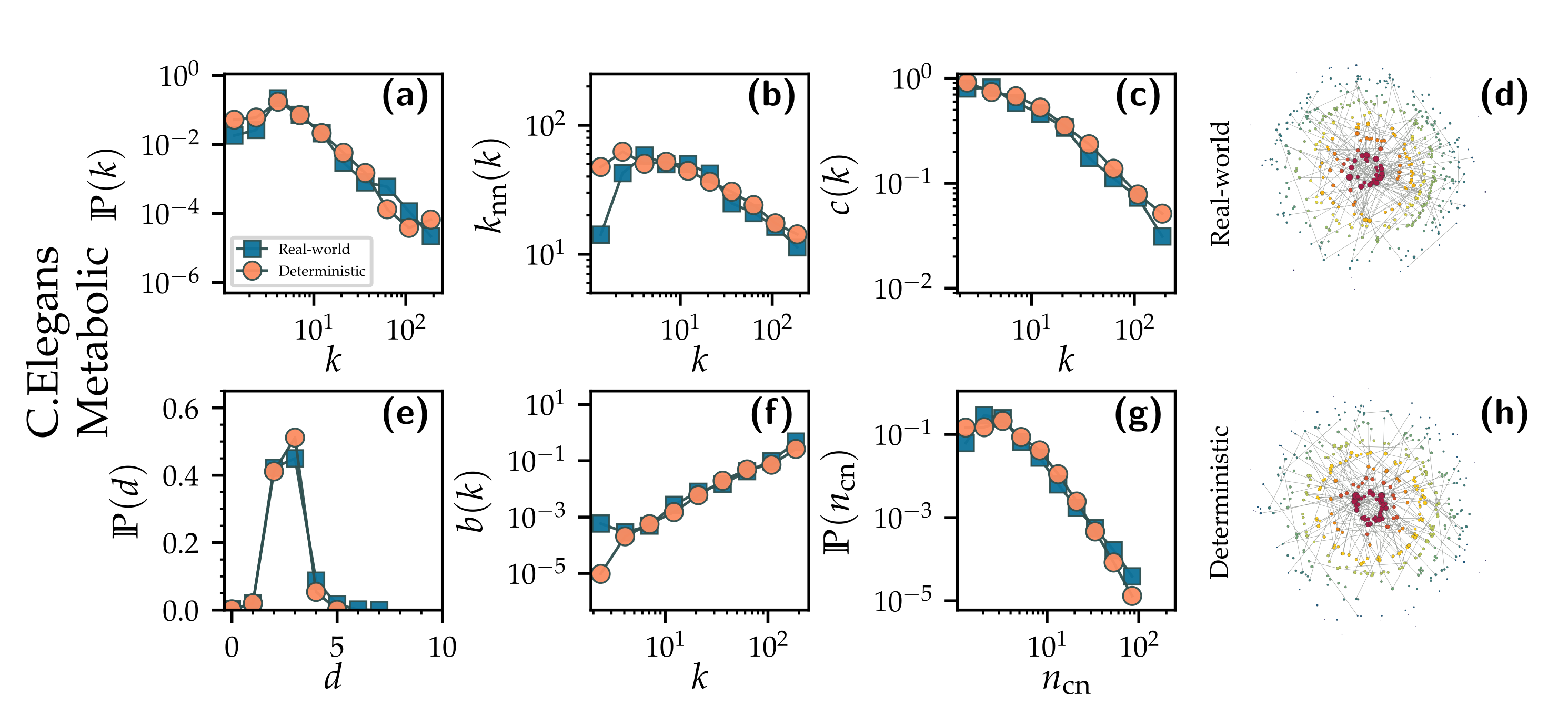}
    \caption{Network properties (degree distribution $\mathbb{P}(k)$ (panel \textbf{(a)}), average nearest neighbor degree $k_\mathrm{nn}(k)$ (panel \textbf{(b)}), local clustering  $c(k)$ (panel \textbf{(c)}), pairwise distance distribution $\mathbb{P}(d)$ (panel \textbf{(e)}), betweenness centrality $b(k)$ (panel \textbf{(f)}), edge multiplicity distribution $\mathbb{P}(n_\mathrm{cn})$ (panel \textbf{(g)})) of the \texttt{celegans\_metabolic} network~\cite{duch2005_communitydetectioncomplex} (blue squares) and a DHG graph (orange circles) with matching parameters, as well as artistic representations of the real-world (panel \textbf{(d)}) and deterministic (panel \textbf{(h)}) network.}
    \label{fig:real_world_celegans_metabolic}
\end{figure*}

The network \texttt{celegans\_metabolic}~\cite{duch2005_communitydetectioncomplex} (Fig.~\ref{fig:real_world_celegans_metabolic}) is the metabolic network of the \textit{caenorhabditis elegans} (\textit{C.~elegans}). The DHG model (orange circles) reproduces many properties of the real-world network (blue squares). This includes the degree distribution (panel \textbf{(a)}), the dependence of average nearest neighbor degree (panel \textbf{(b)}) and clustering (panel \textbf{(c)}) on node degree, as well as the distance distribution (panel \textbf{(e)}), betweenness centrality (panel \textbf{(f)}), and edge multiplicity distribution (panel \textbf{(g)}). We provide visualizations of the real-world (panel \textbf{(d)}) and deterministic (panel \textbf{(h)}) network, showing $10\%$ of edges.

\begin{figure*}[tb]
    \centering
    \includegraphics[width=0.95\textwidth]{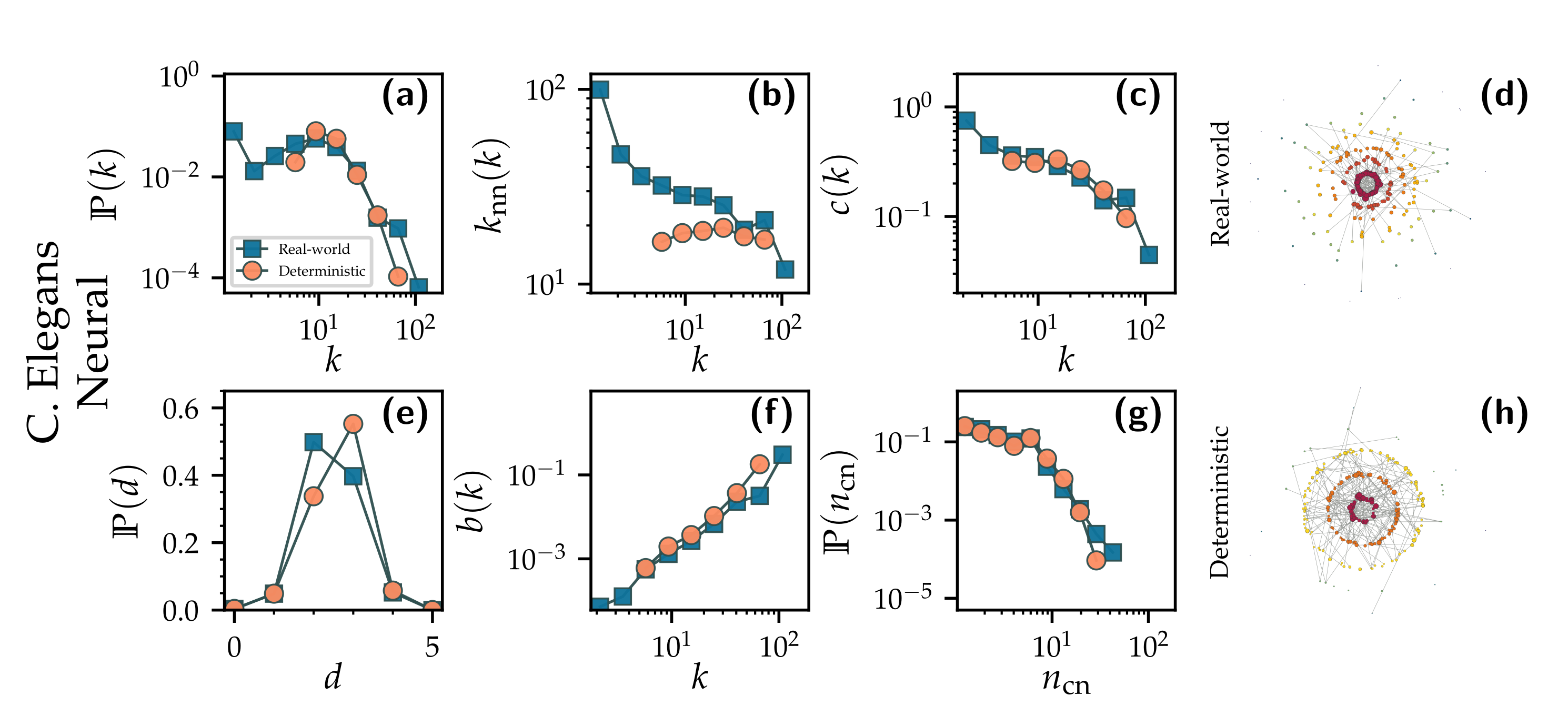}
    \caption{Network properties (degree distribution $\mathbb{P}(k)$ (panel \textbf{(a)}), average nearest neighbor degree $k_\mathrm{nn}(k)$ (panel \textbf{(b)}), local clustering  $c(k)$ (panel \textbf{(c)}), pairwise distance distribution $\mathbb{P}(d)$ (panel \textbf{(e)}), betweenness centrality $b(k)$ (panel \textbf{(f)}), edge multiplicity distribution $\mathbb{P}(n_\mathrm{cn})$ (panel \textbf{(g)})) of the \texttt{celegansneural} network~\cite{white1997_structurenervoussystem, watts1998_wsmodel} (blue squares) and a DHG graph (orange circles) with matching parameters, as well as artistic representations of the real-world (panel \textbf{(d)}) and deterministic (panel \textbf{(h)}) network.}
    \label{fig:real_world_celegansneural}
\end{figure*}

The network \texttt{celegansneural}~\cite{white1997_structurenervoussystem, watts1998_wsmodel} (Fig.~\ref{fig:real_world_celegansneural}) is the neural network of the nematode \textit{C.~elegans}, an important model organism in biology. For this relatively small network ($n=297$) the correspondence between the DHG model (orange circles) and the real-world network (blue squares) is less good: The shape of the degree distribution (panel \textbf{(a)}) differs especially at low degrees. The dependence of the average nearest neighbor degree (panel \textbf{(b)}) and clustering on degree (panel \textbf{(c)}) on node degree are poorly captured. However, the distance distribution (panel \textbf{(e)}), betweenness centrality (panel \textbf{(f)}), and edge multiplicity distribution (panel \textbf{(g)}) are nicely reproduced. We provide artistic visualizations of the real-world (panel \textbf{(d)}) and deterministic (panel \textbf{(h)}) networks, showing $10\%$ of edges.

\subsubsection{Literature \& Words Networks}

\begin{figure*}[tb]
    \centering
    \includegraphics[width=0.95\textwidth]{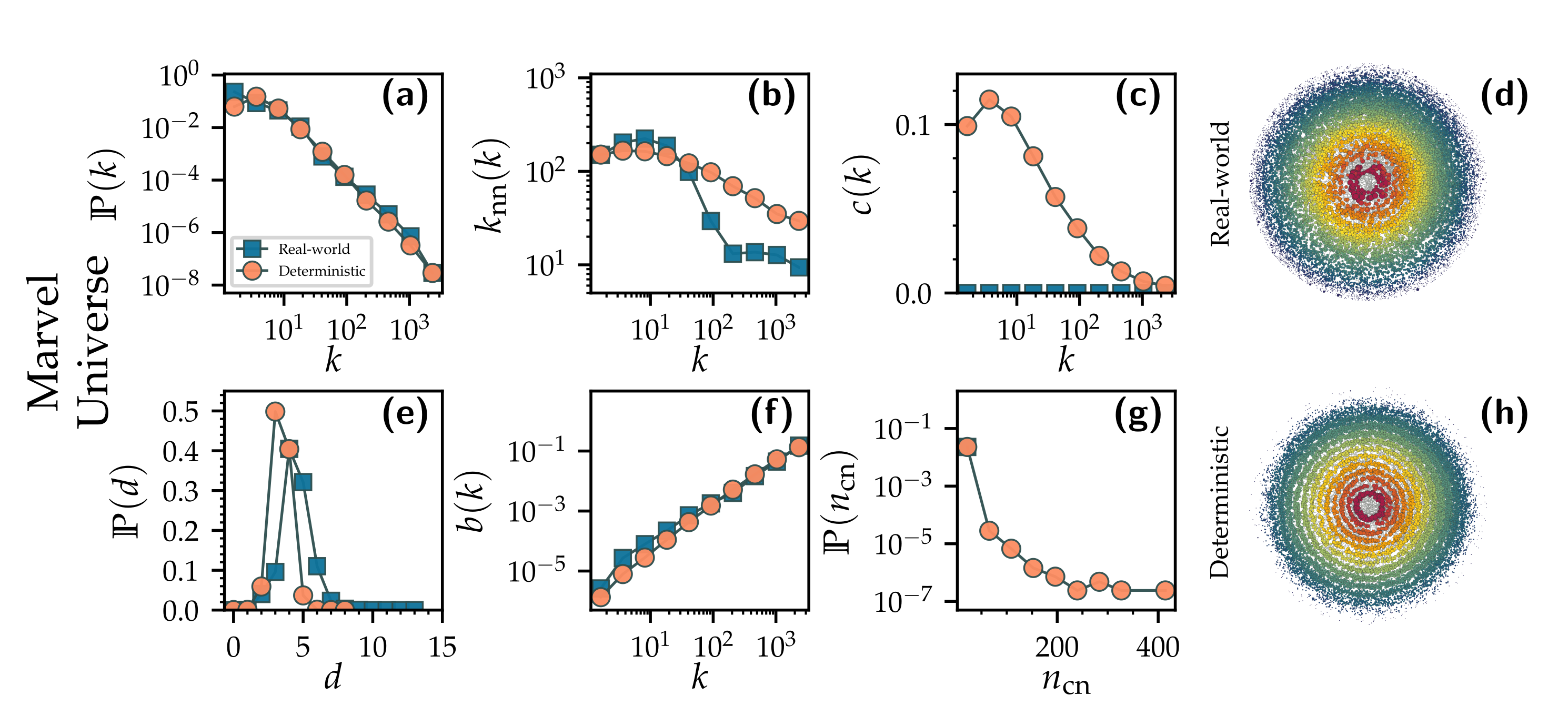}
    \caption{Network properties (degree distribution $\mathbb{P}(k)$ (panel \textbf{(a)}), average nearest neighbor degree $k_\mathrm{nn}(k)$ (panel \textbf{(b)}), local clustering  $c(k)$ (panel \textbf{(c)}), pairwise distance distribution $\mathbb{P}(d)$ (panel \textbf{(e)}), betweenness centrality $b(k)$ (panel \textbf{(f)}), edge multiplicity distribution $\mathbb{P}(n_\mathrm{cn})$ (panel \textbf{(g)})) of the \texttt{marvel\_universe} network~\cite{alberich2002_marveluniverselooks} (blue squares) and a DHG graph (orange circles) with matching parameters, as well as artistic representations of the real-world (panel \textbf{(d)}) and deterministic (panel \textbf{(h)}) network.}\label{fig:real_world_marvel_universe}
\end{figure*}

In the \texttt{marvel\_universe} network~\cite{alberich2002_marveluniverselooks} (Fig.~\ref{fig:real_world_marvel_universe}), nodes are Marvel characters, and edges encode their co-appearance in the same comic book. The DHG model (orange circles) can capture some properties of the real-world network (blue squares) but not others: The degree distribution (panel \textbf{(a)}) and betweenness centrality (panel \textbf{(f)}) are reproduced well. The model also captures the degree-degree correlations (panel \textbf{(b)}) and pairwise distance distribution (panel \textbf{(e)}) to some extent. However, the real-world network seems to be tree-like. As a consequence we observe stark deviations in the clustering (panel \textbf{(c)}) and edge multiplicity distribution (panel \textbf{(g)}). We visualize both the real-world (panel \textbf{(d)}) and deterministic network (panel \textbf{(h)}), showing $0.45\%$ of edges.

\begin{figure*}[tb]
    \centering    \includegraphics[width=0.95\textwidth]{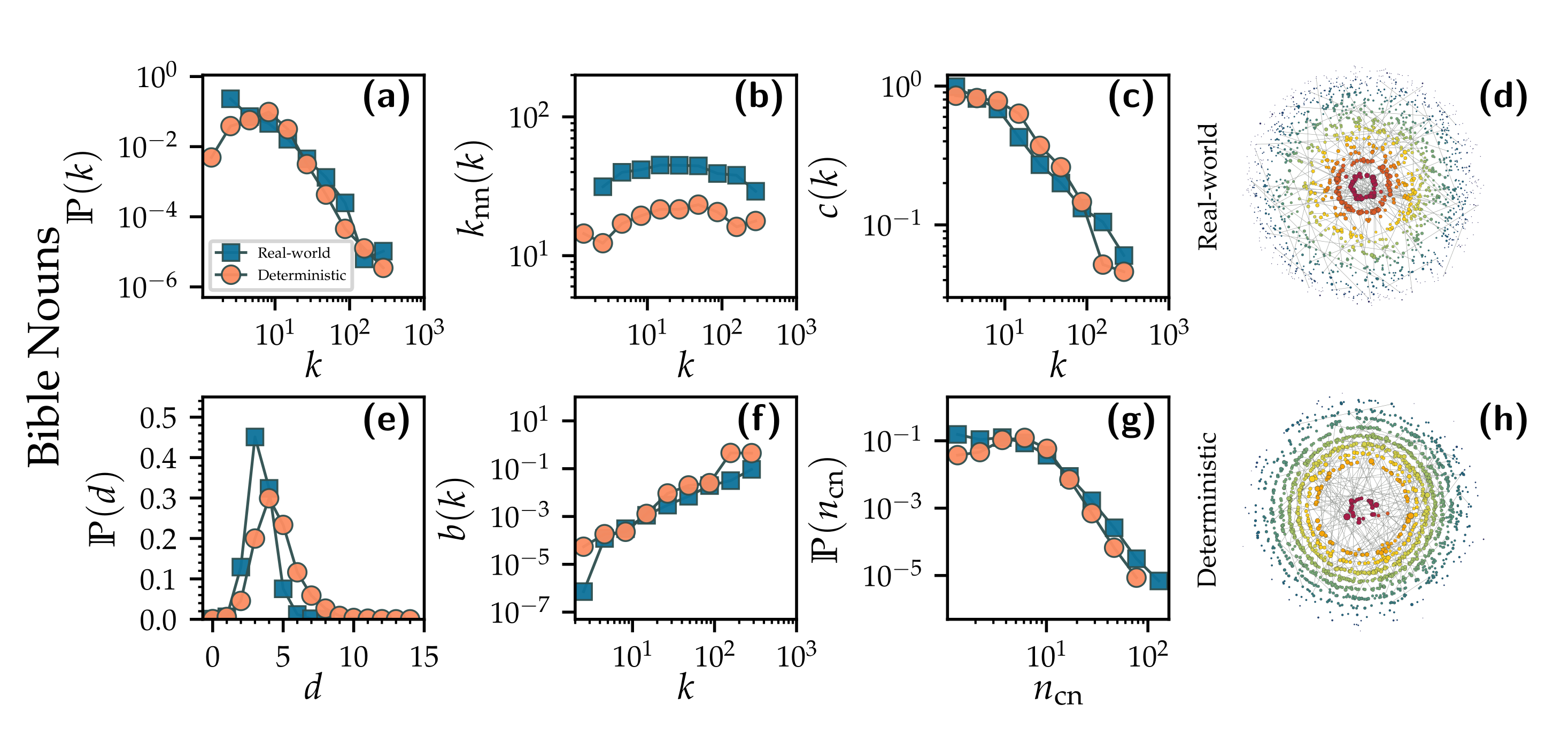}
    \caption{Network properties (degree distribution $\mathbb{P}(k)$ (panel \textbf{(a)}), average nearest neighbor degree $k_\mathrm{nn}(k)$ (panel \textbf{(b)}), local clustering  $c(k)$ (panel \textbf{(c)}), pairwise distance distribution $\mathbb{P}(d)$ (panel \textbf{(e)}), betweenness centrality $b(k)$ (panel \textbf{(f)}), edge multiplicity distribution $\mathbb{P}(n_\mathrm{cn})$ (panel \textbf{(g)})) of the \texttt{bible\_nouns} network~\cite{peixoto2020_netzschleudernetworkcatalogue} (blue squares) and a DHG graph (orange circles) with matching parameters, as well as artistic representations of the real-world (panel \textbf{(d)}) and deterministic (panel \textbf{(h)}) network.}
    \label{fig:real_world_bible_nouns}
\end{figure*}

In the \texttt{bible\_nouns} network~\cite{peixoto2020_netzschleudernetworkcatalogue} (Fig.~\ref{fig:real_world_bible_nouns}), nodes are nouns, and edges encode their co-appearance in the same Bible verse. The real-world network (blue squares) is in many aspects well described by the DHG model (orange circles): This includes the degree distribution (panel \textbf{(a)}), absence of major degree correlations (panel \textbf{(b)}), clustering (panel \textbf{(c)}), small-worldness (panel \textbf{(e)}), betweenness (panel \textbf{(f)}), and edge multiplicity distribution (panel \textbf{(g)}). We visualize the real-world (panel \textbf{(d)}) and deterministic (panel \textbf{(h)}) network, showing $2\%$ of edges.

\begin{figure*}[tb]
    \centering\includegraphics[width=0.95\textwidth]{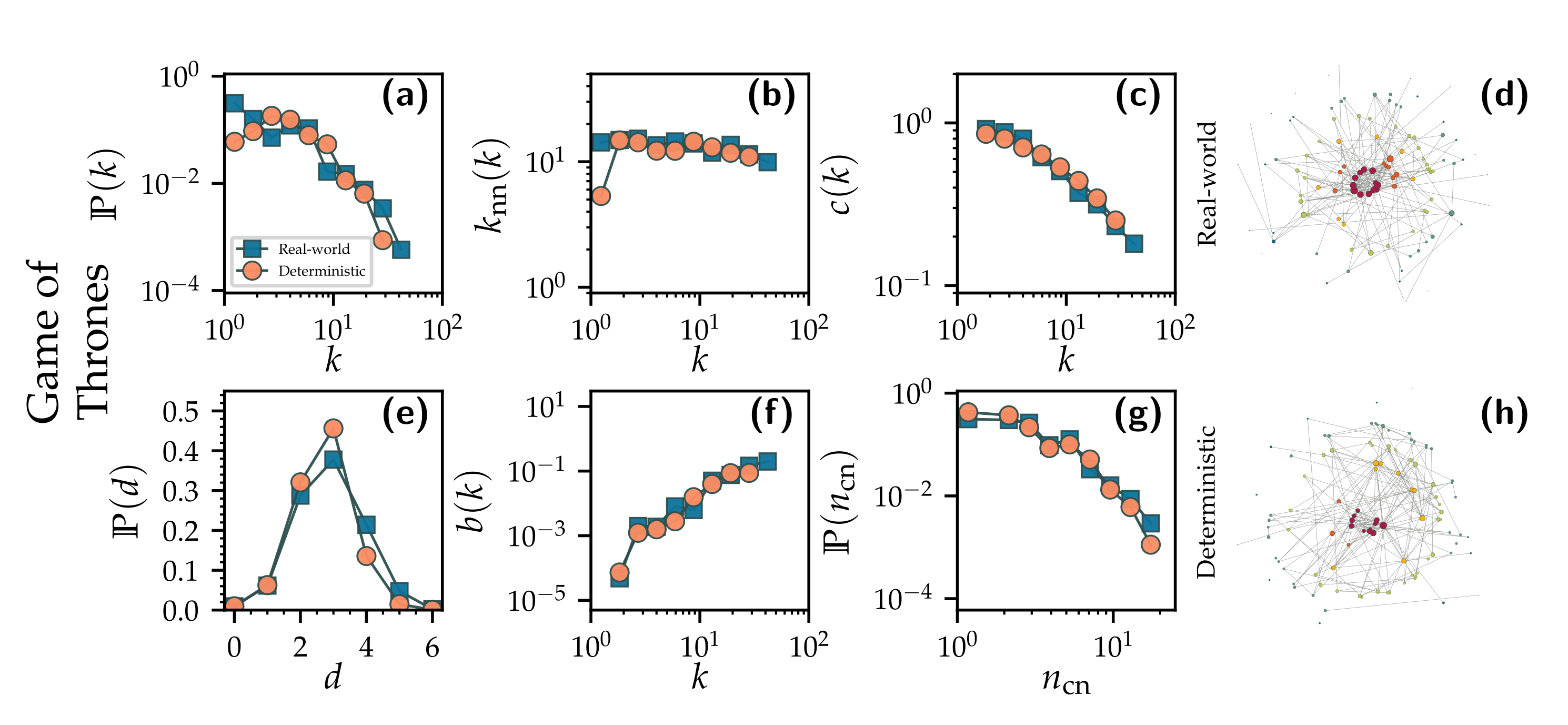}
    \caption{Network properties (degree distribution $\mathbb{P}(k)$ (panel \textbf{(a)}), average nearest neighbor degree $k_\mathrm{nn}(k)$ (panel \textbf{(b)}), local clustering  $c(k)$ (panel \textbf{(c)}), pairwise distance distribution $\mathbb{P}(d)$ (panel \textbf{(e)}), betweenness centrality $b(k)$ (panel \textbf{(f)}), edge multiplicity distribution $\mathbb{P}(n_\mathrm{cn})$ (panel \textbf{(g)})) of the \texttt{game\_thrones} network~\cite{beveridge2016_networkthrones} (blue squares) and a DHG graph (orange circles) with matching parameters, as well as artistic representations of the real-world (panel \textbf{(d)}) and deterministic (panel \textbf{(h)}) network.}
    \label{fig:real_world_game_thrones}
\end{figure*}

In the \texttt{game\_thrones} network~\cite{beveridge2016_networkthrones} (Fig.~\ref{fig:real_world_game_thrones}), nodes are characters in the novel \textit{A Storm of Swords} from the Game of Thrones franchise, and edges encode that both characters are mentioned within 15 words of each other. Given the small size of this network ($n=107$), we observe a very good correspondence between the real-world (blue squares) and deterministic (orange circles) networks in terms of their network properties including degree distribution (panel \textbf{(a)}), absence of degree-degree correlations (panel \textbf{(b)}), clustering (panel \textbf{(c)}), pairwise distance distribution (panel \textbf{(e)}), betweenness centrality (panel \textbf{(f)}), and edge multiplicity distribution (panel \textbf{(g)}). We visualize both the real-world (panel \textbf{(d)}) and deterministic (panel \textbf{(h)}) networks, showing $50\%$ of edges.

\subsubsection{Transportation Networks}

\begin{figure*}[tb]
    \centering
    \includegraphics[width=0.95\textwidth]{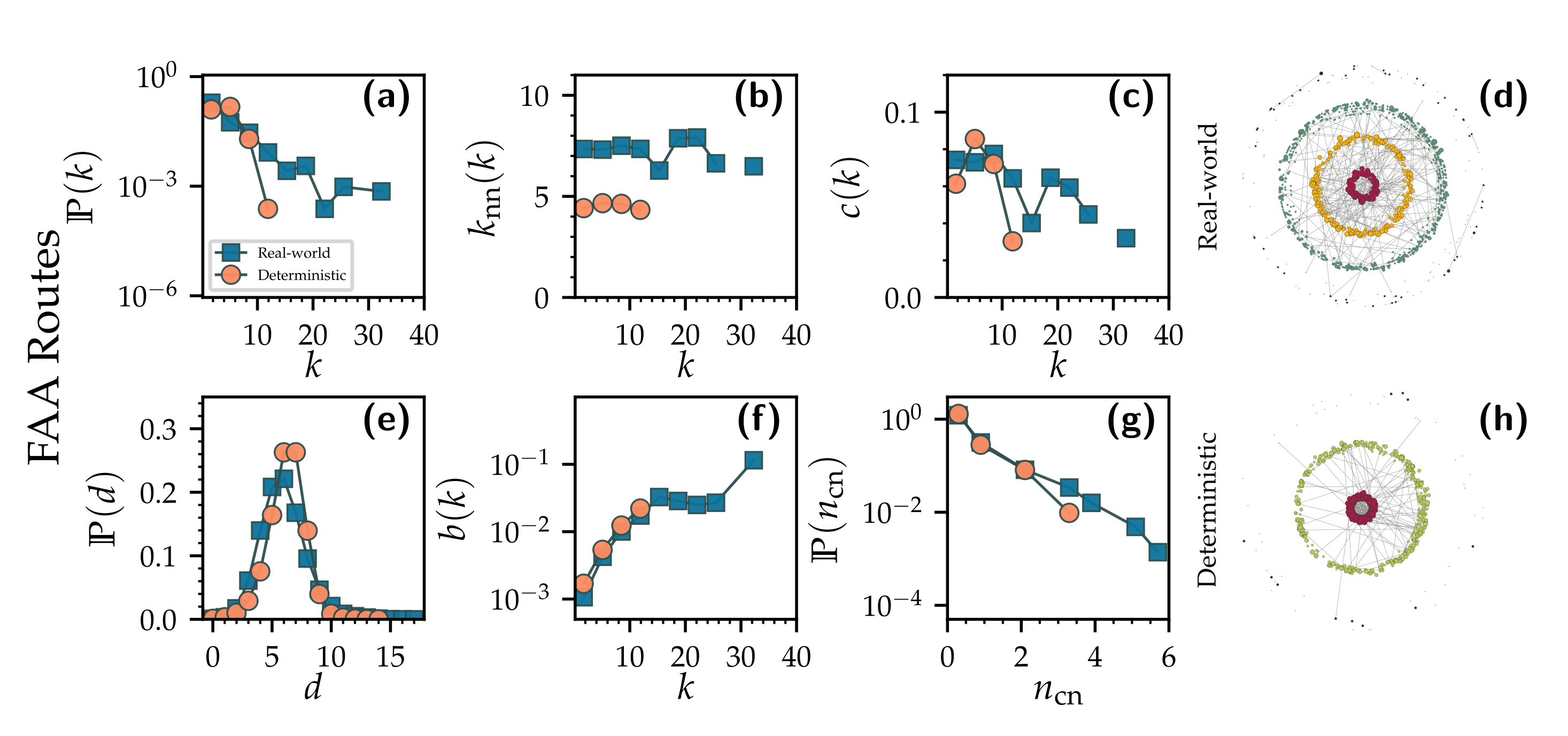}
    \caption{Network properties (degree distribution $\mathbb{P}(k)$ (panel \textbf{(a)}), average nearest neighbor degree $k_\mathrm{nn}(k)$ (panel \textbf{(b)}), local clustering  $c(k)$ (panel \textbf{(c)}), pairwise distance distribution $\mathbb{P}(d)$ (panel \textbf{(e)}), betweenness centrality $b(k)$ (panel \textbf{(f)}), edge multiplicity distribution $\mathbb{P}(n_\mathrm{cn})$ (panel \textbf{(g)})) of the \texttt{faa\_routes} network~\cite{peixoto2020_netzschleudernetworkcatalogue} (blue squares) and a DHG graph (orange circles) with matching parameters, as well as artistic representations of the real-world (panel \textbf{(d)}) and deterministic (panel \textbf{(h)}) network.}
    \label{fig:real_world_faa_routes}
\end{figure*}

In the \texttt{faa\_routes}~\cite{peixoto2020_netzschleudernetworkcatalogue} dataset (Fig.~\ref{fig:real_world_faa_routes}), nodes are airports, and edges encode the preferred routes between them. The degree distribution (panel \textbf{(a)}) of the real-world network (blue squares) lacks heavy tails and is less well described by the DHG model (orange circles). Similarly, the degree-correlations (panel \textbf{(b)}), clustering (panel \textbf{(c)}), and betweenness centrality (panel \textbf{(f)}) are only roughly reproduced. The models edge multiplicity distribution (panel \textbf{(g)}) is slightly less heterogeneous. However, the pairwise distance distribution (panel \textbf{(e)}) is accurately captures by the model. We also show artistic visualizations of the real-world (panel \textbf{(d)}) and deterministic (panel \textbf{(h)}) network, showing $10\%$ of edges.

\begin{figure*}[tb]
    \centering
    \includegraphics[width=0.95\textwidth]{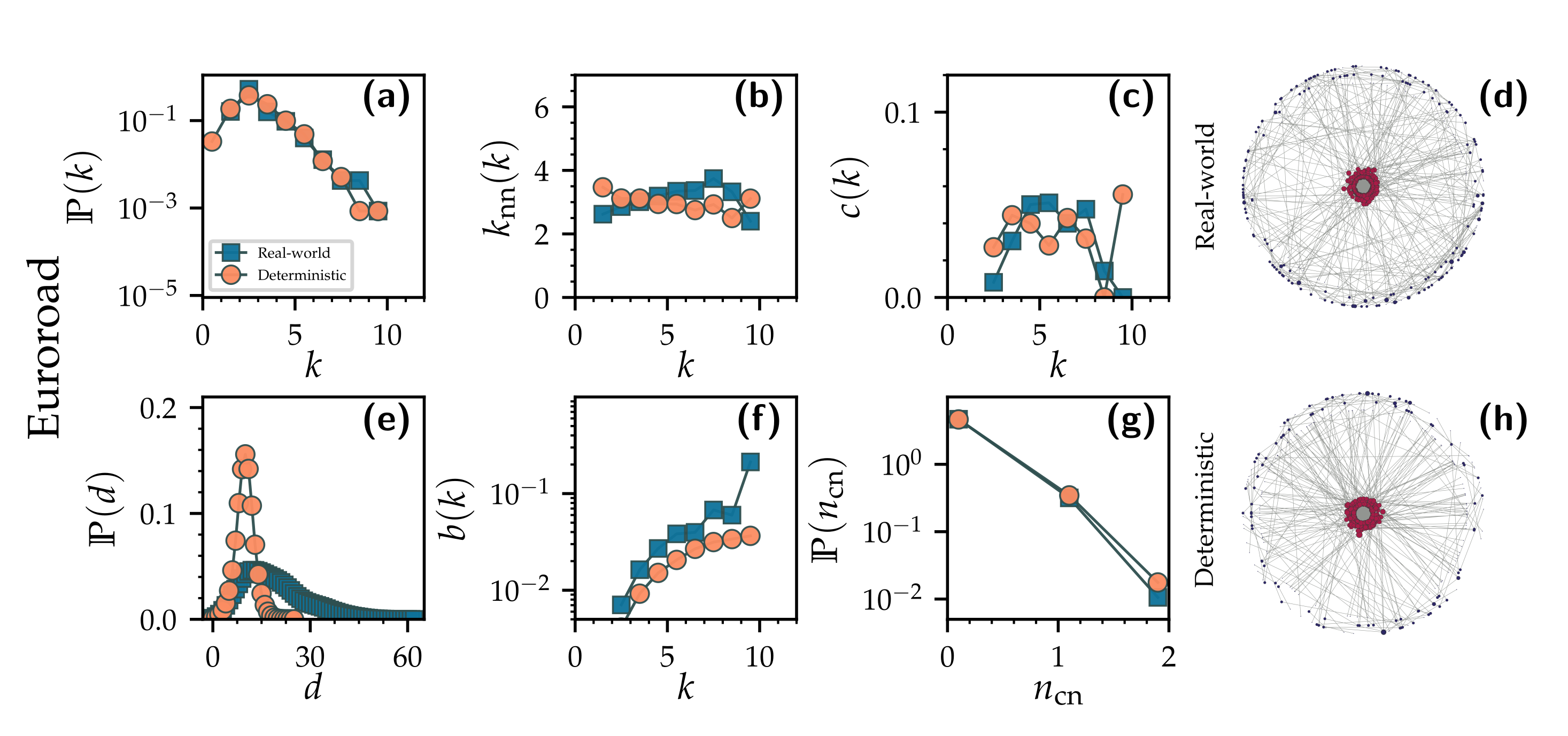}
    \caption{Network properties (degree distribution $\mathbb{P}(k)$ (panel \textbf{(a)}), average nearest neighbor degree $k_\mathrm{nn}(k)$ (panel \textbf{(b)}), local clustering  $c(k)$ (panel \textbf{(c)}), pairwise distance distribution $\mathbb{P}(d)$ (panel \textbf{(e)}), betweenness centrality $b(k)$ (panel \textbf{(f)}), edge multiplicity distribution $\mathbb{P}(n_\mathrm{cn})$ (panel \textbf{(g)})) of the \texttt{euroroad} network~\cite{subelj2011_robustnetworkcommunit} (blue squares) and a DHG graph (orange circles) with matching parameters, as well as artistic representations of the real-world (panel \textbf{(d)}) and deterministic (panel \textbf{(h)}) network.}
    \label{fig:real_world_euroroads}
\end{figure*}

In the \texttt{euroroads} network~\cite{subelj2011_robustnetworkcommunit} (Fig.~\ref{fig:real_world_euroroads}), nodes are cities, and edges encode international E-roads between them. The DHG model (orange circles) can reproduce some aspects of the real-world network (blue squares), e.g., the degree distribution (panel \textbf{(a)}), absence of degree correlations (panel \textbf{(b)}), and the edge multiplicity distribution (panel \textbf{(g)}). The model does not capture the clustering (panel \textbf{(c)}) and the betweenness centrality (panel \textbf{(f)}) well. Also the distance distribution is very different (panel \textbf{(e)}). This is expected as the underlying metric space that governs connections in the road network is Euclidean not hyperbolic. We visualize the real-world (panel \textbf{(d)}) and deterministic (panel \textbf{(h)}) network, showing all edges.

\begin{figure*}[tb]
    \centering
    \includegraphics[width=0.95\textwidth]{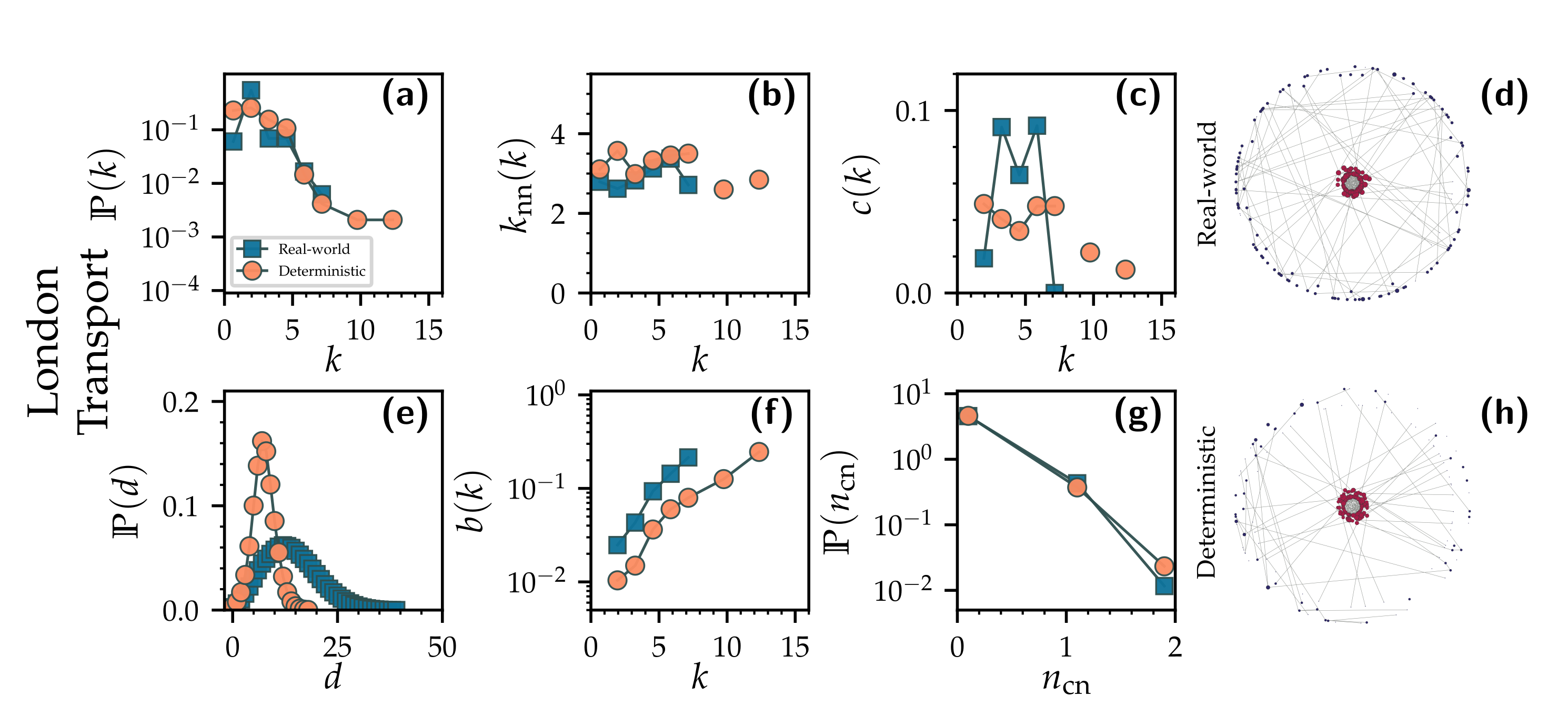}
    \caption{Network properties (degree distribution $\mathbb{P}(k)$ (panel \textbf{(a)}), average nearest neighbor degree $k_\mathrm{nn}(k)$ (panel \textbf{(b)}), local clustering  $c(k)$ (panel \textbf{(c)}), pairwise distance distribution $\mathbb{P}(d)$ (panel \textbf{(e)}), betweenness centrality $b(k)$ (panel \textbf{(f)}), edge multiplicity distribution $\mathbb{P}(n_\mathrm{cn})$ (panel \textbf{(g)})) of the \texttt{london\_transport} network~\cite{dedomenico2014_navigabilityinterconnectednetworks} (blue squares) and a DHG graph (orange circles) with matching parameters, as well as artistic representations of the real-world (panel \textbf{(d)}) and deterministic (panel \textbf{(h)}) network.}
    \label{fig:real_world_london_transport}
\end{figure*}

In the \texttt{london\_transport} network~\cite{dedomenico2014_navigabilityinterconnectednetworks} (Fig.~\ref{fig:real_world_london_transport}), nodes are public transportation stations in London, and edges encode connections between them via one of three means of transportation. The DHG model (orange circles) appears to be a poor model of the real-world network (blue squares): While it can capture qualitative features of the degree distribution (panel \textbf{(a)}), average nearest neighbor degree (panel \textbf{(b)}), clustering (panel \textbf{(c)}), betweenness centrality (panel \textbf{(f)}), and edge multiplicity distribution (panel \textbf{(g)}), there are substantial quantitative difference. Moreover, the pairwise distance distribution (panel \textbf{(e)}) of the real-world network differs from the one under the model. We provide visualizations of the real-world (panel \textbf{(d)}) and deterministic (panel \textbf{(h)}) network, showing $50\%$ of edges.

\section{Matching sufficient statistics in RHGs with atypical coordinates}\label{sec:atypical-ppp}
In this section, we provide an example of how RHGs' sufficient statistics can be set to their peak values with an atypical assignment of coordinates in an atypical graph. This example illustrates the importance of typicality of our PPP derandomization in Sec.~\ref{sec: hypercanonical models}. 

Let the expected number of links in an RHG be $\bar{m}=n-1$, so the most typical graph must have $m^\star=n-1$ links as well. Yet let us consider an atypical graph with $n-1$ links, an $n$-star with the hub at node~$1$ whose adjacency matrix is $G_{1j}=1$ for $j=2,3,\ldots,n$, and $G_{ij}=0$ for any $i\neq1$.

We now outline an atypical coordinate assignment that will match exactly any desired value of the second sufficient statistic, energy~$\bar{\e}$, in this atypical graph:
\begin{enumerate}
  \item fix the coordinates of node~$1$ to the origin, $(\tilde{x}_{1}, \tilde{y}_{1}) = (0,0)$,
  \item set the $\tilde{y}$-coordinates (see Appendix~\ref{sec:DHG}) of all other nodes $i\in[2,3,\cdots, n]$ in to lattice values $\tilde{y}_i= (i-0.5)/n$.
  \item determine $i$'s $\kappa_i$ as in DHGs, i.e., according to~\eqref{eq:DHG-kappa}, and finally,
  \item set each $\tilde{x}_j$ for $j=2,3,\ldots,n$ to such values that energy $\e_{1j}$ of all links in the graph---given by~\eqref{eq: S1 connection probability}---is the same~$\e_{1j}=\bar{\e}/(n-1)$.
\end{enumerate}

\begin{figure}
    \centerline{\includegraphics[width=0.97\columnwidth]{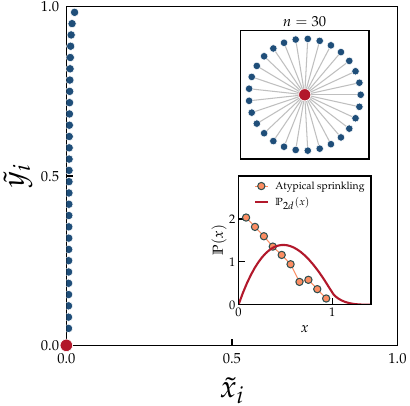}}
    \caption{\textbf{Atypical sprinkling.} A realization of the atypical node sprinkling procedure described in the text for $n=30$, $\bar{m}=29$, $\beta=4$, and $\gamma = 4$. The inset compares the interpoint distance distribution in this sprinkling with that of the PPP in Eq.~\eqref{eq: distance distribution pdfs}.
    }
    \label{fig:atypical sprinkling}
\end{figure}

This coordinates assignment matches~$\bar{\e}$ exactly, but it is extremely atypical, as can be seen in Fig.~\ref{fig:atypical sprinkling}. It definitely does not look---cf.~Fig.~\ref{fig:distance_distribution}---like a typical sample from the PPP on the unit square, since the points are not evenly spread out through the square but squeezed to its left border, resulting in an interpoint distance distribution which is far from the distance distribution between two random points on the unit square. Yet this atypical sprinkling matches exactly the sufficient statistics values in our atypical graph.

\FloatBarrier
\balance

\end{document}